\documentclass[twocolumn]{emulateapj}

\usepackage{subfigure}

\usepackage[usenames,dvipsnames,svgnames,table]{xcolor}
\usepackage{graphicx,color,soul}
\usepackage{latexsym}
\usepackage{amsmath,amssymb}      
\usepackage[draft=false]{hyperref}
\hypersetup{
     colorlinks   = true,
     citecolor    = blue
}

\def\apj{ApJ}
\def\apjl{ApJ}
\def\apjs{ApJS}
\def\apss{Ap\&SS}
\def\aap{A\&A}
\def\jcap{J. Cosmology Astropart. Phys.}
\def\mnras{MNRAS}
\def\prc{Phys.~Rev.~C}
\def\prd{Phys.~Rev.~D}

\begin{document}

\title{Numerical General Relativistic MHD with Magnetically Polarized Matter}

\author{Oscar M. Pimentel}

\author{F. D. Lora-Clavijo}

\author{Guillermo A. Gonz\'alez}

\affiliation{Grupo de Investigaci\'on en Relatividad y Gravitaci\'on,
Escuela de F\'isica, Universidad Industrial de Santander,
\\A. A. 678, Bucaramanga 680002, Colombia.}

\email{OMP: oscar2127821@correro.uis.edu.co;  \\ FDLC: fadulora@uis.edu.co; \\ GAG: guillermo.gonzalez@saber.uis.edu.co}

\date{\today}

\begin{abstract}

The magnetically polarized matter in astrophysical systems may be relevant in some magnetically dominated regions. For instance, the funnel that is generated in some highly magnetized disks configurations whereby relativistic jets are thought to spread, or in pulsars where the fluids are subject to very intense magnetic fields. With the aim of dealing with magnetic media in the astrophysical context, we present for the first time the conservative form of the ideal general relativistic magnetohydrodynamics (GRMHD) equations with a non-zero magnetic polarization vector $m^{\mu}$. Then, we follow the Anile method to compute the eigenvalue structure in the case where the magnetic polarization is parallel to the magnetic field, and it is parametrized by the magnetic susceptibility $\chi_m$. This approximation allows us to describe diamagnetic fluids, for which $\chi_m<0$, and paramagnetic fluids where $\chi_m>0$. The theoretical results were implemented in the CAFE code to study the role of the magnetic polarization in some 1D Riemann problems. We found that independently of the initial condition, the first waves that appear in the numerical solutions are faster in diamagnetic materials than in paramagnetic ones. Moreover, the constant states between the waves change notably for different magnetic susceptibilities. All these effects are more appreciable if the magnetic pressure is much bigger than the fluid pressure. Additionally, with the aim of analysing a magnetic media in a strong gravitational field, we carry out for the first time the magnetized Michel accretion of a magnetically polarized fluid. With this test, we found that the numerical solution is effectively maintained over time ($t>4000$), and that the global convergence of the code is $\gtrsim$ 2 for $\chi_{m}\lesssim 0.005$, for all the magnetic field strength $\beta$  we considered. Finally, when $\chi_m=0.008$ and $\beta\geq 10$, the global convergence of the code is reduced to a value between first and second order. 
\end{abstract}


\maketitle

\section{Introduction}

Nowadays, it is widely known that the strong magnetic and gravitational fields are fundamental for giving a correct description of some systems involving compact objects. For example, it is believed that an accreting fluid onto a black hole is the main power source in active galactic nuclei and microquasars (\cite{2002apa..book.....F}). In those systems it can be observed highly collimated radio jets, which suggests the existence of strong magnetic fields (\cite{1977MNRAS.179..433B, 2008ApJ...678.1180B, 2005ApJ...620..878D}). Additionally, the interaction of the magnetic field and the differential rotation of the disk produces turbulence via the magnetorotational instability (MRI) (\cite{1991ApJ...376..214B}). This turbulence acts on the disk as an effective viscosity that transports angular momentum and dissipates energy. Such a mechanism is fundamental for explaining how the accretion process begins and maintains in time. The first observations of the structure of those magnetic fields near the event horizon have been reported recently in \cite{2015Sci...350.1242J}, by resolving the linearly polarized emission of Sagittarius $A^{*}$. On the other hand, some models that try to explain the central engine of the Gamma-Ray Burst include a millisecond-highly-magnetized pulsar or magnetar, whose magnetic field is about $\sim 10^{14}$ G (\cite{2001ApJ...552L..35Z}). 

Now, as it was pointed out by \cite{2008LRR....11....7F}, some high-energy astrophysical processes, involving the dynamics of a fluid around a compact object, can be successfully modeled by the general relativistic ideal magnetohydrodynamics (GRMHD) in the test fluid approximation. In this theoretical context, the gravitational field of the fluid is neglected in comparison with that of the compact object, and the fluid is considered as a perfect conductor (\cite{1989rfmw.book.....A}). Nevertheless, since the GRMHD is described by a non-linear, time-dependent, and multidimensional system of equations, it is very difficult to find analytic solutions (\cite{2003ApJ...589..444G}). Indeed, the only solution with a magnetic field in the context of the relativistic accretion disk theory was obtained by \cite{2006MNRAS.368..993K} and describe the stationary state of a non-self-gravitating disk-like fluid with a toroidal magnetic field in the Kerr spacetime. Fortunately, nowadays we can study realistic astrophysical systems by solving numerically the GRMHD equations. For example, \cite{2003ApJ...589..444G} simulated the accretion of a toroidal disk with an {\em ad hoc} poloidal magnetic field and found that the MRI increases the magnetic strength sufficiently to distort the disk and droop the fluid into the black hole. In \cite{2017PhRvL.119w1102S}, the authors carried out three-dimensional GRMHD simulations of toroidal accretion disks, which resulted from the merger of neutron stars. They found that in the state where the heating due to MRI turbulence is balanced by the neutrino cooling, the disk outflows launch more mass than those simulations with hydrodynamical $\alpha$-viscosity. These results strongly suggest that the outflows of a post-merger disk are possible scenarios for the nucleosynthesis of heavy elements. 

Another interesting astrophysical system that can be analyzed in the context of the GRMHD is the accretion disk around a black hole, whose symmetry plane is misaligned with the midplane of the disk. This feature is observed for example in the active galactic nuclei NGC4258 (\cite{2007MNRAS.379..135C}). The first simulations of these kind of systems (\cite{2007ApJ...668..417F}) exhibit remarkable differences with the accretion of a aligned disk. For instance, the inner most stable circular orbit is located at a larger radius than in the untilted disks, and the disk precesses uniformly with a frequency that could explain some observed quasi-periodic oscillations.

In a slightly different context, the GRMHD is useful for studying the acceleration and collimation of relativistic jets. Currently, it is widely believed that the jets are accelerated and self-collimated by a large scale magnetic field through the Blandford-Znajek (\cite{1977MNRAS.179..433B}) or the Blandford-Payne (\cite{1982MNRAS.199..883B}) mechanisms. Those models have been tested by observations (\cite{2012MNRAS.419L..69N}) and by numerical GRMHD simulations (\cite{2005MNRAS.359..801K}, \cite{2005ApJ...630L...5M}, \cite{2012MNRAS.420.2020L}). Additionally, \cite{2008ApJ...678.1180B} have shown, through magnetized disk simulations, that the jet-launching strongly depends on the initial configuration of the magnetic field. Furthermore, as suggested in \cite{2011MNRAS.418L..79T}, it is possible to explain the apparent observed efficiency to generate energy in jets, within some active galactic nuclei, through a magnetically arrested accretion. 

In all the works mentioned before, the magnetic fields are due only to the motion of free charges. Nevertheless, at a microscopic level, all substances contain spinning electrons that move around orbits, forming small dipoles. Those dipoles can generate a magnetic field due to a net alignment of themselves, or as a response to an external magnetic field by means of magnetic torques (paramagnetism) or induced magnetic moments (diamagnetism) (\cite{citeulike:4033945}). Therefore, in order to study the response of the fluid to an applied magnetic field, it is necessary to include magnetic polarization terms in the energy-momentum tensor of the fluid. Fortunately, this tensor was computed by \cite{Maugin:1978tu} through the relativistic invariant Lorentz force in a particle aggregate. Recently, \cite{2015MNRAS.447.3785C} obtained the same tensor but starting from the Lagrangian density of a fermion system in the presence of a magnetic field.

The magnetic polarization of an astrophysical fluid has already been considered for studying the equilibrium structure of neutron stars. For example, \cite{1982JPhC...15.6233B} computed the magnetic susceptibility of the degenerate free electrons in the star crust and concluded that the magnetization do not contribute significantly to the surface properties, but it probably may be coupled indirectly to observable effects. In fact, in \cite{2010ApJ...717..843S} the authors investigate the possibility that the Soft Gamma-Ray Repeaters and the Anomalous X-ray Pulsars might be observational evidences for magnetic domain formation in magnetars. However, as far as we know, in the ideal GRMHD context there are not previous research dealing with the evolution of a magnetically polarized test fluid with magnetic field in the vicinity of a relativistic astrophysical object. Therefore, the aim of this paper is to present the theoretical and numerical background to study the evolution of a magnetic media in the fixed gravitational field of a compact object. With this background, we can analyze the role of the magnetic polarization in some standard tests in literature. 

This paper is organized as follows: we first present in Sec. \ref{sec2} the GRMHD equations for a magnetically polarized fluid as a conservative system, in order to obtain numerical solutions through Godunov-type shock-capturing schemes (\cite{toro2009riemann}). In Sec. \ref{sec3}, we obtain the eigenvalue structure of the system of equations by following the Anile method (\cite{1989rfmw.book.....A}). The eigenvalues are computed in the linear approximation, where the magnetic polarization vector is in the same direction as the applied magnetic field. The eigenvalues are physically relevant, not only because they are fundamental in the Godunov-type schemes, but also because they provide important information about the wave propagation speeds. Next, in Sec. \ref{sec4} we give a brief summary of the numerical methods implemented in the CAFE code (\cite{2015ApJS..218...24L}), where we incorporate the new theoretical results presented in previous sections. In Sec. \ref{sec5}, we present some magnetized Riemann problems with magnetic polarization in order to analyze the structure of the solutions when we consider the fluid, either as a diamagnetic or as a paramagnetic media. Additionally, we use the stationary magnetized Michel solution to estimate the values of $\chi_m$ for which the code converge to second order, in the case where a strong gravitational field is present in the system. Finally, in Sec \ref{conclutions} we summarize the main results of this work. 

Throughout the paper, the Greek indices run from 0 to 3, while the Latin ones run from 1 to 3. Additionally, the equations are written in units where the gravitational constant $G$ and the speed of light $c$ are equal to one.


\section{The GRMHD equations for a magnetically polarized fluid as a conservative system}
\label{sec2}

The evolution of a test fluid with magnetic polarization in the presence of a strong gravitational field and a magnetic field can be obtained from the local conservation laws, {\em i.e}, from the conservation of the baryon number, the conservation of the energy and momentum,
\begin{eqnarray}
\nabla_{\mu}\left(\rho u^{\mu}\right)&=&0, \label{mass_conservation}\\
\nabla_{\mu}T^{\mu\nu}&=&0, \label{momentum_conservation}
\end{eqnarray}
and from the homogeneous Maxwell equations
\begin{equation}
\nabla_{[\kappa}F_{\mu\nu]}=0, \label{maxwell_homogeneas}
\end{equation}
where $\nabla_{\mu}$ is the covariant derivative, $u^{\mu}$ is the four-velocity vector of the fluid, $\rho$ is the rest mass density, $T^{\mu\nu}$ is the energy-momentum tensor, and $F_{\mu\nu}$ is the Faraday tensor. 

Now, within the ideal GRMHD approximation, where the fluid is assumed as a perfect conductor, the Faraday tensor takes the form
\begin{equation}
F^{\mu\nu}=\epsilon^{\mu\nu\kappa\delta}b_{\kappa}u_{\delta},
\label{faraday_tensor}
\end{equation}
where $b^{\mu}$ is the magnetic field measured by a comoving observer with the fluid, and $\epsilon^{\mu\nu\kappa\delta}$ is the Levi-Civita tensorial density,
\begin{equation}
\epsilon^{\mu\nu\kappa\delta}=\frac{1}{\sqrt{-g}}\eta^{\mu\nu\kappa\delta},
\label{levi_civita}
\end{equation}
being $g$ the determinant of the metric tensor $g_{\mu\nu}$, and $\eta^{\mu\nu\kappa\delta}$ the Levi-Civita tensor. Substituting (\ref{faraday_tensor}) in (\ref{maxwell_homogeneas}) and multiplying the resulting equation by $\eta^{\mu\nu\kappa\delta}$ one obtains
\begin{equation}
\nabla_{\mu}(u^{\mu}b^{\nu}-b^{\mu}u^{\nu})=0,
\label{relevant_maxwell}
\end{equation}
which are known as the relevant Maxwell equations (\cite{1989rfmw.book.....A}).

On the other hand, the energy-momentum tensor for a magnetically polarized fluid in a magnetic field can be written as the following superposition
\begin{equation}
T^{\mu\nu}=T^{\mu\nu}_{f}+T^{\mu\nu}_{em}
\label{energy_momen_terms}
\end{equation}
where $T^{\mu\nu}_{f}$ and $T^{\mu\nu}_{em}$ are the fluid and electromagnetic field energy-momentum tensors, respectively. Now, following \cite{2010PhRvD..81d5015H} and \cite{2015MNRAS.447.3785C},  with the aim of giving a first approximation to the fluid description, we will consider that the thermodynamic pressure is isotropic, and that the heat flux is zero. It is worth mentioning that the total thermodynamic pressure is actually the sum of the kinetic pressure and an additional contribution due to the Lorenz force density related to the magnetization currents \citep{2012PhRvC..85c9801P}. Thus the kinetic pressure is anisotropic in a magnetically polarized fluid, but the total thermodynamic pressure is isotropic.
These assumptions enable us to treat the fluid as a perfect one. Thus, the energy-momentum tensor of the fluid takes the form,
\begin{equation}
T^{\mu\nu}_{f}=\rho hu^{\mu}u^{\nu}+pg^{\mu\nu},
\label{tpf}
\end{equation}
where $p$ is the thermodynamic pressure and $h=(e+p)/\rho$ is the specific enthalpy, being $e$ the energy density. On the other hand, the energy-momentum associated with the electromagnetic field and its interaction with matter can be described, in the ideal GRMHD approximation, through the tensor (\cite{Maugin:1978tu, 2015MNRAS.447.3785C})
\begin{eqnarray}
T_{em}^{\mu\nu}=&&\left(u^{\mu}u^{\nu}+\frac{1}{2}g^{\mu\nu}\right)\frac{b^{2}}{\mu_0}-\frac{1}{\mu_0}b^{\mu}b^{\nu}\nonumber\\
&&-\left(u^{\mu}u^{\nu}+g^{\mu\nu}\right)b^{\kappa}m_{\kappa}+m^{(\mu}b^{\nu)}
\label{tem}
\end{eqnarray}
where $b^{2}=b^{\mu}b_{\mu}$, $\mu_0$ is the vacuum permeability, and $m^{\mu}$ is the magnetic polarization four-vector in the comoving observer. It is important to mention that since the electric field in the comoving frame is zero for a perfect conductor, then it is reasonable to assume that the electric polarization of the matter is also zero in the same frame. With these two contributions, $T^{\mu\nu}$ takes the form \citep{2016GReGr..48..124P,2017CQGra..34g5008P}
\begin{equation}
T^{\mu\nu}=\rho h^{*}u^{\mu}u^{\nu}+p^{*}g^{\mu\nu}-\Pi^{\mu\nu},
\label{energy_momen}
\end{equation}
with,
\begin{eqnarray}
\rho h^{*}&=&\rho h+\frac{b^{2}}{\mu_{0}}-b^{\kappa}m_{\kappa},\\
p^{*}&=&p+\frac{b^{2}}{2\mu_{0}}+b^{\kappa}m_{\kappa},\\
\Pi^{\mu\nu}&=&\frac{1}{\mu_{0}}b^{\mu}b^{\nu}-m^{(\mu}b^{\nu)}.
\end{eqnarray}
This is the energy-momentum tensor of a magnetically polarized perfect fluid endowed with a magnetic field.

Once we have defined the total energy-momentum tensor, the next step is to write (\ref{mass_conservation}), (\ref{momentum_conservation}), and (\ref{relevant_maxwell}) as a conservative system of equations, in which the time derivatives are separated from the space ones, so numerical methods can be applied to compute the evolution of the system from an initial state. To do this, we use the well known 3+1 formalism (\cite{2004rtmb.book.....P}) in which the spacetime is foliated with spacelike hypersurfaces of $t$-constant, $\Sigma_t$. These hypersurfaces have a normalized timelike normal vector $n^{\mu}$, and a set of three spacelike tangent vectors, $e_{i}^{\mu}$, which define the induced metric $\gamma_{ij}=g_{\mu\nu}e_{i}^{\mu}e_{j}^{\nu}$ on $\Sigma_t$. Now, taking $x^{i}$ as a set of three spatial coordinates on the hypersurface, it is possible to use the coordinate transformations $x^{\mu}=x^{\mu}(t,x^{i})$ to write the line element as
\begin{equation}
ds^{2}=-(\alpha^{2}-\beta_{i}\beta^{i})dt^{2}+2\beta_{i}dx^{i}dt+\gamma_{ij}dx^{i}dx^{j},
\label{line}
\end{equation}
where $\alpha$ is the lapse function, and $\beta^{i}$ is the shift vector. 

In addition to the comoving observer, it is convenient to introduce another one that moves perpendicular to $\Sigma_t$ with four-velocity $n_{\mu}=(-\alpha,0,0,0)$, which is called Eulerian observer. The spatial velocity of the fluid $v^{i}$, measured by this observer is tangent to the hypersurfaces, and satisfies the equation
\begin{equation}
v^{i}=\frac{u^{i}}{W}+\frac{\beta^{i}}{\alpha},
\label{Eulerian_velocity}
\end{equation}
where $W=-u^{\mu}n_{\mu}=\alpha u^{0}$ is the Lorentz factor between the eulerian and comoving observers. Now, the GRMHD equations with magnetic polarization (\ref{mass_conservation}), (\ref{momentum_conservation}), and (\ref{relevant_maxwell}) can be written as a conservative system of the form
\begin{equation}
\partial_{0}(\sqrt{\gamma}\ \vec{U})+\partial_{i}(\sqrt{-g}\ \vec{F}^{i})=\sqrt{-g}\ \vec{S},
\label{conservative}
\end{equation}
\begin{equation}
\partial_{i}(\sqrt{\gamma}B^{i})=0.
\label{B_constrain}
\end{equation}
The equation (\ref{B_constrain}) is the divergence-free constrain for the magnetic field, $\gamma$ is the determinant of the induced metric, and $\vec{U}$ is the state vector
\begin{equation}
\vec{U}=[D,S_j,\tau,B^{k}]^{T},
\label{state_vector}
\end{equation}
whose components are the mass density $D=-\rho u^{\mu}n_{\mu}$, the energy flux vector $S_{i}=-T^{\mu\nu}n_{\mu}e_{i\nu}$, the energy density (minus D) $\tau=T^{\mu\nu}n_{\mu}n_{\nu}-D$, and the magnetic field $B^{k}$, all of them measured in the eulerian observer. By doing the projections along the basis $\{n^{\mu},e_{i}^{\mu}\}$, we obtain
\begin{eqnarray}
D&=&\rho W, \label{restmass_euler}\\
S_{j}&=&\rho h^{*}W^{2}v_{j}-\frac{\alpha}{\mu_{0}}b_{j}b^{0}+\frac{\alpha}{2}(m^{0}b_{j}+b^{0}m_{j}), \label{momentum_euler}\\
\tau&=&\rho h^{*}W^{2}-p^{*}-\frac{\alpha^{2}}{\mu_{0}}(b^{0})^{2}+\alpha^{2}m^{0}b^{0}-D. \label{energy_euler}
\end{eqnarray}

Now, it is possible to show that when the matter is magnetically polarized, the flux vector $\vec{F}^{i}$ takes the form
\begin{equation}
\vec{F}^{i}=\left[
\begin{array}{c}
\vspace{2mm}
D \tilde{v}^{i}\\
\vspace{2mm}
S_{j}\tilde{v}^{i}+p^{*}\delta_{j}^{i}-\frac{b_{j}B^{i}}{\mu_{0}W}+\frac{1}{2W}(m_{j}B^{i}+b_{j}M^{i})\\
\vspace{2mm}
\tau\tilde{v}^{i}+p^{*}v^{i}-\frac{\alpha}{\mu_{0}W}b^{0}B^{i}+\frac{\alpha}{2W}(m^{0}B^{i}+b^{0}M^{i})\\
\tilde{v}^{i}B^{k}-\tilde{v}^{k}B^{i}
\end{array}
\right],
\end{equation}
being $M^{i}$ the magnetic polarization vector in the eulerian frame and $\tilde{v}^{i}=v^{i}-\beta^{i}/\alpha$. It is worth mentioning that $b^{\mu}$ and $m^{\mu}$ are related to the same quantities in the eulerian frame through the equations 
\begin{eqnarray}
b^{i}&=&\frac{B^{i}+\alpha b^{0}u^{i}}{W}, \hspace{7mm} b^{0}=\frac{W}{\alpha}B^{i}v_{i},\label{transformations}\\
m^{i}&=&\frac{M^{i}+\alpha m^{0}u^{i}}{W}, \hspace{3mm} m^{0}=\frac{W}{\alpha}M^{i}v_{i}.
\end{eqnarray}
Finally, the terms that do not contain derivatives of the state variables are included in the source vector
\begin{center}
\begin{equation}
\vec{S}=\left[
\begin{array}{c}
\vspace{2mm}
0\\
\vspace{2mm}
T^{\mu\nu}(\partial_{\mu}g_{j\nu}-g_{j\delta}\Gamma_{\mu\nu}^{\delta})\\
\vspace{2mm}
T^{\mu 0}\partial_{\mu}\alpha-\alpha T^{\mu\nu}\Gamma_{\mu\nu}^{0}\\
0^{k}
\end{array}
\right].
\end{equation}
\end{center}
The conservative equations (\ref{conservative}) become a closed system once we introduce an equation of state $p=p(\rho,e)$, and a constitutive relation between the magnetic polarization vector and the magnetic field, $m^{\mu}=m^{\mu}(b^{\nu})$.

\section{The eigenvalue structure}
\label{sec3}

The system of equations (\ref{conservative}-\ref{B_constrain}) do not have a general exact solution, so it is necessary to solve them numerically. Some current numerical methods devoted to solve the system of evolution equations in conservative form are based on the eigenvalue structure, {\em i.e.} on the propagation speeds of the signals in the fluid. Therefore, with the aim of computing the eigenvalues of (\ref{conservative}), we follow a similar procedure as the one described in \cite{1989rfmw.book.....A}, whose starting point is to write the equations (\ref{mass_conservation}), (\ref{momentum_conservation}), and (\ref{relevant_maxwell}) in the covariant form
\begin{equation}
A^{\mu \mathcal{K}}_{\mathcal{P}}	\nabla_{\mu}V^{\mathcal{P}}=0,
\label{covariant_system}
\end{equation}
where $A^{\mu \mathcal{K}}_{\mathcal{P}}$ are $N\times N$ coefficient matrices, $V^{\mathcal{P}}$ is a column vector whose $N$ components are state variables, and the indices $\mathcal{P}, \mathcal{K}=1,2,...,N$. We recall that the eigenvalue structure we want to compute in this paper differs from the usual GRMHD eigenvalues in that we are including the magnetic polarization of the fluid.

To write the equations in the covariant form (\ref{covariant_system}) we project (\ref{relevant_maxwell}) along $u_{\nu}$ and $b_{\nu}$ to obtain
\begin{eqnarray}
\nabla_{\mu}b^{\mu}+u^{\mu}u^{\nu}\nabla_{\mu}b_{\nu}=0,\label{cons_1}\\
b^{2}\nabla_{\mu}u^{\mu}+\frac{1}{2}u^{\mu}\nabla_{\mu}(b^{2})-b^{\mu}b^{\nu}\nabla_{\mu}u_{\nu}=0\label{cons_2},
\end{eqnarray}
respectively. Now, we obtain the energy conservation equation by projecting (\ref{momentum_conservation}) along $u_{\nu}$ and by using (\ref{cons_1}) in the resulting expression, so we get
\begin{equation}
u^{\mu}\nabla_{\mu}e+(e+p-b^{\kappa}m_{\kappa})\nabla_{\mu}u^{\mu}+m^{(\mu}b^{\nu)}\nabla_{\mu}u_{\nu}=0.
\label{energy_prev}
\end{equation}
Nevertheless, from an equation of state $e=e(p,s)$, being $s$ the specific entropy, we can rewrite the first term in the last equation, such that the energy conservation takes the following form,
\begin{eqnarray}
e'_{p}u^{\mu}\nabla_{\mu}p+e'_{s}u^{\mu}\nabla_{\mu}s
+K\nabla_{\mu}u^{\mu}\nonumber\\+m^{(\mu}b^{\nu)}
\nabla_{\mu}u_{\nu}=0,
\label{energy}
\end{eqnarray}
where $K=e+p-b^{\kappa}m_{\kappa}$, $e'_{p}=\partial e/\partial p$, and $e'_{s}=\partial e/\partial s$.

The second equation in the Anile method is the one that describes the entropy production. This equation is computed from the first law of thermodynamics for a magnetically polarized material (\cite{groot1972foundations})
\begin{equation}
T\nabla_{\mu}s=\nabla_{\mu}\epsilon-\frac{p}{\rho^{2}}\nabla_{\mu}\rho+\frac{1}{\rho}m_{\kappa}\nabla_{\mu}b^{\kappa},
\label{first_law_thermo}
\end{equation}
being $T$ the temperature. Now, from the conservation of the baryon number (\ref{mass_conservation}) and the conservation of the energy (\ref{energy_prev}), we can rewrite (\ref{first_law_thermo}) as
\begin{eqnarray}
\rho Tu^{\mu}\nabla_{\mu}s&=&b^{\kappa}m_{\kappa}\nabla_{\mu}u^{\mu}+u^{\mu}m_{\kappa}
\nabla_{\mu}b^{\kappa}\nonumber\\
&&-m^{(\mu}b^{\nu)}\nabla_{\mu}u_{\nu}.
\label{entropy}
\end{eqnarray}
Note that unlike in the GRMHD case, where the adiabaticity condition $u^{\mu}\nabla_{\mu}s=0$ is satisfied, when the magnetic polarization of the matter is considered, the right hand side of the last equation does not vanish in general.

Another relevant equation in the Anile method is the conservation of the momentum. We also compute this equation in the case of a magnetically polarized fluid; first of all, we contract (\ref{momentum_conservation}) with $b_{\nu}$ in order to obtain the auxiliary relation 
\begin{eqnarray}
(e+p-\frac{1}{2}b^{\kappa}m_{\kappa})\nabla_{\mu}b^{\mu}+b^{\mu}\nabla_{\mu}p-b^{\mu}m^{\nu}\nabla_{\mu}b_{\nu}\nonumber\\
-\frac{1}{2}b^{\mu}b^{\nu}\nabla_{\mu}m_{\nu}+\frac{1}{2}m^{\mu}b^{\nu}\nabla_{\mu}b_{\nu}+\frac{1}{2}b^{2}\nabla_{\mu}m^{\mu}=0.
\label{third_projection}
\end{eqnarray}
Then, we project (\ref{momentum_conservation}) on the hypersurface $\Sigma_{t}$, by contracting it with the projector tensor $h_{\nu\kappa}=g_{\nu\kappa}+u_{\nu}u_{\kappa}$. In the resulting expression we replace $\nabla_{\mu}u^{\mu}$ and $\nabla_{\mu}b^{\mu}$, which can be obtained from (\ref{energy}) and (\ref{third_projection}), respectively, to get the conservation of momentum equation
\begin{eqnarray}
\mathcal{A}^{\kappa\mu}\nabla_{\mu}p+\mathcal{B}^{\kappa\mu}_{~~~\nu}
\nabla_{\mu}b^{\nu}+\mathcal{C}^{\kappa\mu}_{~~~\nu}
\nabla_{\mu}u^{\nu}\nonumber\\
+\mathcal{D}^{\kappa\mu}_{~~~\nu}
\nabla_{\mu}m^{\nu}+\mathcal{E}^{\kappa\mu}\nabla_{\mu}s=0,
\label{momentum}
\end{eqnarray}
where
\begin{align*}
\mathcal{A}^{\kappa\mu}=&h^{\kappa\mu}-\frac{b^{2}}{\mu_{0}}K^{-1}e'_{p}u^{\kappa}u^{\mu}-\zeta\eta^{\kappa}b^{\mu},\nonumber\\
\mathcal{B}^{\kappa\mu}_{~~~\nu}=&\frac{1}{\mu_{0}}(h^{\kappa\mu}+u^{\kappa}u^{\mu})b_{\nu}+\zeta\eta^{\kappa}\left(b^{\mu}m_{\nu}-\frac{1}{2}m^{\mu}b_{\nu}\right)\nonumber\\&-h^{\kappa\mu}m_{\nu}+\eta^{\mu}\delta^{\kappa}_{\nu},\nonumber\\
\mathcal{C}^{\kappa\mu}_{~~~\nu}=&\left(K+\frac{b^{2}}{\mu_{0}}\right)[u^{\mu}\delta^{\kappa}_{\nu}-K^{-1}u^{\kappa}m^{(\mu}b^{\gamma)}g_{\gamma\nu}],\nonumber\\
\mathcal{D}^{\kappa\mu}_{~~~\nu}=&\frac{1}{2}\zeta\eta^{\kappa}b^{\mu}b_{\nu}-h^{\kappa\mu}b_{\nu}-\frac{b^{2}}{2}\zeta\eta^{\kappa}\delta^{\mu}_{\nu}+\frac{1}{2}b^{\kappa}\delta^{\mu}_{\nu}+\frac{1}{2}b^{\mu}\delta^{\kappa}_{\nu},\nonumber\\
\mathcal{E}^{\kappa\mu}=&-\frac{b^{2}}{\mu_{0}}K^{-1}e'_{s}u^{\kappa}u^{\mu},\nonumber
\end{align*}
with, $\zeta=\left(K+\frac{1}{2}b^{\sigma}m_{\sigma}\right)^{-1}$, and $\eta^{\mu}=\frac{1}{2}m^{\mu}-\frac{1}{\mu_{0}}b^{\mu}$. 

Finally, the relevant Maxwell equations (\ref{relevant_maxwell}) can be rewritten as
\begin{eqnarray}
\mathcal{F}^{\kappa\mu}\nabla_{\mu}p+\mathcal{G}^{\kappa\mu}_{~~~\nu}
\nabla_{\mu}b^{\nu}+\mathcal{H}^{\kappa\mu}_{~~~\nu}
\nabla_{\mu}u^{\nu}\nonumber\\
+\mathcal{I}^{\kappa\mu}_{~~~\nu}
\nabla_{\mu}m^{\nu}+\mathcal{J}^{\kappa\mu}\nabla_{\mu}s=0,
\label{maxwell}
\end{eqnarray} 
where,
\begin{eqnarray}
\mathcal{F}^{\kappa\mu}&=&\zeta b^{\mu}u^{\kappa}-K^{-1}e'_{p}u^{\mu}b^{\kappa},\nonumber\\
\mathcal{G}^{\kappa\mu}_{~~~\nu}&=&u^{\mu}\delta_{\nu}^{\kappa}-u^{\kappa}\zeta\left(b^{\mu}m_{\nu}-\frac{1}{2}m^{\mu}b_{\nu}\right),\nonumber\\
\mathcal{H}^{\kappa\mu}_{~~~\nu}&=&-b^{\mu}\delta^{\kappa}_{\nu}-K^{-1}g_{\sigma\nu}m^{(\mu}b^{\sigma)}b^{\kappa},\nonumber\\
\mathcal{I}^{\kappa\mu}_{~~~\nu}&=&\frac{1}{2}(b^{2}\delta_{\nu}^{\mu}u^{\kappa}-b^{\mu}u^{\kappa}b_{\nu})\zeta,\nonumber\\
\mathcal{J}^{\kappa\mu}&=&-K^{-1}e'_{s}b^{\kappa}u^{\mu}\nonumber.
\end{eqnarray}
It is important to mention that when $m^{\mu}=0$, the conservation of the energy (\ref{energy}), the entropy production equation (\ref{entropy}), the conservation of the momentum (\ref{momentum}), and the relevant Maxwell equations (\ref{maxwell}) reduce to the equations of the ideal GRMHD as expected (cf. \cite{1989rfmw.book.....A}).

On the other hand, as it was pointed out at the end of Sec. \ref{sec2}, it is necessary to introduce a constitutive relation between $m^{\mu}$ and $b^{\mu}$ to form a closed set of equations. In this work we concentrate our attention to the special case in which the magnetic polarization is always in the same direction as the magnetic field, so we can relate them through the equation
\begin{equation}
m^{\mu}=\frac{\chi}{\mu_{0}}b^{\mu},
\label{constituvive_relation}
\end{equation} 
where $\chi=\chi_m/(1+\chi_m)$, and $\chi_m$ is the magnetic susceptibility of the fluid. This last physical parameter is negative for diamagnets and positive for paramagnets.  As it is argued by \cite{1999JChPh.110.7403F}, this particular relation is useful to describe polar fluids in equilibrium, or systems in which the constituent particles do not have permanent magnetic dipole moments. When a polar fluid is not in equilibrium, it is necessary to introduce a phenomenological relaxation equation that describe the evolution of $m^{\mu}$ in the presence of an applied magnetic field \citep{schumacher2010effects}. 

For the particular choice of $m^{\mu}$ presented in (\ref{constituvive_relation}), it is possible to see that the adiabaticity condition, which implies a zero heat transfer between two adjacent elements of the fluid, is satisfied. With this condition, all the terms proportional to $u^{\mu}\nabla_{\mu}s$ in the equations vanish. On the other hand, when using the constitutive relation (\ref{constituvive_relation}) to reduce the number of variables, and therefore, to close the system of equations, it is important to take into account that $b^{\mu}$ has to satisfy (\ref{cons_1}) and (\ref{cons_2}). With these considerations, the GRMHD equations with magnetization can be set in the covariant form (\ref{covariant_system}), with ${\bf V}=({u^{\mu},b^{\mu},p,s})^{T}$ and 
\begin{widetext}
\begin{equation}
A^{\mu}=\begin{bmatrix}
[\eta+(1-\chi)b^{2}]u^{\mu}\delta_{\nu}^{\kappa} & \tilde{P}^{\kappa\mu}b_{\nu}-(1-\chi)b^{\mu}\delta_{\nu}^{\kappa} &  \tilde{1}^{\kappa\mu} & 0^{\kappa\mu} \\
b^{\mu}\delta_{\nu}^{\kappa} & -u^{\mu}\delta_{\nu}^{\kappa}+2\chi\eta^{-1}b_{\nu}u^{(\kappa}b^{\mu)} & f^{\mu\kappa} & 0^{\kappa\mu} \\
\eta\delta^{\mu}_{\nu}+\chi\tilde{\gamma}_{\ \nu}^{\mu} & 0^{\mu}_{\nu} & e'_{p}u^{\mu} & 0^{\mu} \\
0_{\nu}^{\mu} & 0^{\mu}_{\nu} & 0^{\mu} & u^{\mu}
\label{matrix}
\end{bmatrix},
\end{equation}
\end{widetext}
where $\nu$ indicates the columns and $\kappa$ the rows. In this matrix we have introduced the following functions: $\eta=\rho h=e+p$, $\tilde{\gamma}^{\mu}_{\ \nu}=b^{\mu}b_{\nu}-b^{2}\delta_{\nu}^{\mu}$,
\begin{eqnarray}
\tilde{P}^{\kappa\mu}&=&(1-2\chi)h^{\kappa\mu}-\chi(1-\chi)\eta^{-1}b^{\kappa}b^{\mu}
\nonumber\\&&+(1-\chi)(\eta-\chi b^{2})\eta^{-1}u^{\kappa}u^{\mu},\\
\tilde{1}^{\kappa\mu}&=&h^{\kappa\mu}+\eta^{-1}(1-\chi)(b^{\kappa}b^{\mu}-e'_{p}b^{2}u^{\kappa}u^{\mu}),\\
f^{\mu\kappa}&=&-\eta^{-1}(-e'_{p}u^{\mu}u^{\kappa}+u^{\kappa}b^{\mu}),
\end{eqnarray}
with the aim of reducing the expressions.

Now, to compute the characteristic structure of the system of equations (\ref{covariant_system}), we introduce an hypersurface given by $\phi(x^{\nu})=0$ with normal vector $\phi_{\mu}=\nabla_{\mu}\phi$. This hypersurface is said to be characteristic if $\text{det}(A^{\mu}\phi_{\mu})=0$. Therefore, by contracting $A^{\mu}$ with the vector $\phi_\mu$, we obtain the characteristic matrix
\begin{widetext}
\begin{equation}
A^{\mu}\phi_{\mu}=\begin{bmatrix}
[\eta+(1-\chi)b^{2}]a\delta_{\nu}^{\kappa} & m^{\kappa}_{\ \nu} & \tilde{1}^{\kappa} & 0^{\kappa} \\
B\delta_{\nu}^{\kappa} & -a\delta_{\nu}^{\kappa}+\chi\eta^{-1}b_{\nu}(Bu^{\kappa}+ab^{\kappa}) & f^{\kappa} & 0^{\kappa} \\
\eta\phi_{\nu}+\chi(Bb_{\nu}-b^{2}\phi_{\nu}) & 0_{\nu} & ae'_{p} & 0 \\
0_{\nu} & 0_{\nu} & 0 & a
\label{matrix_characteristic}
\end{bmatrix},
\end{equation}
\end{widetext}
where $a=u^{\mu}\phi_{\mu}$, $B=b^{\mu}\phi_{\mu}$, and
\begin{equation}
m^{\kappa}_{\ \nu}=\tilde{P}^{\kappa\mu}\phi_{\mu}b_{\nu}-(1-\chi)B\delta_{\nu}^{\kappa},
\end{equation}
\begin{equation}
f^{\kappa}=f^{\kappa\mu}\phi_{\mu}=-\eta^{-1}(-e'_{p}au^{\kappa}+Bu^{\kappa}),
\end{equation}
\begin{eqnarray}
\tilde{1}^{\kappa}&=&\tilde{1}^{\kappa\mu}\phi_{\mu},\nonumber\\
&=&\phi^{\kappa}+au^{\kappa}+\eta^{-1}(1-\chi)(Bb^{\kappa}-e'_{p}b^{2}au^{\kappa}).
\end{eqnarray}


Hence, the characteristic equation takes the simple form, 
\begin{equation}
\text{det}(A^{\mu}\phi_{\mu})=\eta^{-1}a^{2}\tilde{A}^{2}\tilde{N}_{4}Q=0,
\label{characteristic_equation}
\end{equation}
with
\begin{eqnarray}
\tilde{A}&=&[\eta+(1-\chi)b^{2}]a^{2}-(1-\chi)B^{2},\\
\tilde{N}_{4}&=&f_{1}a^{4}-f_{2}a^{2}G-f_{3}a^{2}B^{2}+f_{4}B^{2}G,\label{n4}\\
Q&=&b^{4}\chi^{2}-b^{2}(b^{2}+2\eta)\chi+\eta(\eta+b^{2}),
\end{eqnarray}
being
\begin{eqnarray}
f_1&=&(e'_{p}-1)\eta+(e'_{p}+1)\chi b^{2},\\
f_2&=&\eta+(1-2\chi)b^{2}e'_{p}-\chi b^{2},\\
f_3&=&(e'_{p}+1)\chi,\\
f_4&=&1-\chi,
\end{eqnarray}
and $G=\phi_{\mu}\phi^{\mu}$ the functions that define $\tilde{N}_{4}$. The characteristic equation (\ref{characteristic_equation}) depends on the vector $\phi_\mu$, which represents the local velocity of the travelling wave (\cite{Ruggeri1981}). The characteristic wave speeds can be computed by equaling to zero the functions $a$, $\tilde{A}$, and $\tilde{N}_4$, since $\eta^{-1}$ and $Q$ do not depend on $\phi_{\mu}$. Now, a wave that propagates with speed $\lambda$ in the first spatial direction (let's call $x$) has a vector $\phi_{\mu}=(-\lambda,1,0,0)$. Then, the eigenvalue that results from the equation $a=0$ is
\begin{equation}
\lambda_e=\alpha v^{x}-\beta^{x},
\label{entropic}
\end{equation} 
and correspond to the propagation speed of disturbances in the mass density (material waves). $\lambda_e$ is commonly known as the entropic eigenvalue and appears as a double root in the system of equations (\cite{1999MNRAS.303..343K}).

On the other hand, the Alfv$\grave{\text{e}}$n wave speeds are the roots of the equation $\tilde{A}=0$. These waves are produced from the interaction of a conducting fluid and the magnetic field. When one consider the matter as a magnetically polarized fluid, the Alfv$\grave{\text{e}}$n eigenvalues become dependent on the magnetic susceptibility of the material in the form
\begin{equation}
\lambda_a=\frac{b^{x}\pm\sqrt{H}u^{x}}{b^{0}\pm\sqrt{H}u^{0}},
\label{alfven}
\end{equation}
where $H=b^{2}+\eta/(1-\chi)$. Now, in order to preserve the real character of the Alfv$\grave{\text{e}}$n eigenvalues, and therefore to ensure the hyperbolicity of the system of equations, we need to have $H=\eta(1+\chi_{m})+b^{2}\geq0$. This condition constrains the magnetic susceptibility to the interval:
\begin{equation}
\chi_m\geq-\left(1+\frac{b^{2}}{\eta}\right).
\label{restiction}
\end{equation}
Additionally, we can see from (\ref{alfven}) that when $\chi=1$, the Alfv$\grave{\text{e}}$n waves travel at the same speed as the entropic waves, but with a non-zero magnetic field. Nevertheless, this value for $\chi$ correspond to an infinite magnetic susceptibility. 

The remaining eigenvalues are solutions to the quartic equation $\tilde{N}_4=0$, and correspond to the magnetosonic waves. Unfortunately, the analytic expressions for these eigenvalues are not simple, so one of the best options is to compute them numerically with a Newton-Raphson method. Nevertheless, as it was pointed out by \cite{2005A&A...436..503L}, this solution makes the solver unstable for high Lorentz factors. Hence, the authors implemented in their code the analytical expressions for the fast magnetosonic waves with $B^{i}v_{i}=0$, in the degenerate case $B^{x}=0$, and achieved very similar results to those obtained by the Newton-Raphson algorithm. In our paper, we followed the same method presented by Leismann {\em et al.} to obtain the boundary values for the magnetosonic speeds. These values have the following form
\begin{equation}
\lambda_{bms}=\frac{a-\alpha\sqrt{b^{2}+cd}}{c},
\end{equation}
where,
\begin{eqnarray}
a&=&b^{0}b^{x}q_{4}\chi\alpha^{2}-H[q_{3}\chi\beta^{x}+q_{2}W^{2}(\alpha v^{x}-\beta^{x})]\nonumber\\&&+q_{1}H[W^{2}(\alpha v^{x}-\beta^{x})+\beta^{x}]\Omega^{2},\\
b&=&\alpha b^{0}q_{4}\chi(b^{x}+b^{0}\beta^{x})-Hv^{x}W^{2}(q_{2}-\Omega^{2}q_{1}),\\
c&=&(b^{0})^{2}\alpha^{2}\chi q_{4}+H[\chi q_{3}+q_{1}(W^{2}-1)\Omega^{2}-W^{2}q_{2}],\\
d&=&H[q_{2}(v^{x})^{2}W^{2}+q_{3}\chi\gamma^{xx}-q_{1}\Omega^{2}((v^{x})^{2}W^{2}+\gamma^{xx})]\nonumber\\
&&-\chi q_{4}(b^{x}+b^{0}\beta^{x})^{2},
\end{eqnarray}
and the q-functions are defined in terms of the sound speed and the magnetic susceptibility as
\begin{eqnarray}
q_{1}&=&C_{s}^{2}-1+2x,\\
q_{2}&=&q_{1}+\chi q_{4},\\
q_{3}&=&C_{s}^{2}(1+C_{s}^{2}),\\
q_{4}&=&C_{s}^{4}-1.
\end{eqnarray}
With this eigenvalue structure, we can solve the equations by using a numerical method that only requires information about the wave speeds, such as the HLLE Riemann solver. In the next section we will describe briefly the numerical methods that we use for solving the GRMHD equations with magnetic polarization.

\section{Numerical Methods}
\label{sec4}

The conservative system of equations of the GRMHD with magnetically polarized matter (\ref{conservative}) will be solved numerically with a new extension of the CAFE code, where we implement the theoretical formalism presented in Sec. \ref{sec2} and Sec. \ref{sec3}. CAFE was depeloped by \cite{2015ApJS..218...24L} and its first version was designed to solve the equations of the ideal RMHD in three dimensions. The hydrodynamic version of this code has been used to study the shock cone that is produced when a relativistic ideal fluid is accreted on to a rotating Kerr black hole (\cite{2012MNRAS.426..732C}), and when the accretion is produced on to a fixed Schwarzschild spacetime (\cite{2013MNRAS.429.3144L}). Additionally, in \cite{2013JCAP...12..015L} and \cite{2014MNRAS.443.2242L} the authors used a new version of CAFE, capable of solving the hydrodynamic equations coupled with the Einstein equations, to analyze the horizon growth of black hole seeds due to the accretion of radiation fluids and collisional dark matter. More recently, CAFE has been used to model the Bondi-Hoyle accretion of an ideal supersonic gas with velocity gradients (\cite{2016MNRAS.460.3193C}), and with density gradients (\cite{2015ApJS..219...30L}) onto a Kerr black hole. Finally, \cite{2017MNRAS.471.3127C} also studied this kind of accretion in the Schwarzschild background, when some rigid and small bodies are randomly distributed around the black hole.

CAFE uses the method of lines with a Runge-Kutta algorithm to solve the discretized version of the system (\ref{conservative}), which is obtained with a finite volume scheme. The evolution of the state vector is computed with a high resolution shock capturing scheme (HRSC), together with a spatial reconstructor. In this work, we use the approximate Riemann solver proposed by Harten, Lax, van Leer, and Einfeld or HLLE (\cite{Harten1997,einfeldt1988godunov}), because it is a robust method that only requires information about the eigenvalue structure of the system of equations. More exactly, it uses the maximum and minimum wave speeds to estimate an approximate value for the fluxes in the middle of the mesh points (intercells). Finally, we use the MINMOD linear piecewise reconstructor to compute the state vector on both sides of the intercells. 

On the other hand, the code guarantees the constraint of the magnetic field, given in equation (\ref{B_constrain}), with the flux constraint transport method (\cite{1966ITAP...14..302Y}), first implemented by \cite{1988ApJ...332..659E} in an artificial-viscosity scheme, and then modified by \cite{1999JCoPh.149..270B}, in order to use the flux formulas from the HRSC schemes. With this method, the divergence free condition for the magnetic field is satisfied to machine precision, by evolving the integral form of the induction equation. Hence, if $\nabla\cdot\vec{B}=0$ at the initial time, then this value will remain to the machine precision order in the subsequent time steps.

Now, the numerical methods implemented in CAFE evolve in time the state vector $\vec{U}$, whose components are the conservative variables. Nevertheless, the fluxes are written in terms of both $\vec{U}$ and the set of primitive variables $\vec{\mathcal{W}}=[\rho,v^{i},p,B^{k}]^T$; so we need to find, at each time step, the primitive variables in terms of the conservative ones. Unfortunately, from the equations (\ref{restmass_euler}-\ref{energy_euler}), we can see that it is not possible to find explicitly the functions $\vec{\mathcal{W}}(\vec{U})$, so the code uses a Newton-Raphson algorithm to recover the primitive variables in all the points of the spacetime grid for an ideal gas equation of state. The mathematical procedure to compute $\vec{\mathcal{W}}$ when the magnetic polarization vector is different from zero, is described with more detail in Appendix \ref{AppendixA}.

\section{Numerical Experiments}
\label{sec5}

Once we have obtained the GRMHD equations with magnetization in conservative form, and its corresponding eigenvalue structure, we can implement these results into CAFE code to carry out some physically relevant numerical tests, which will be presented in the next two subsections. The fluids that we are going to study in all the problems obey the ideal equation of state,
\begin{equation}
p=\rho\epsilon(\Gamma-1),
\label{ideal-EOS}
\end{equation}
where $\epsilon$ is the specific internal energy and $\Gamma$ is the adiabatic index.

\subsection{1D Magnetized Riemann Problems with Magnetic Polarization}

In this subsection, we analyze the consequences of considering the magnetic polarization of the matter in some standard shock-tube problems in the Minkowsky spacetime. In each simulation, we assume the magnetic susceptibility of the fluids to be constant in all the spacetime domain. These 1D tests represent the physical analogues of the Riemann problems because the initial configurations consist of two different constant states $\vec{\mathcal{W}}_L$ and $\vec{\mathcal{W}}_{R}$, separated by a diaphragm or interface at the possition $x=0$. Here the subscripts $L$ and $R$ denote the left ($x\leq 0$) and right ($x> 0$) states, respectively. The first two tests that we present here are the shock tube and the collision of two fluids, both proposed by \cite{1999MNRAS.303..343K}. We carry out these simulations in the domain $x\in [-2,2]$, with a spatial resolution $\Delta x=1/800$, and a Courant factor of $0.25$. Then, we present the problems proposed by \cite{1993JCoPh.105..339V} and \cite{2001ApJS..132...83B}, for which we obtain the numerical solution in a domain $x\in [-0.5,0.5]$, with a spatial resolution $\Delta x=1/1600$, and a Courant factor of $0.25$. Finally, we present the generic Alfv$\grave{\text{e}}$n wave test, for which we use a domain $x\in [-0.5,0.5]$, a spatial resolution $\Delta x=1/1600$, and a Courant factor of $0.25$. The parameters used in these tests are given in table \ref{table1}.

Each one of the numerical tests mentioned in the last paragraph was carried out for some constant values of magnetic susceptibility, so we can determine the differences in the evolution of a diamagnetic ($\chi_m<0$) and a paramagnetic ($\chi_m>0$) fluid. Now, we also evolve the initial states with a zero magnetic susceptibility in order to reproduce the original RMHD solutions (\cite{2006JFM...562..223G}) and compare them with our results. In this respect, we anticipate that, although the wave structure of the solutions turns out to be the same as in the RMHD case (with $\chi_m=0$), there are remarkable differences in the evolution when the magnetic polarization of the fluid is considered.

\subsubsection{Komissarov shock tube test}

\begin{figure*}
\begin{tabular}{cc}
\includegraphics[scale=0.7]{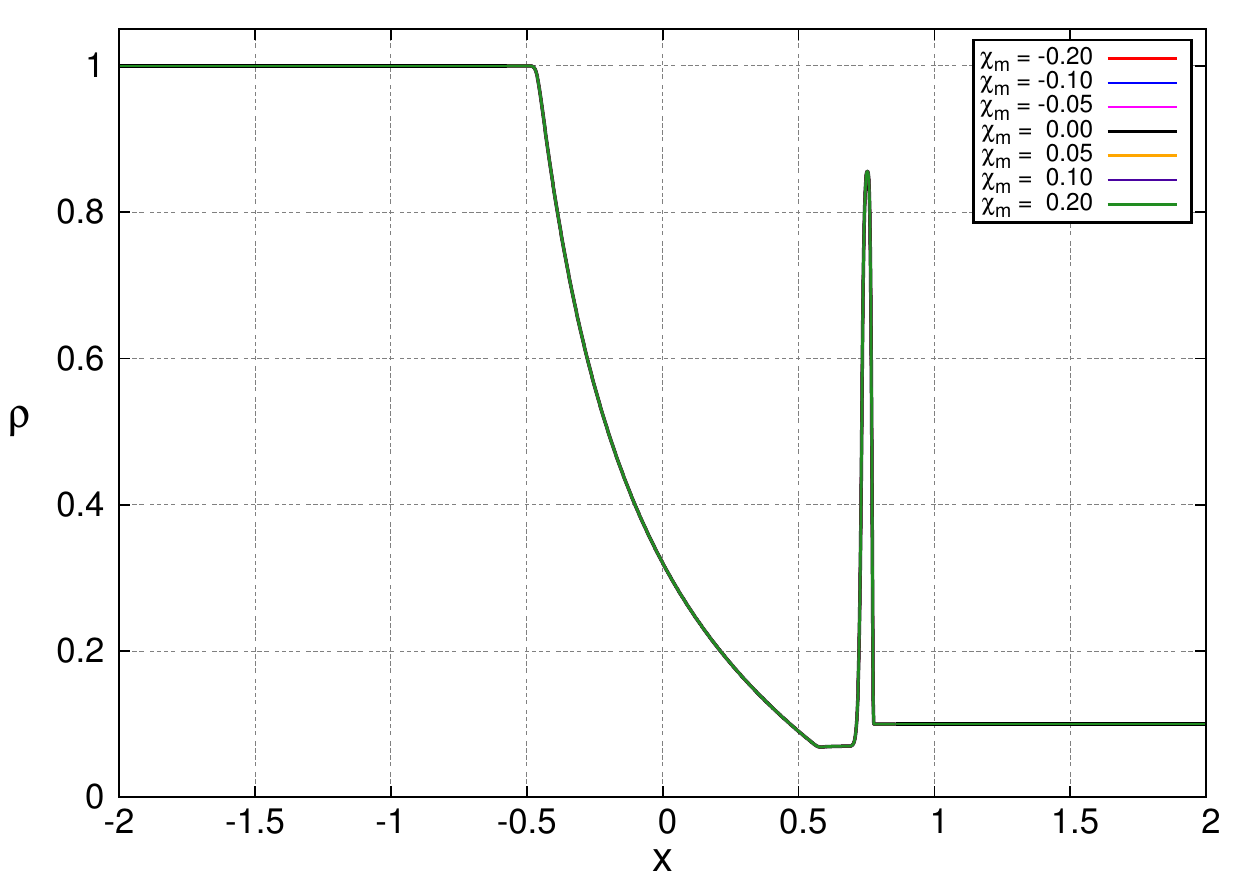} & \includegraphics[scale=0.7]{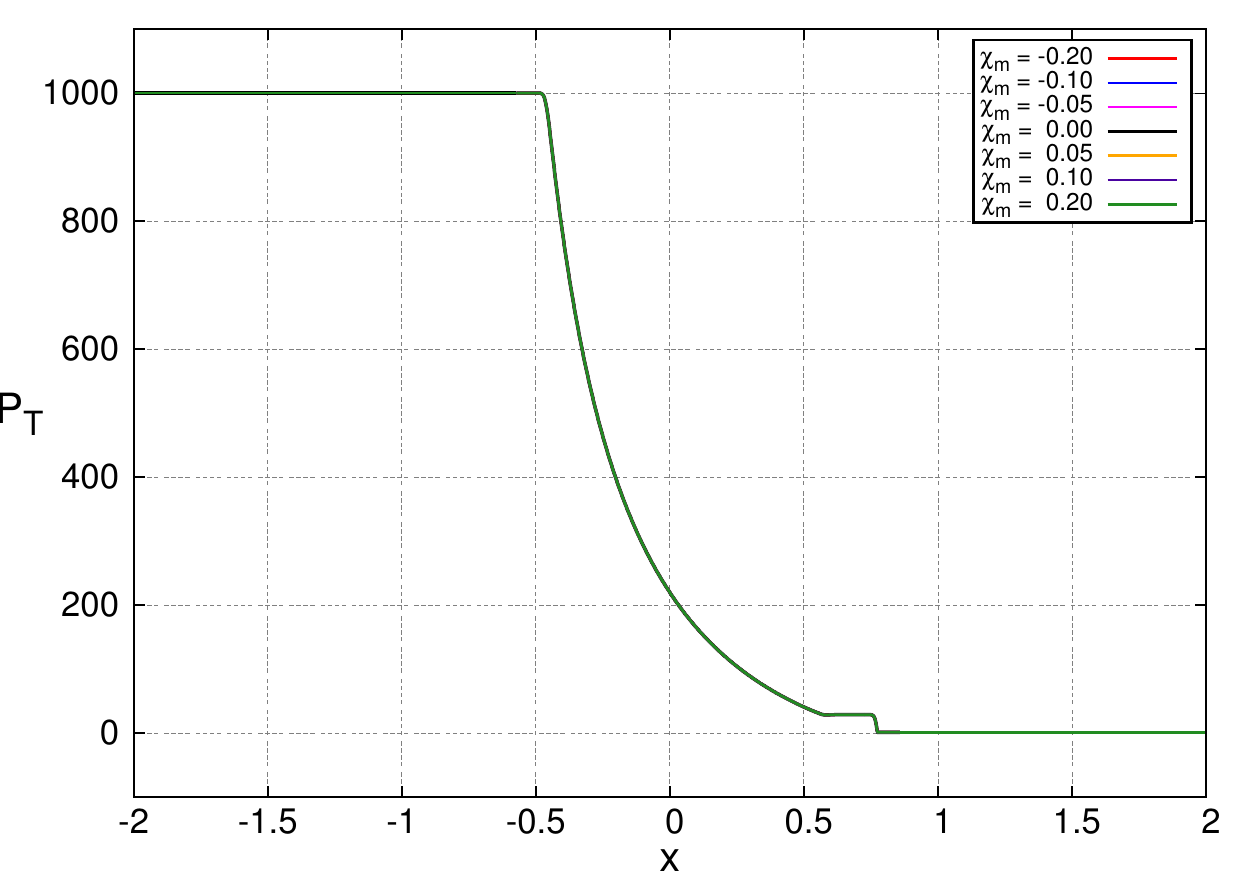}\\
\includegraphics[scale=0.7]{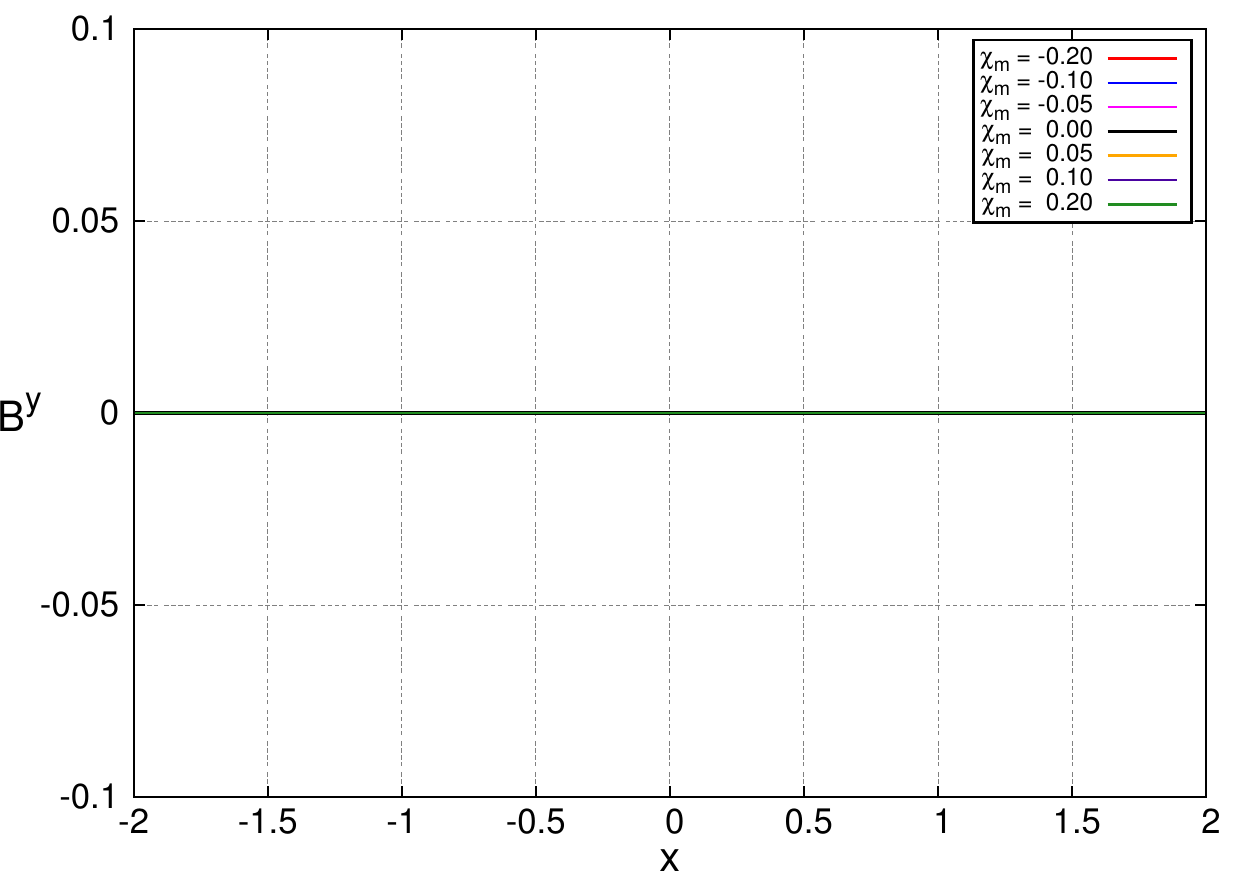} &
\includegraphics[scale=0.7]{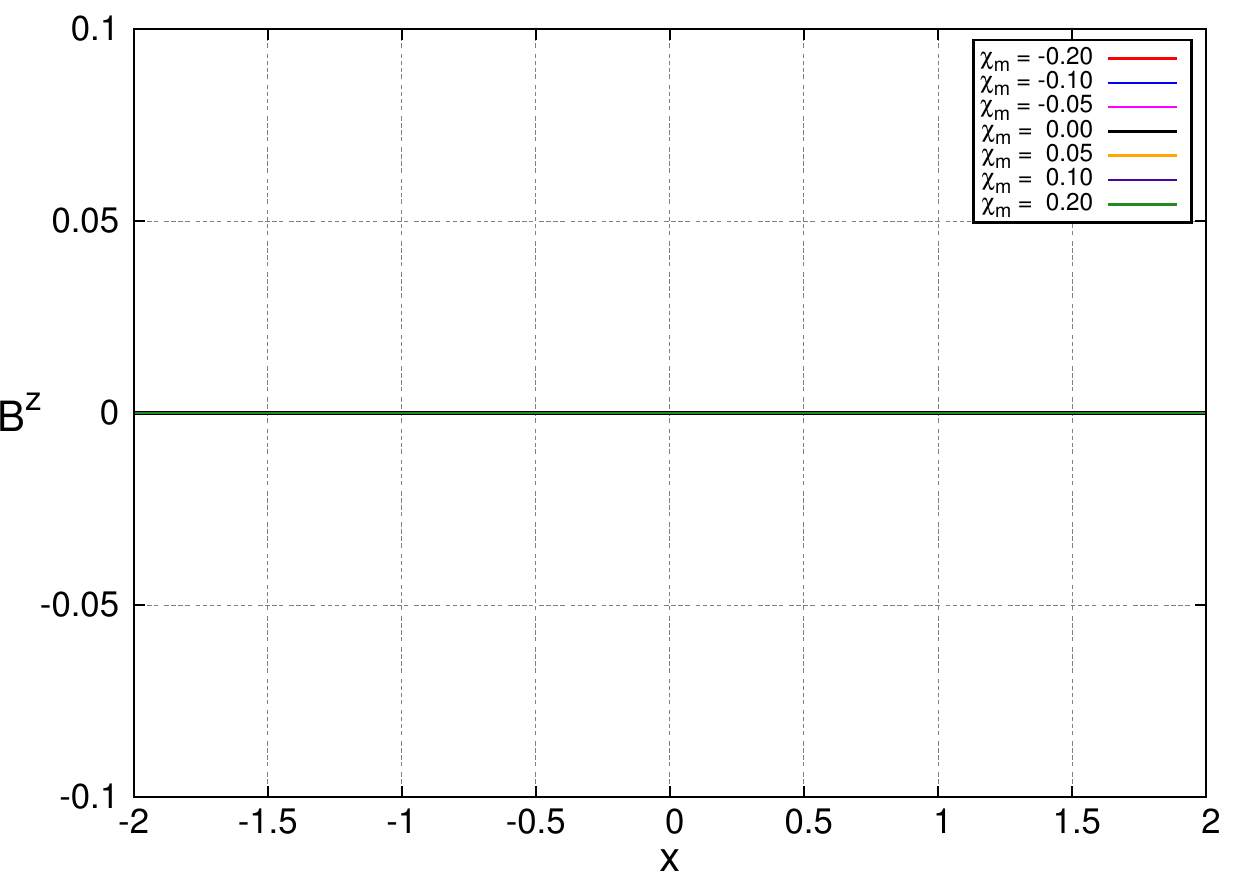}\\
\includegraphics[scale=0.7]{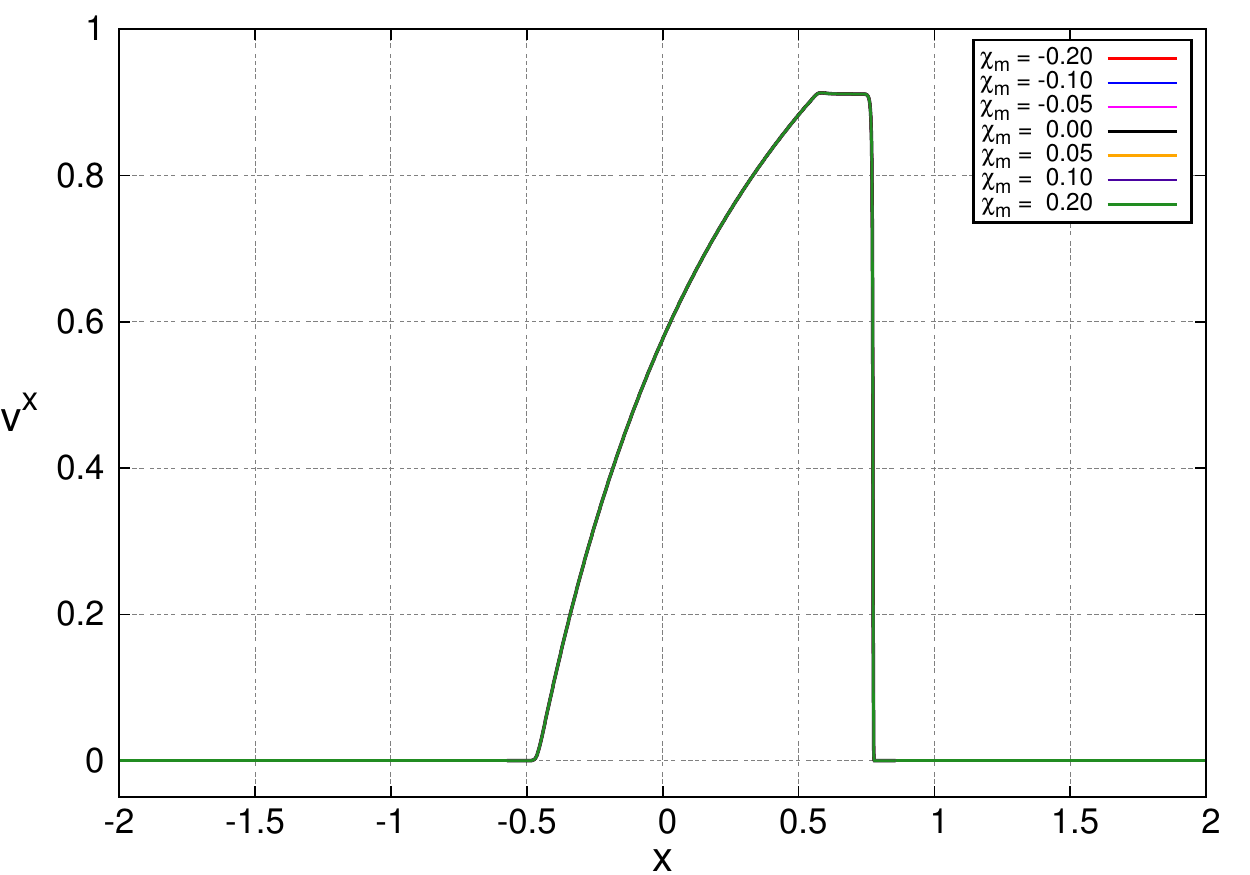} &
\includegraphics[scale=0.7]{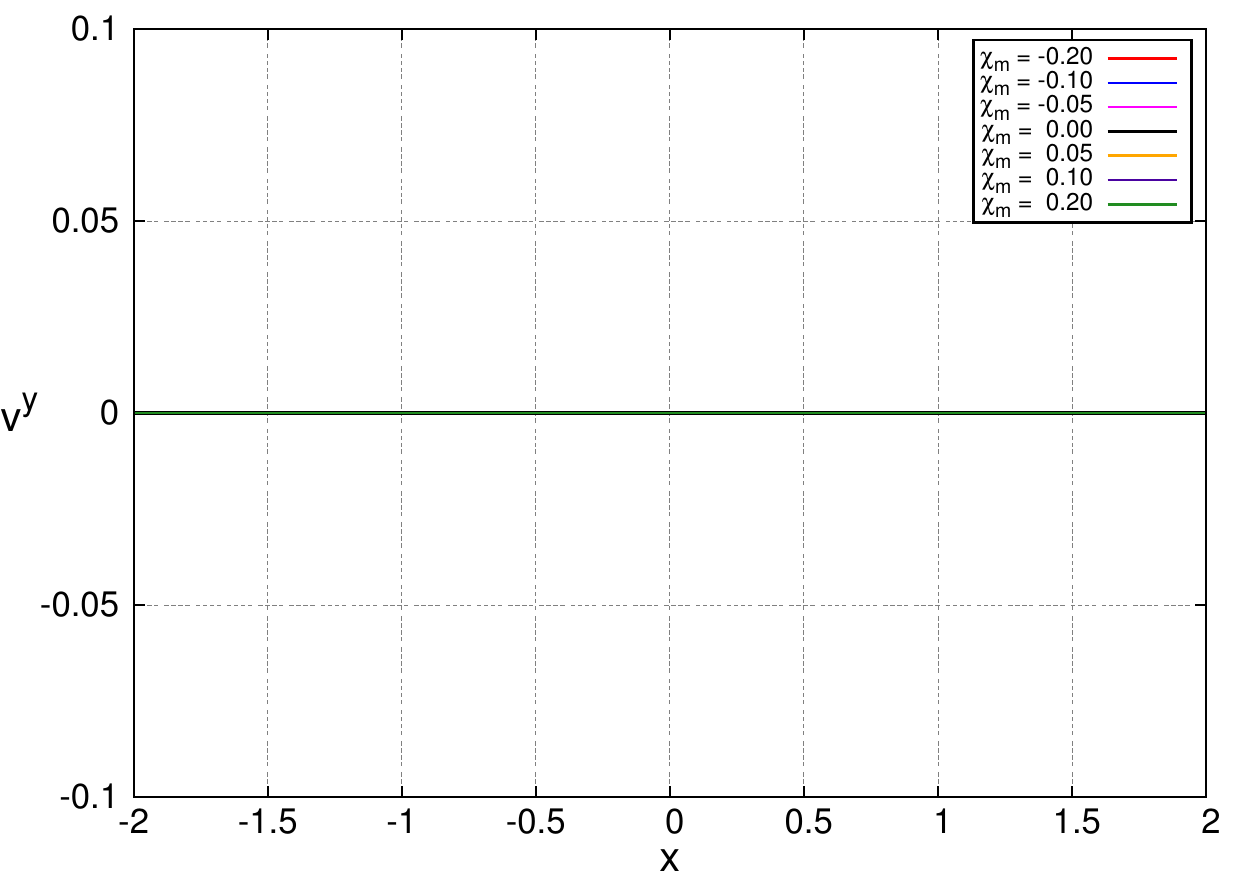}
\end{tabular}
\caption{Komissarov shock tube test at time $t=0.8$. We use a spatial resolution of $\Delta x=1/800$ and a Courant factor of 0.25.}
\label{test1}
\end{figure*}

The first shock tube test consists of a fluid with $\Gamma=4/3$, initially at rest, and with a constant magnetic field, which is normal to the initial discontinuity. The rest mass density and pressure on the left state are different from those on the right. However, the main feature of this test is that the left pressure is $10^{3}$ times the right one. Such a gradient produces a thin shell in the density that moves with a relativistic velocity of approximately $0.9$ times the speed of light. This behaviour is showed in Figure \ref{test1} at time $t=0.8$, for the magnetic susceptibilities $\chi_m=-0.20, -0.10, -0.05, 0.00, 0.05, 0.10, 0.20$. We can also distinguish a contact discontinuity and a shock wave both moving to the right, and a rarefaction zone in the central region of the domain. This test is very interesting because the magnetic field is normal to the initial discontinuity and therefore, as it is pointed out in \cite{1999MNRAS.303..343K}, the solution is purely hydrodynamical {\em i.e.} the magnetic field does not have a dynamical effect on the evolution. As a consequence of this fact, the solution showed in Figure \ref{test1} does not change with the magnetic susceptibility. 

\subsubsection{Komissarov collision test}

\begin{figure*}
\begin{tabular}{cc}
\includegraphics[scale=0.7]{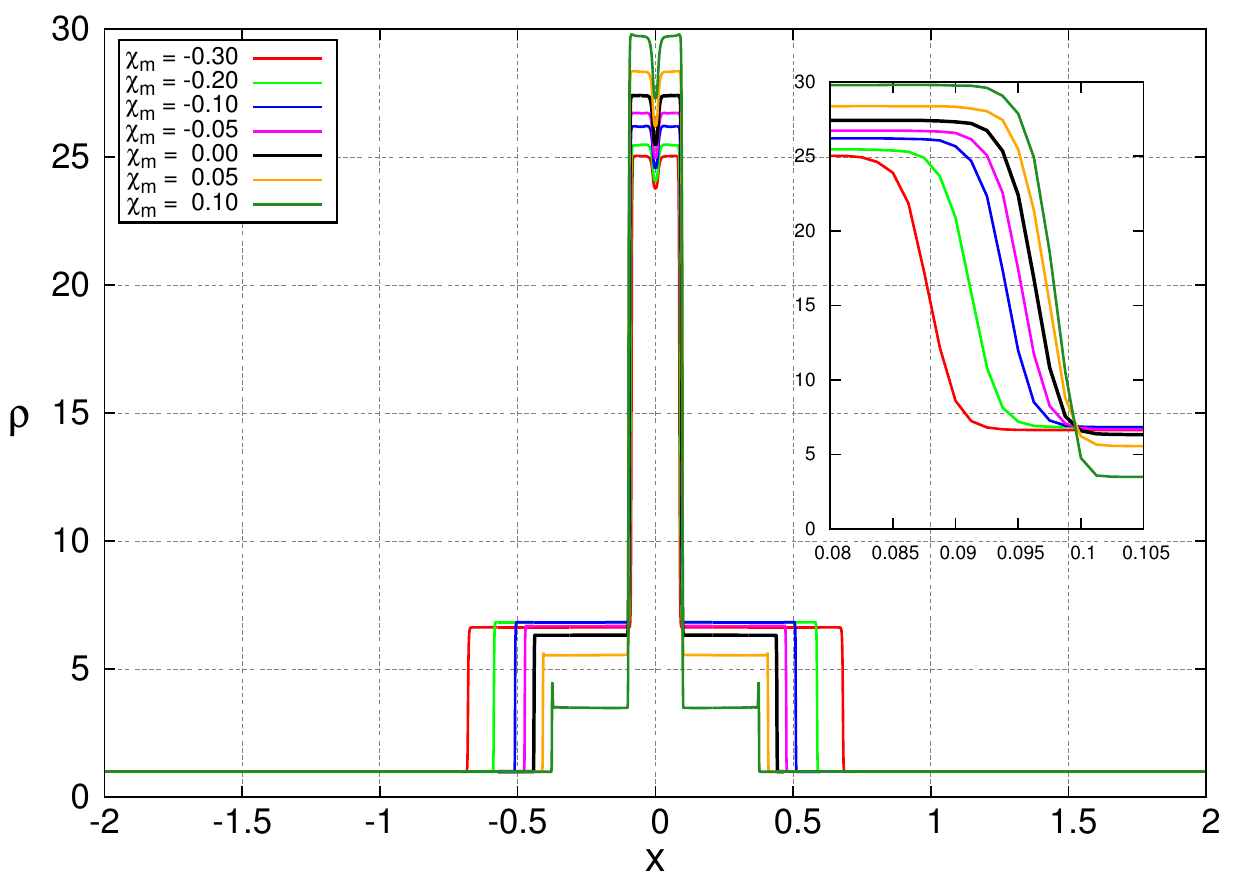} & \includegraphics[scale=0.7]{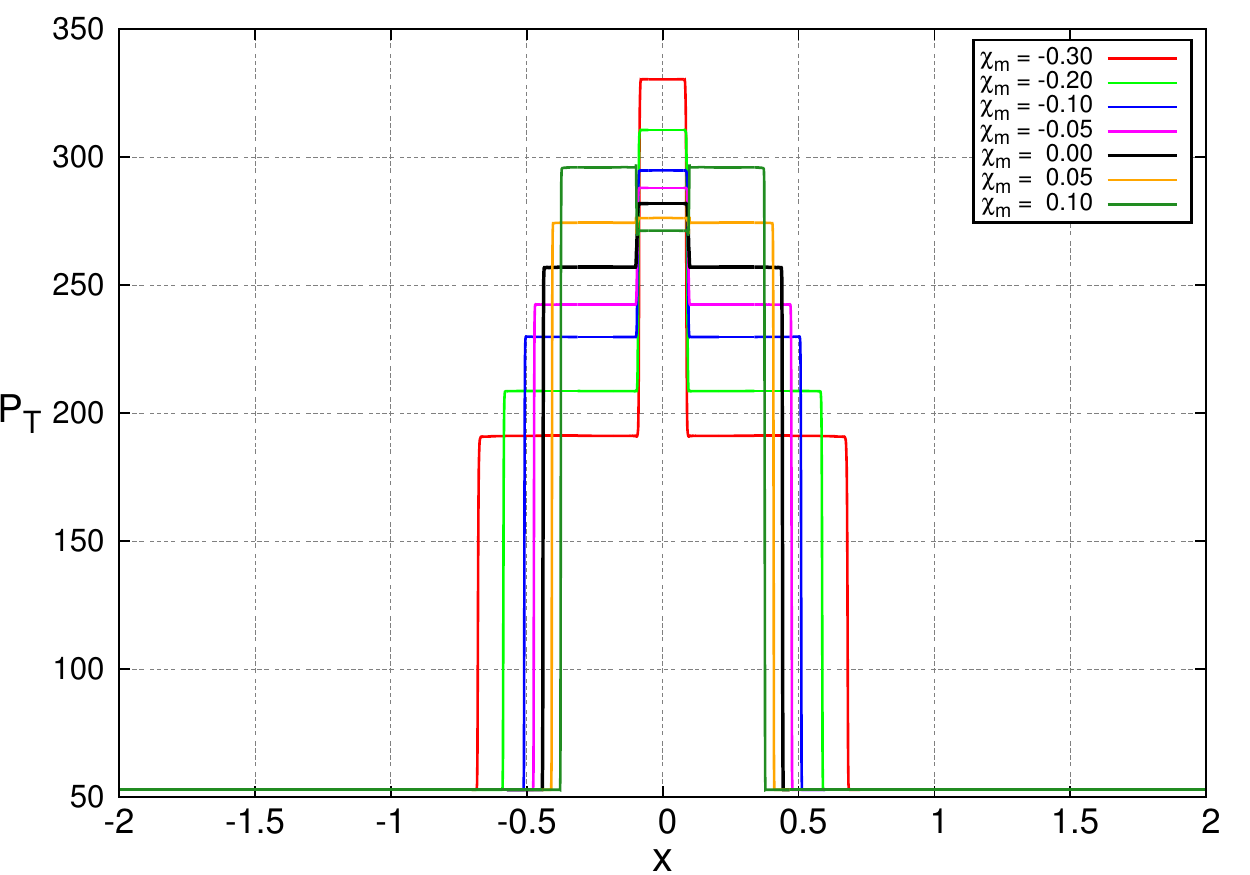}\\
\includegraphics[scale=0.7]{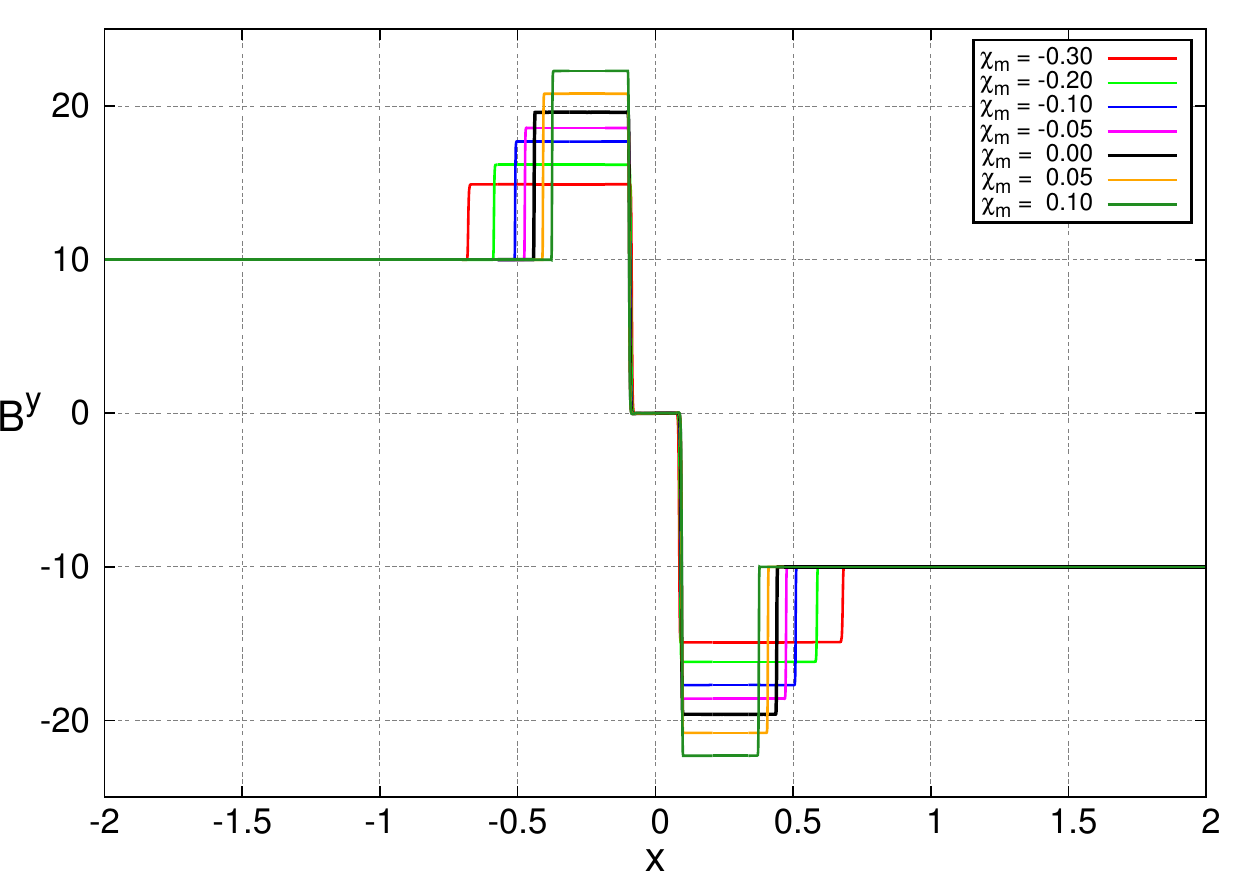} &
\includegraphics[scale=0.7]{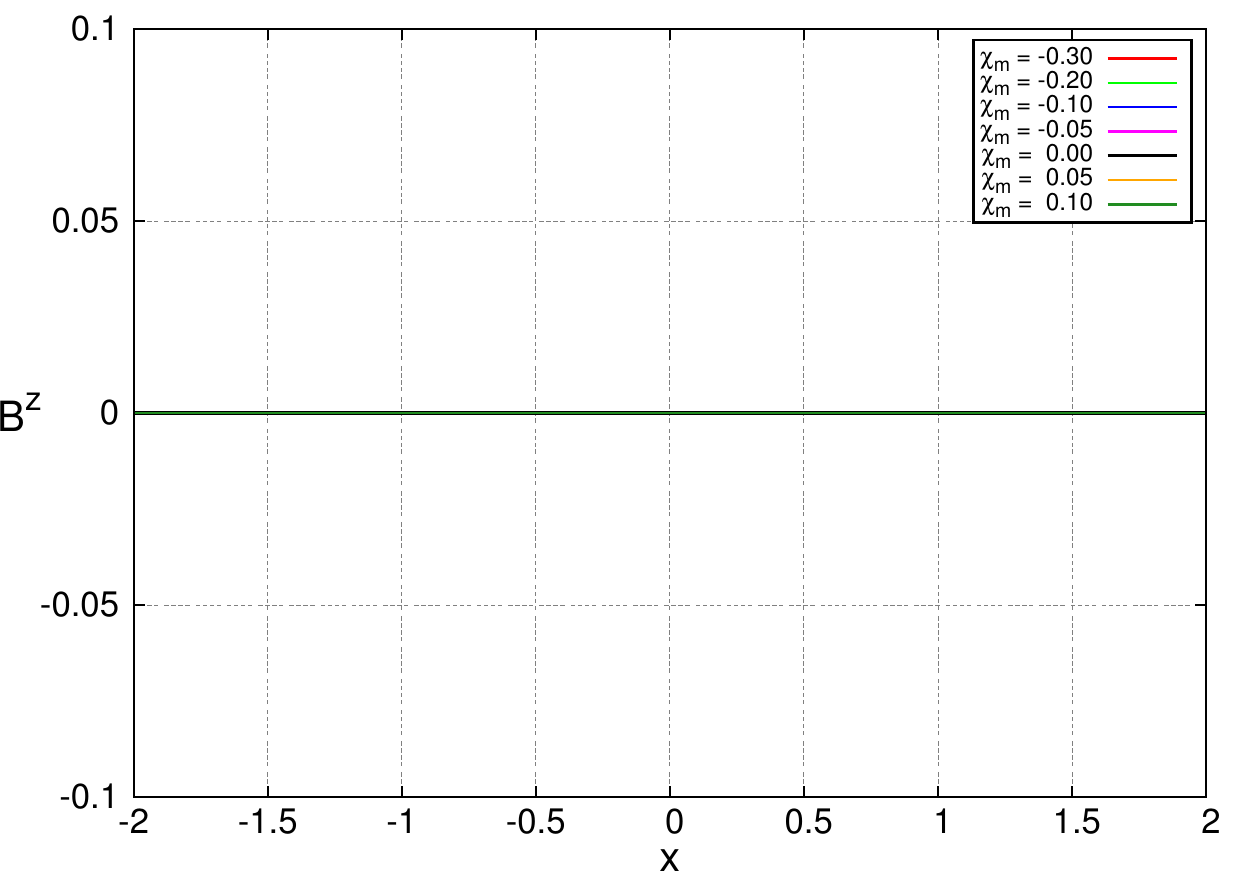}\\
\includegraphics[scale=0.7]{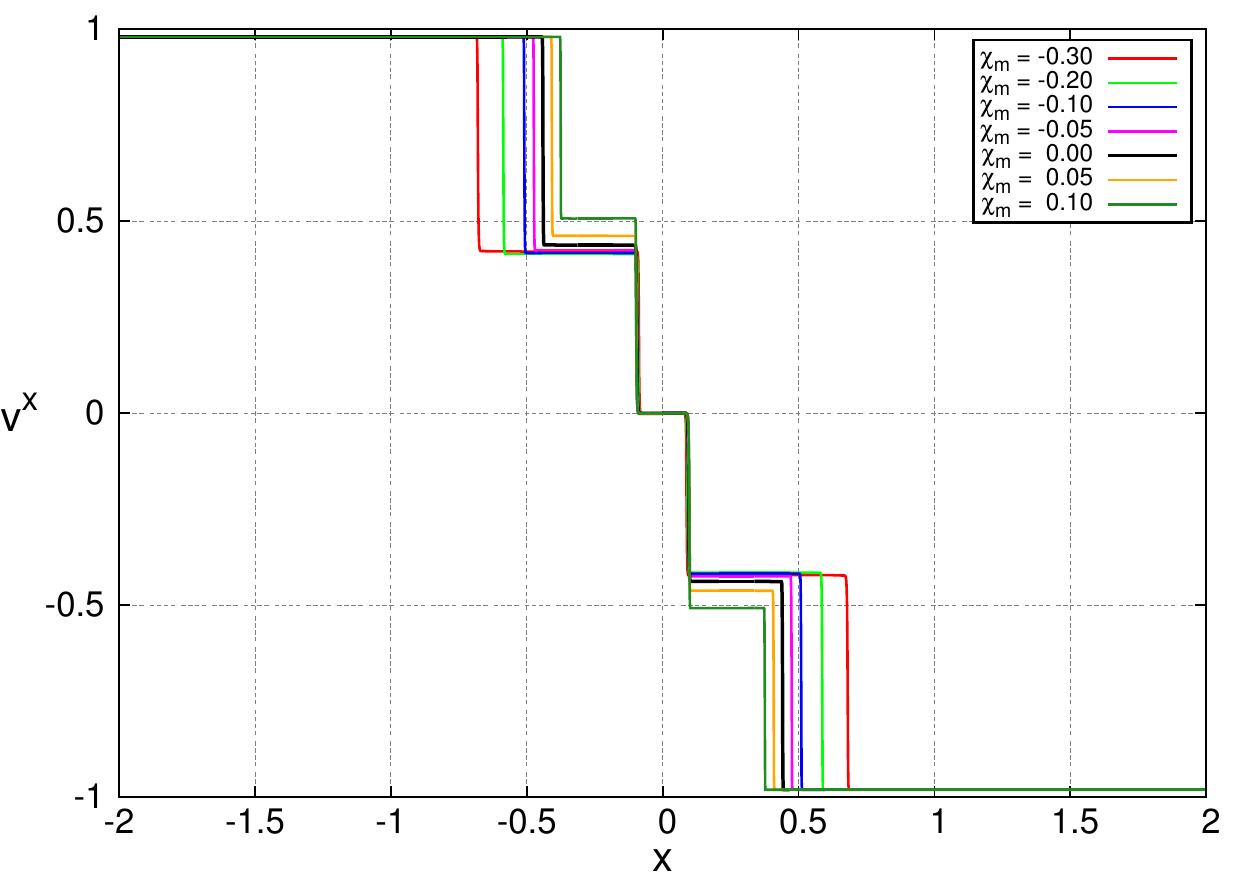} &
\includegraphics[scale=0.7]{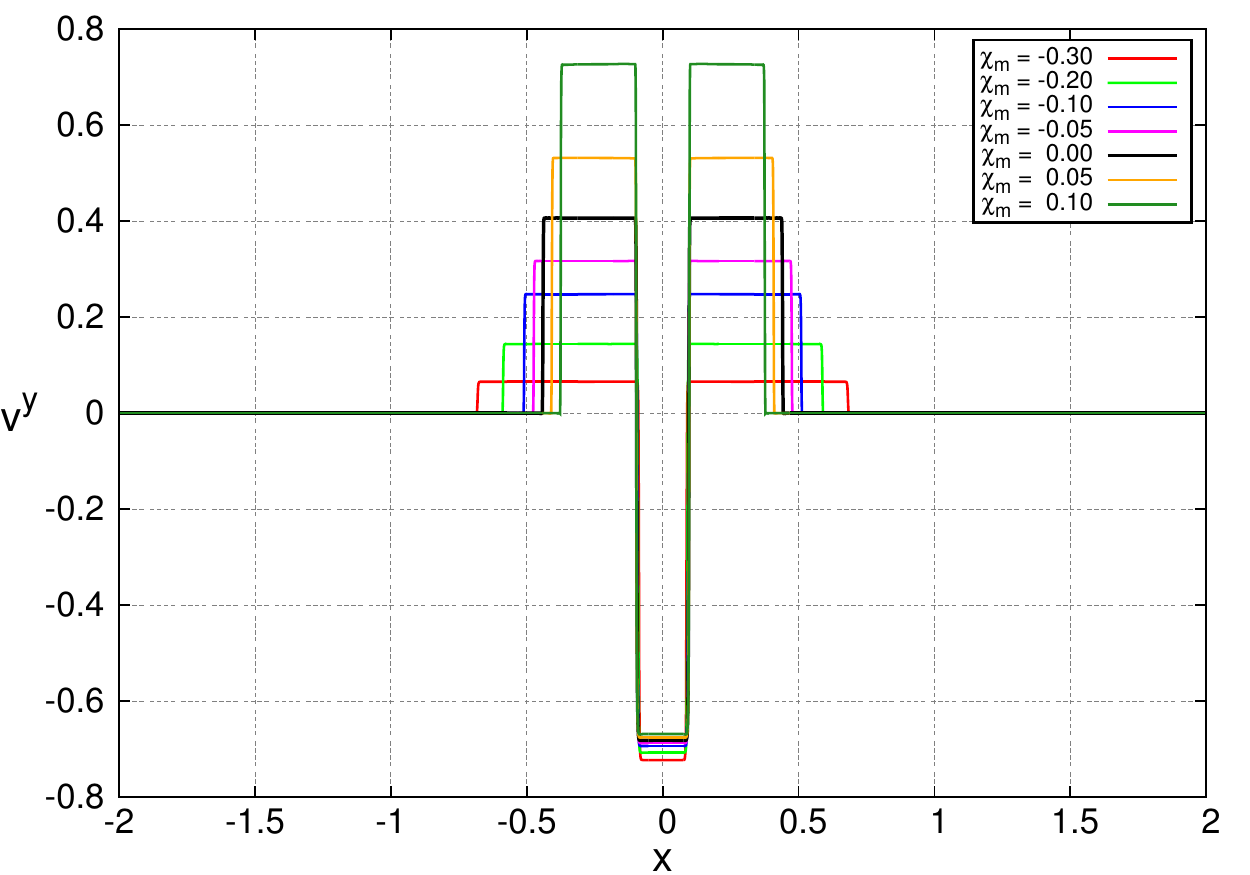}
\end{tabular}
\caption{Komissarov collision test at time $t=0.8$. We use a spatial resolution of $\Delta x=1/800$ and a Courant factor of 0.25.}
\label{test2}
\end{figure*}

The second test describes the head-on-one-dimensional collision of two fluids with the same rest mass density and pressure. Both fluids, characterized by an adiabatic index $\Gamma=4/3$, collide each other with a relativistic velocity $v^{x}=0.98058$ at the position $x=0$. In this test, the tangential components of the magnetic field $B^{y}$ on both states have opposite directions. The numerical solution of this initial value problem at time $t=0.8$ is presented in Figure \ref{test2}, for the magnetic susceptibilities $\chi_m=-0.30, -0.20, -0.10, -0.05, 0.00, 0.05, 0.10$. We can observe two fast shock waves moving in opposite directions, with two slow shocks behind each one of them. The solutions for different magnetic susceptibilities show that the fast shocks move faster in diamagnetic materials, and the slow shocks move slightly faster in paramagnetic ones.

Due to the symmetry of the collision, the tangential magnetic field and the normal velocity vanishes between the slow shocks (\cite{1999MNRAS.303..343K}), independently of $\chi_m$. Nevertheless, when the magnetic polarization in the fluid changes, the constant states between the waves take different values. For instance, the maximum value of the rest mass density increases with $\chi_{m}$, the transverse component of the magnetic field in the constant states between the slow and fast shocks changes considerably from the usual RMHD case ($\chi_{m}=0$), and the total pressure gradient through the slow shocks is reversed from a particular positive value of $\chi_{m}$. Finally, it is worth mentioning that in the paramagnetic materials here considered, a transversal relativistic flow was developed. The relativistic character of this flow is significantly reduced in diamagnetic fluids.	

\subsubsection{Balsara 1}

\begin{figure*}
\begin{tabular}{cc}
\includegraphics[scale=0.7]{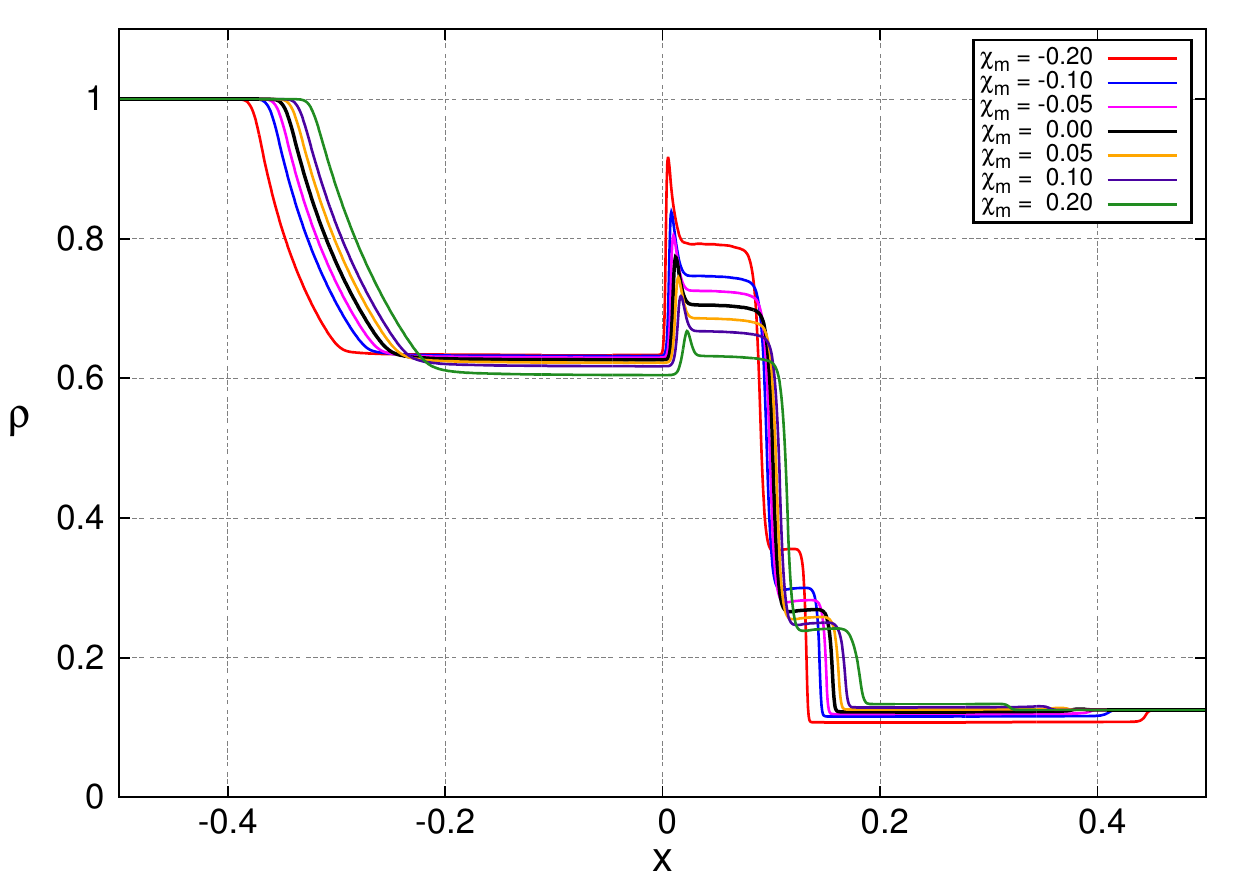} & \includegraphics[scale=0.7]{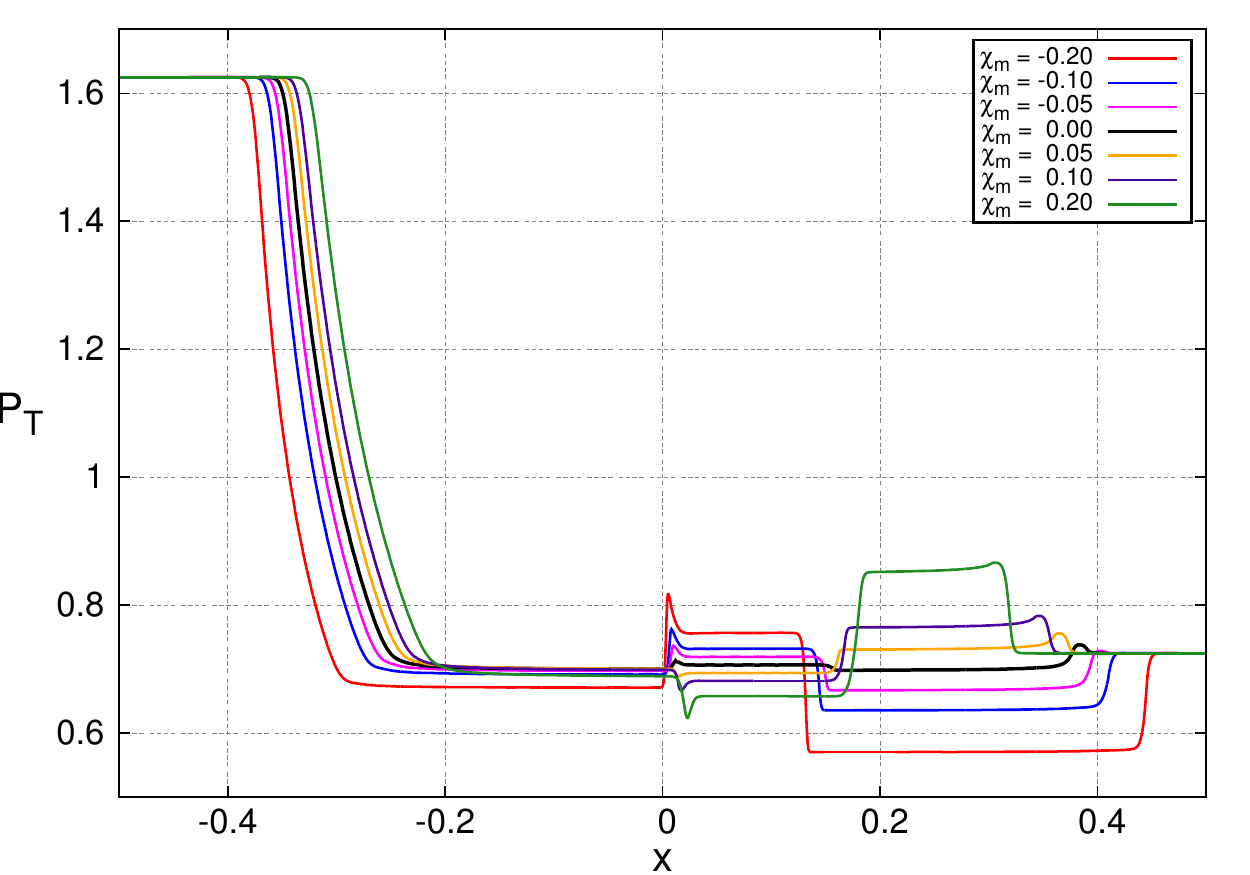}\\
\includegraphics[scale=0.7]{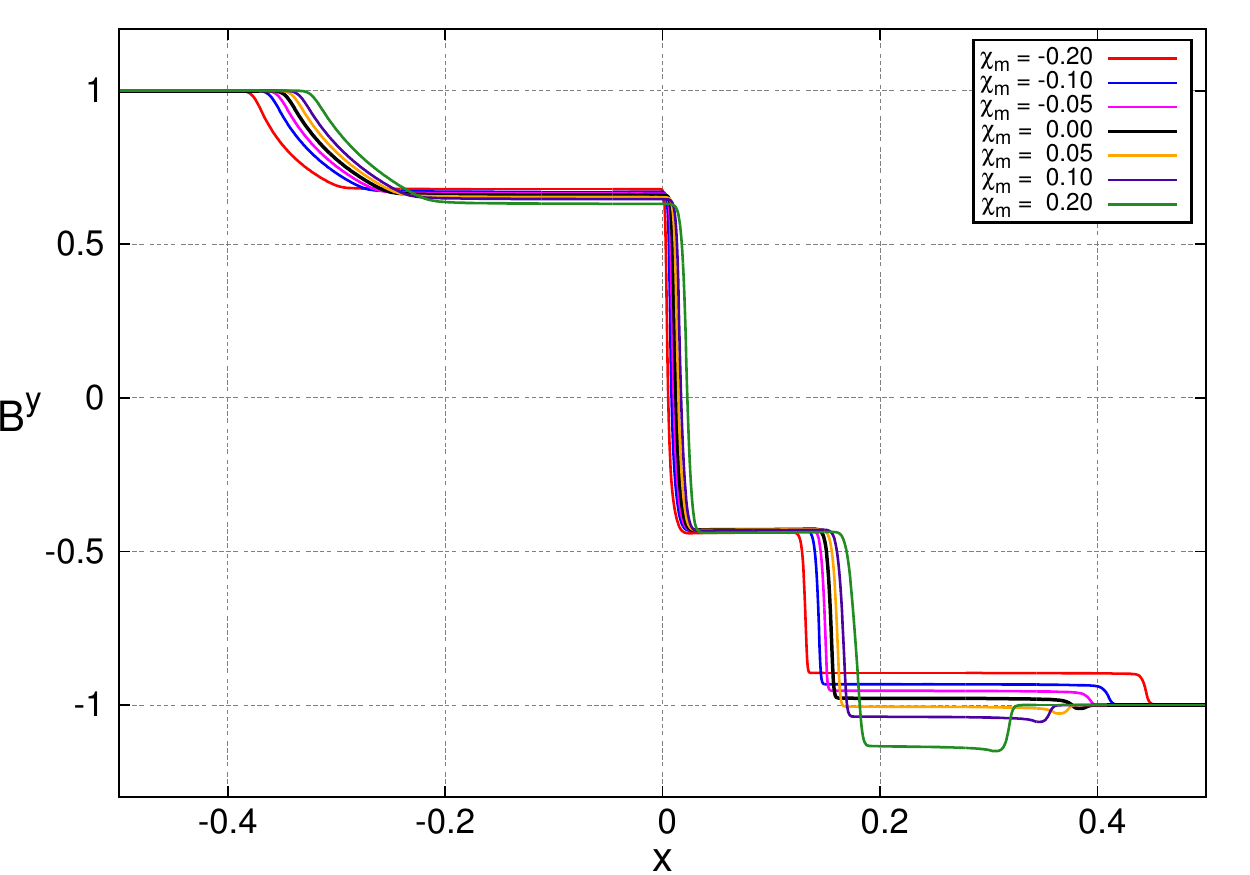} & \includegraphics[scale=0.7]{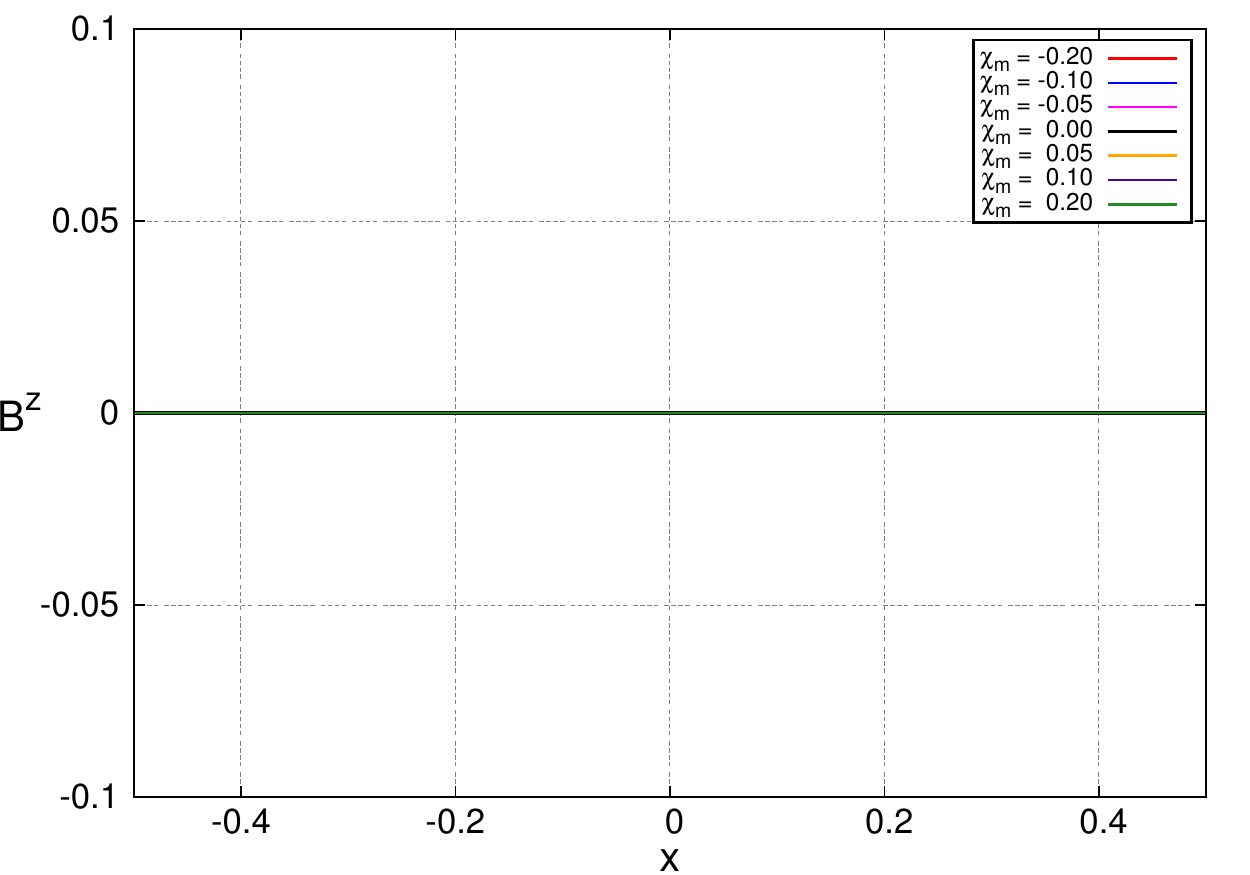}\\
\includegraphics[scale=0.7]{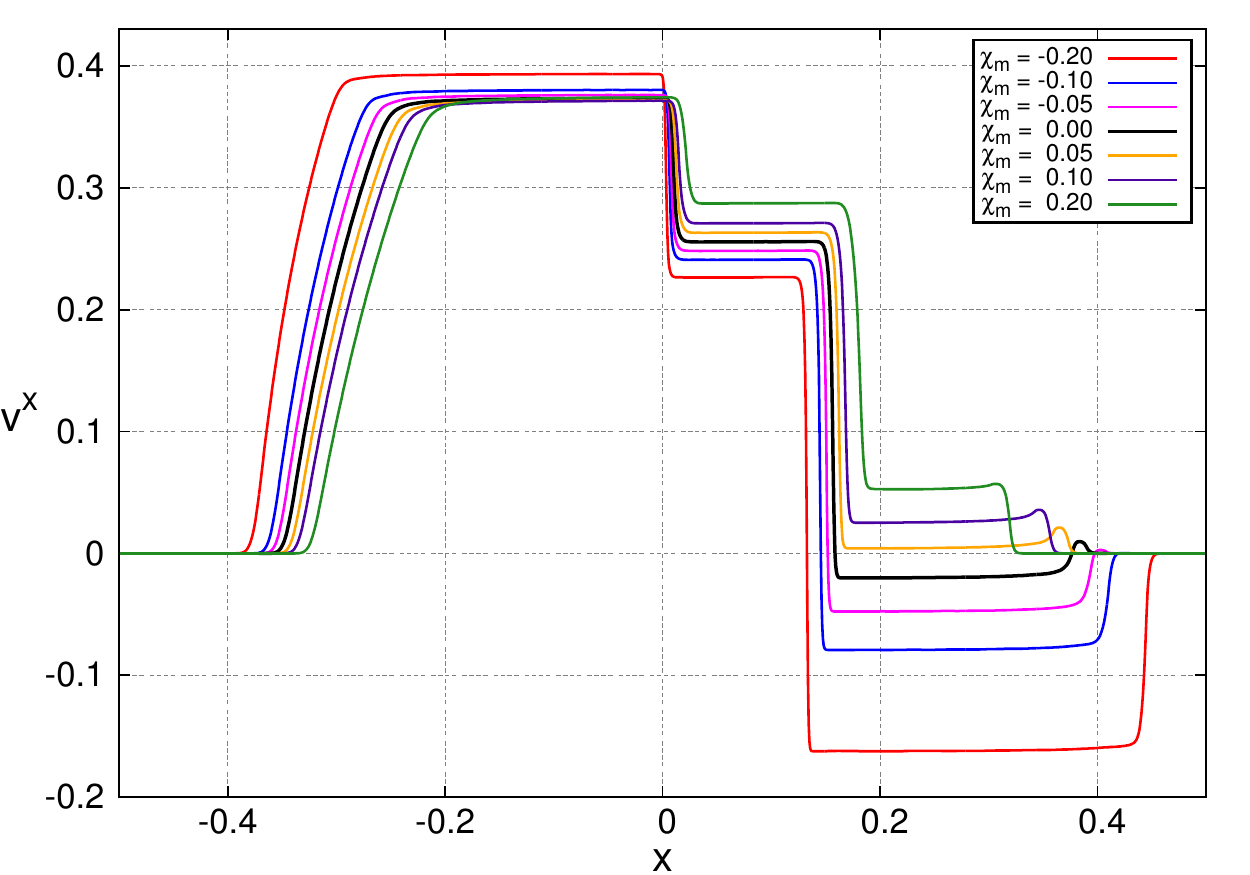} & \includegraphics[scale=0.7]{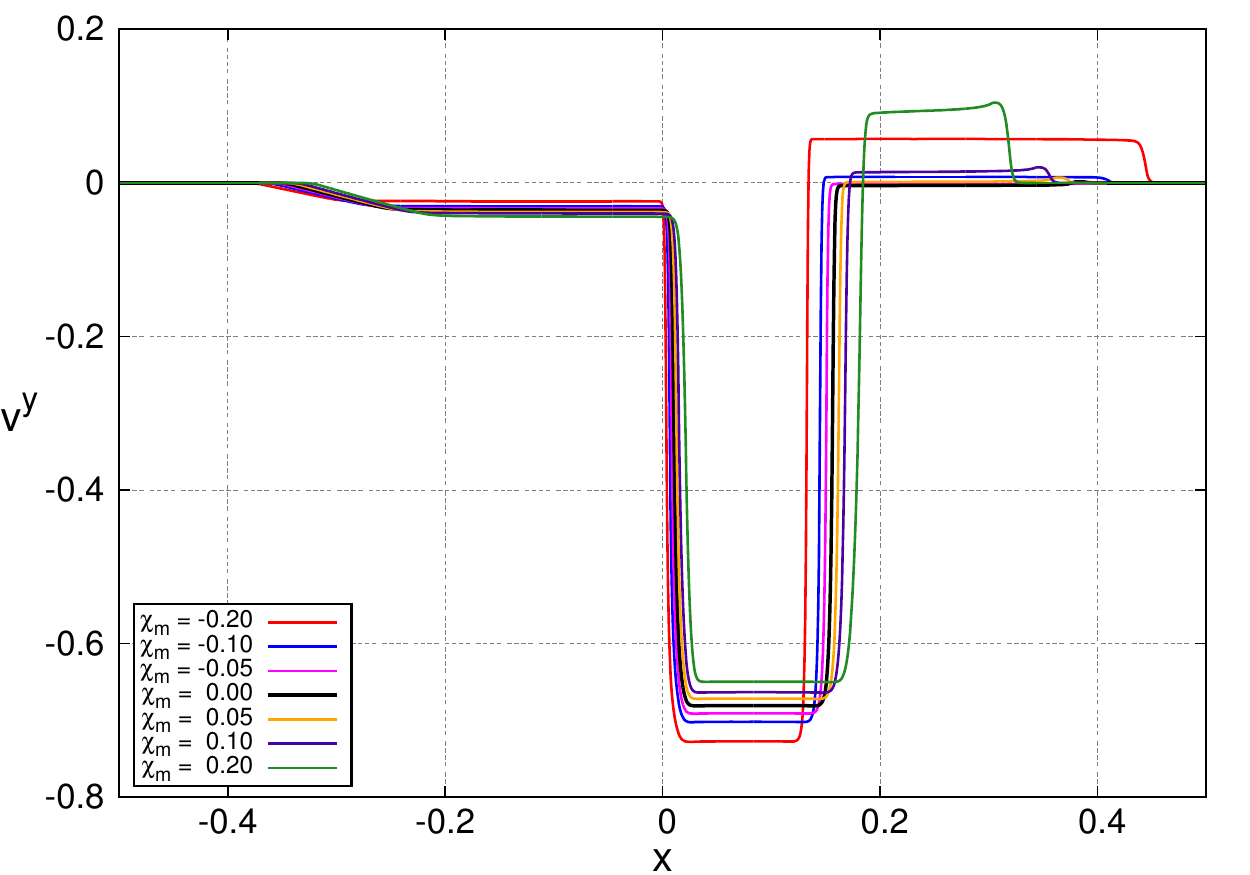}
\end{tabular}
\caption{Balsara 1 test at time $t=0.4$. We use a spatial resolution of $\Delta x=1/1600$ and a Courant factor of 0.25.}
\label{test3}
\end{figure*}

This test is the relativistic generalization of the Brio-Wu problem in Newtonian MHD (\cite{1988JCoPh..75..400B}). The initial data consists of a two-state fluid with $\Gamma=2$. On both sides of the diaphragm the fluid is initially at rest, but with different mass densities and pressures. Moreover, the normal component of the magnetic field to the interface is the same in both states, while the tangential component has opposite direction. Figure {\ref{test3}} shows the structure of the solution at $t=0.4$ for the values of magnetic susceptibility $\chi_m=-0.20, -0.10, -0.05, 0.00, 0.05, 0.10, 0.20$. For all these values, we obain solutions that consists of a left-going fast rarefaction, a left-going compound wave, a contact discontinuity, a right-going slow shock, and a right-going fast rarefaction. These waves are the same as those obtained in the usual relativistic MHD simulations (\cite{2015ApJS..218...24L}, \cite{2007CQGra..24S.235G}). Nevertheless, the rarefaction waves move considerably faster for diamagnetic materials than for those with paramagnetic properties, while the compound, contact, and shock waves move slower in diamagnetic fluids than in paramagnetic ones.

On the other hand, the rest mass density gradient through the shock becomes greater when the diamagnetic character of the fluid increases. The solution for the total pressure shows that the magnetic polarization tends to increment the gradients of this state variable through the waves. In particular, we obtain an interesting behavior across the slow shock, because after a certain positivive value of $\chi_{m}$ the total pressure gradient changes sign, so the dynamcis of the fluid becomes differet. For example, when $\chi_m\leq 0$ the fluid moves in opposite directions across the shock, but in the paramagnetic cases, all the fluid moves to the right. Furthermore, it is interesting to see that, between the shock and the right-going rarefaction, any magnetic polarization tends to generate a transversal flow in the opposite direction to the rest of the fluid. Finally, the magnetic field strength increases in paramagnetic fluids, specially in the constant state between the shock and the right-going rarefaction.

\subsubsection{Balsara 2}

\begin{figure*}
\begin{tabular}{cc}
\includegraphics[scale=0.7]{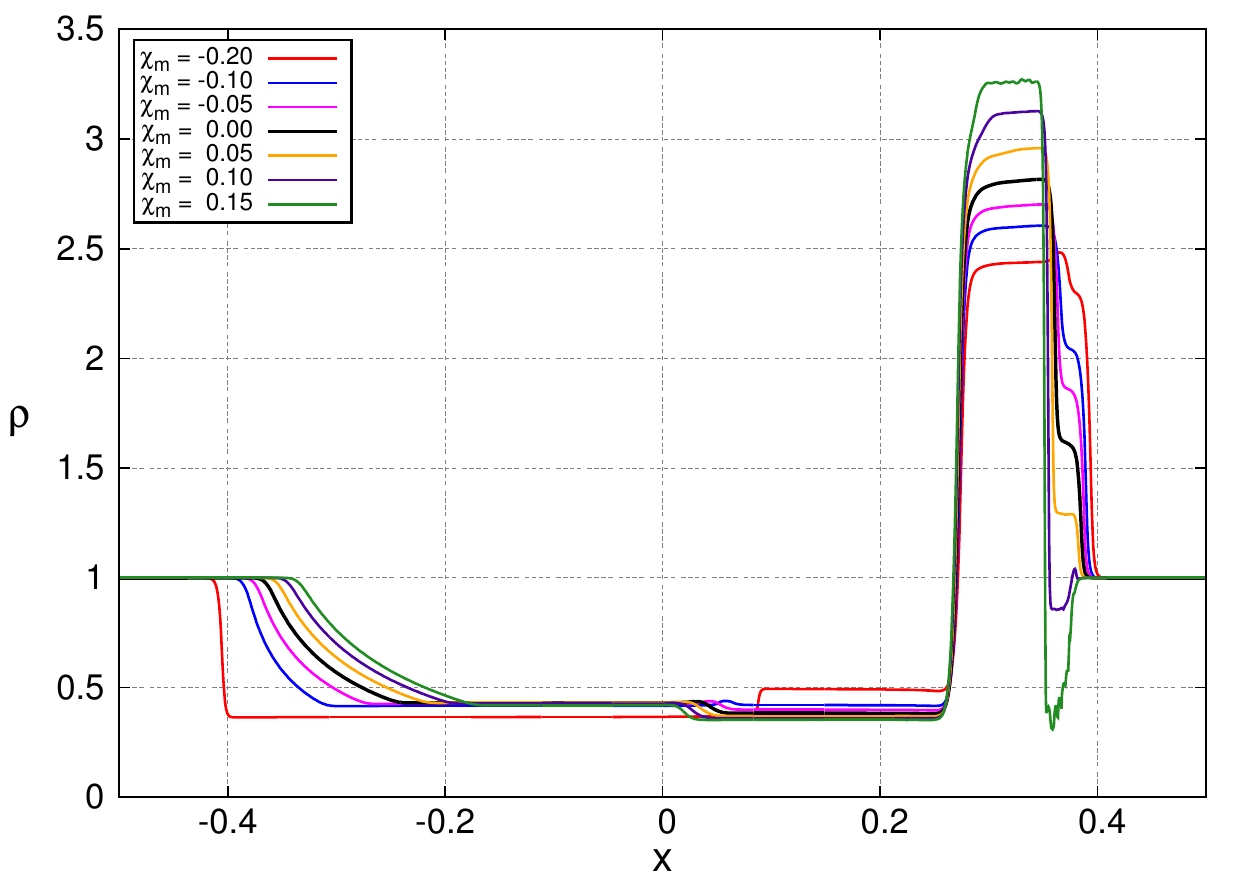} & \includegraphics[scale=0.7]{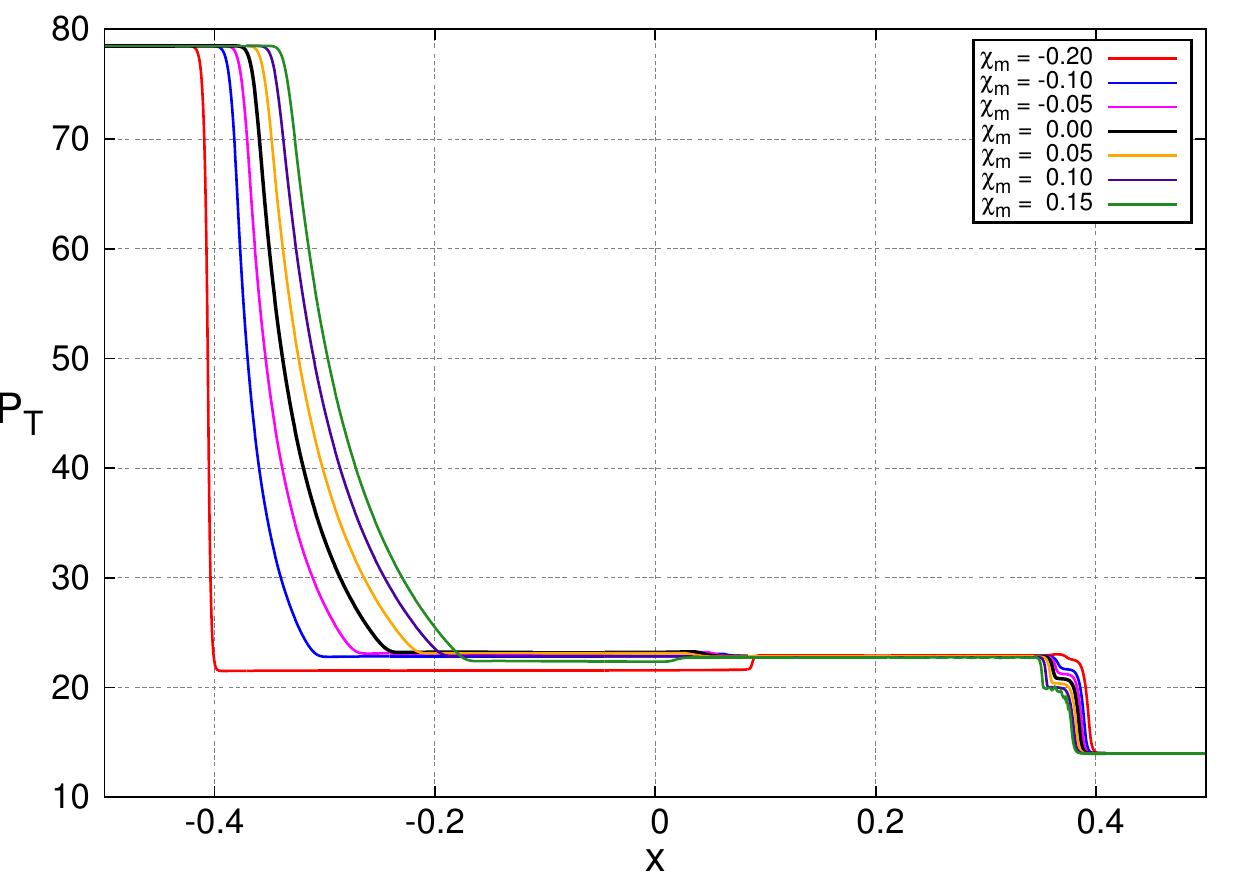}\\
\includegraphics[scale=0.7]{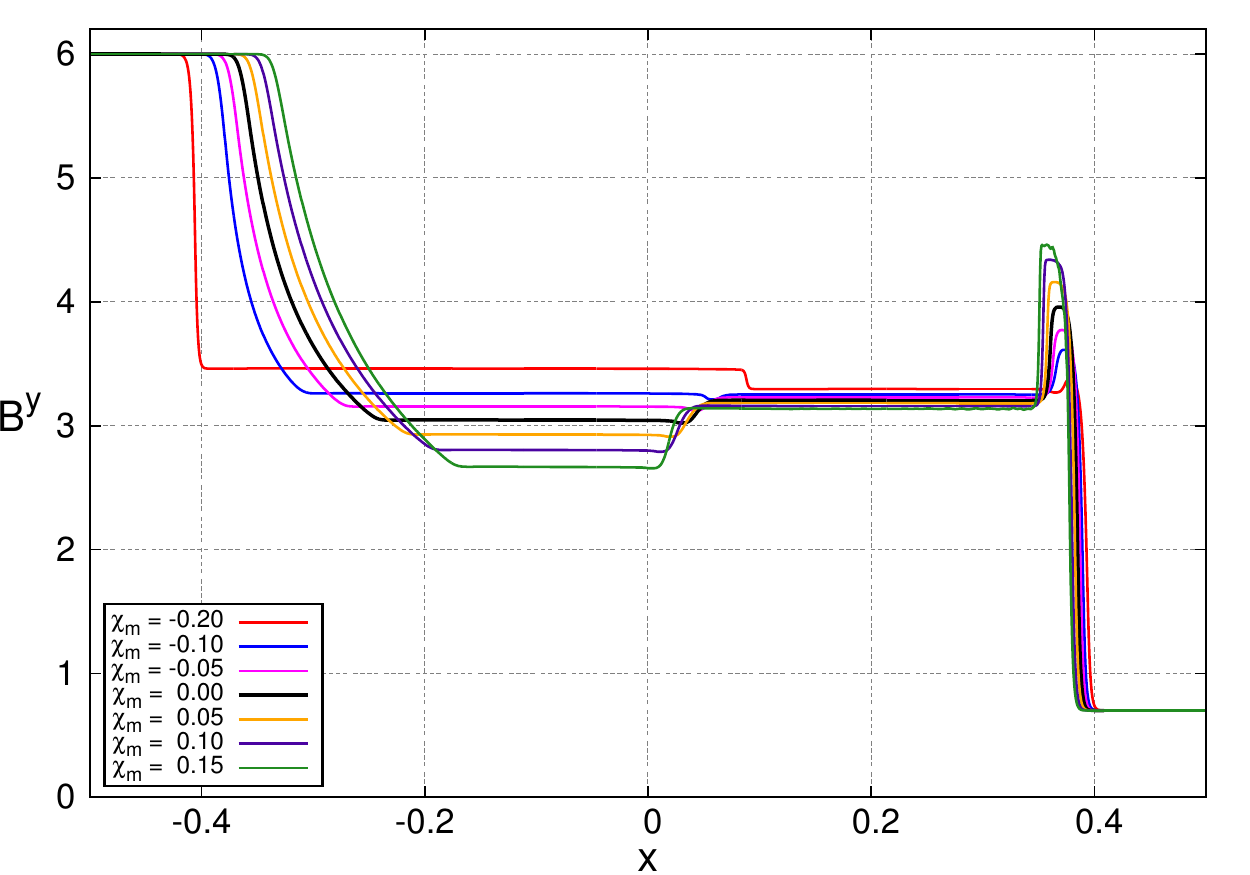} & \includegraphics[scale=0.7]{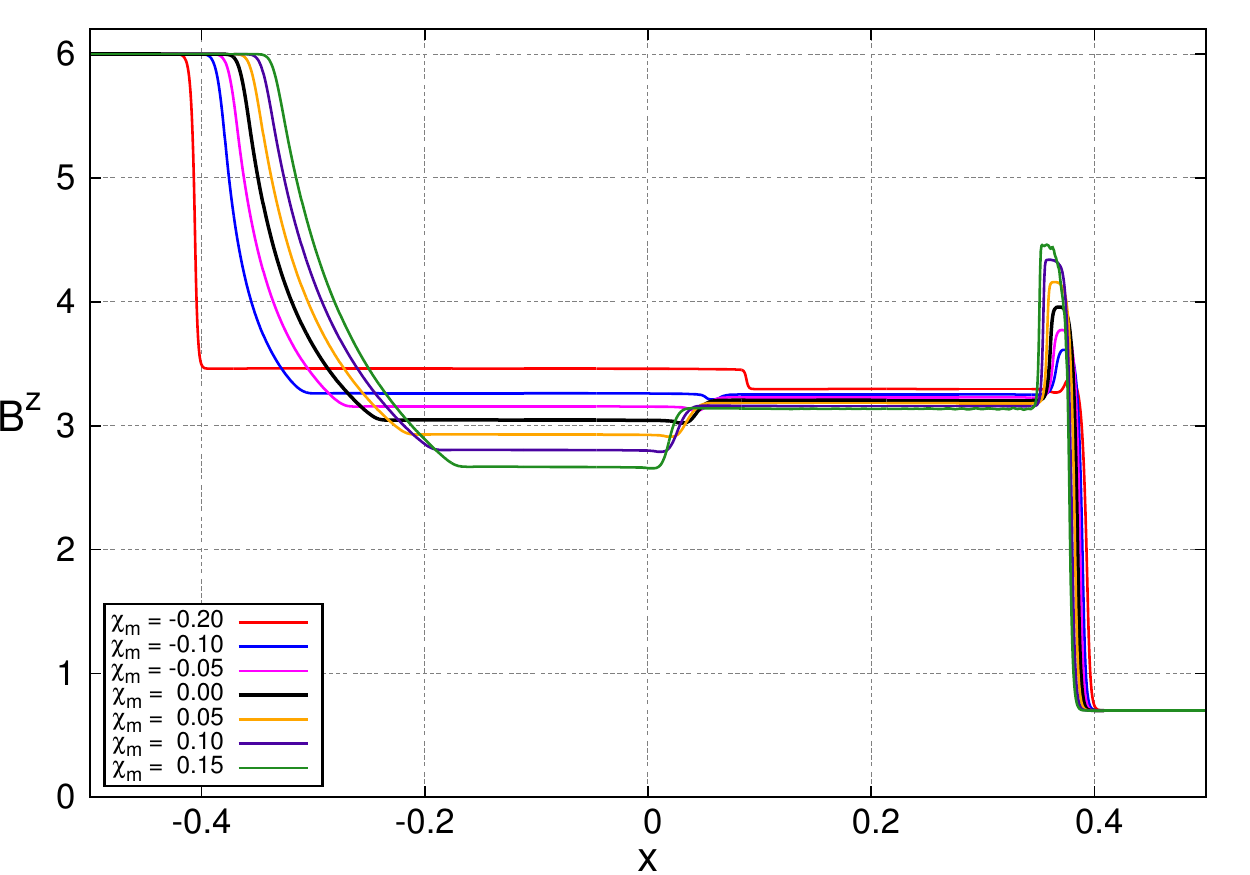}\\
\includegraphics[scale=0.7]{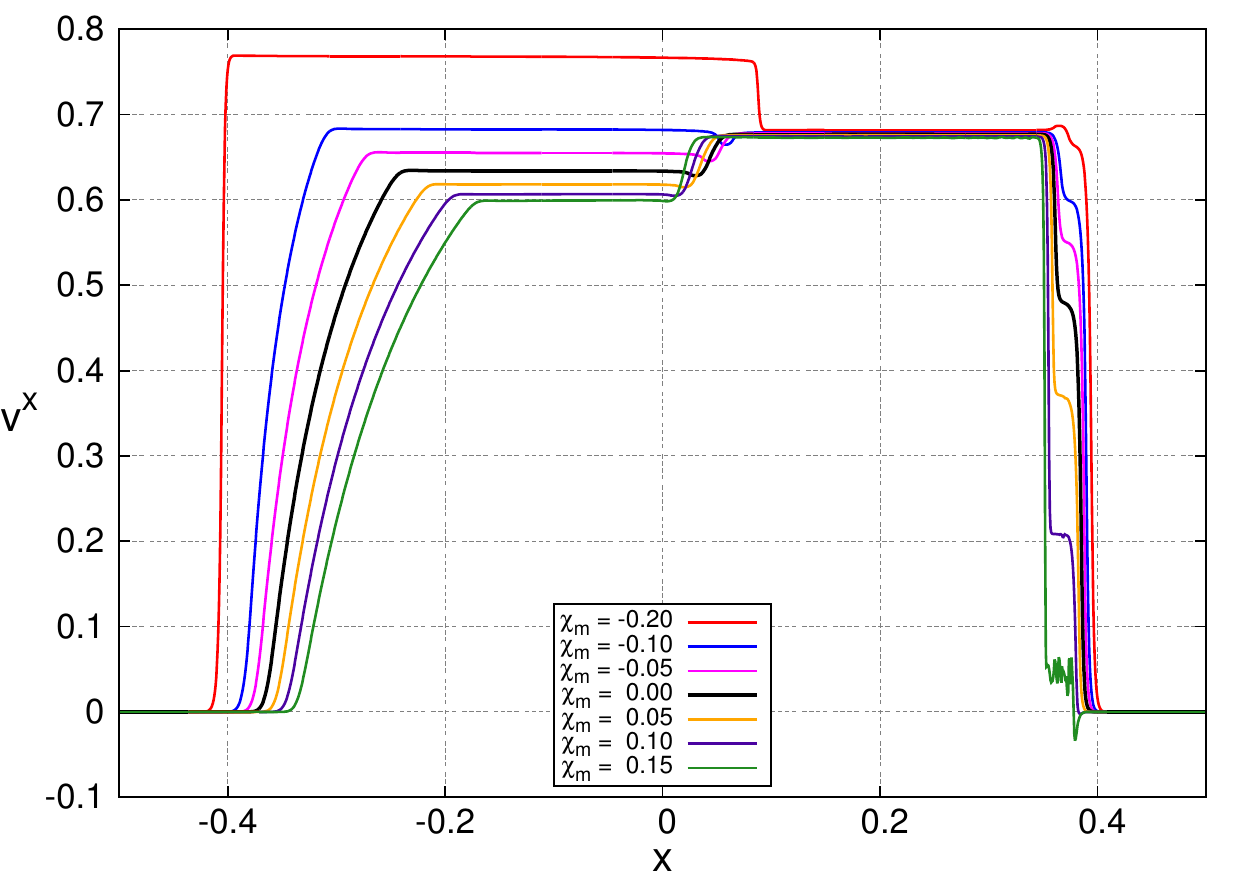} & \includegraphics[scale=0.7]{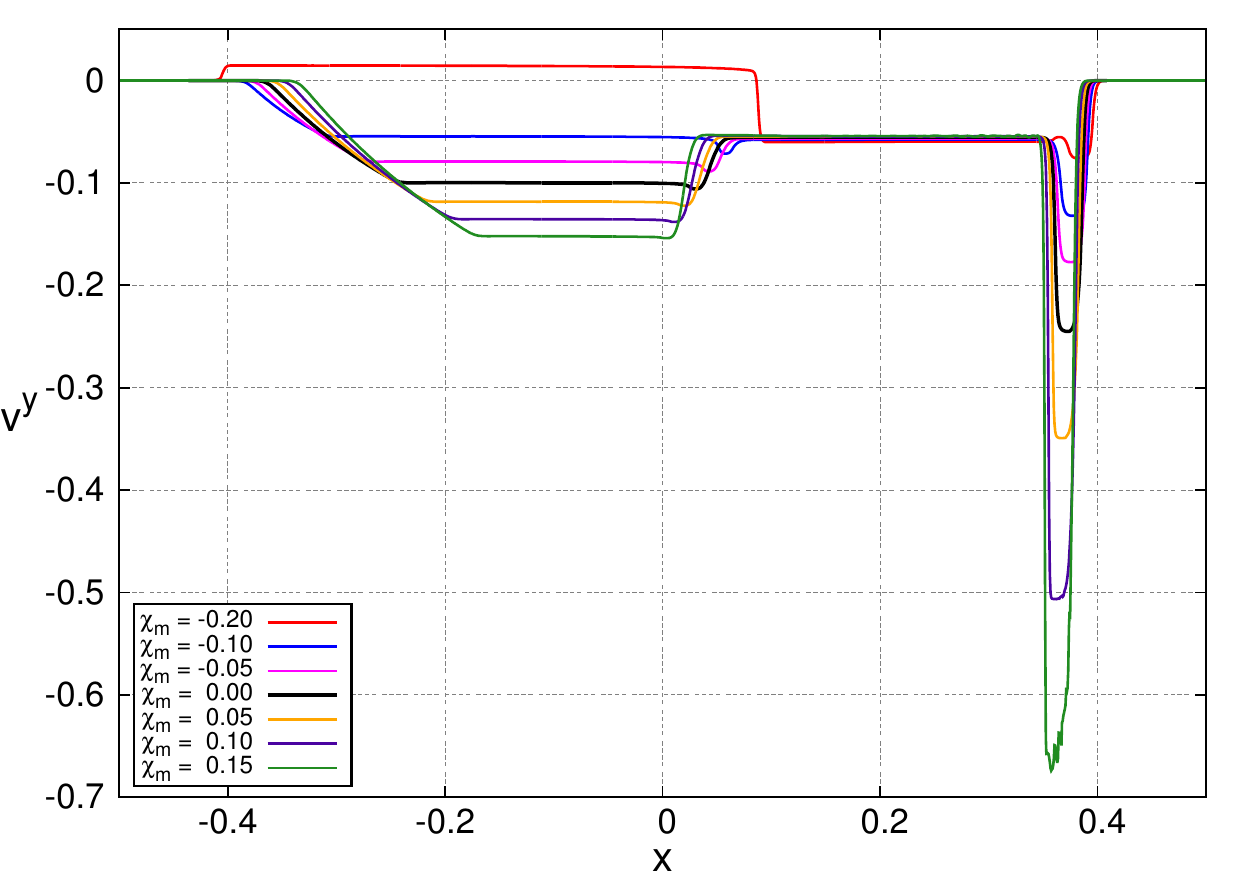}
\end{tabular}
\caption{Balsara 2 test at time $t=0.4$. We use a spatial resolution of $\Delta x=1/1600$ and a Courant factor of 0.25.}
\label{test4}
\end{figure*}

The second test of Balsara consists of a fluid with $\Gamma=5/3$ initially at rest, in which the left state pressure is 30 times the pressure on the right, and the transverse magnetic field is discontinuous in $x=0$. The numerical solution to this problem at time $t=0.4$, is presented in Figure \ref{test4} for the magnetic susceptibilities $\chi_m=-0.20, -0.10, -0.05, 0.00, 0.05, 0.10, 0.15$. The different plots shows a fast rarefaction wave moving to the left, a slow rarefaction wave moving to the right, and a contact right-going discontinuity that moves along with two shock waves. As we can see, all the waves in this solution move faster in diamagnetic materials than in paramagnetic ones, but the rarefaction waves presents the most significant differences.

In this test, it is important to mention that the rarefaction zones are reduced when we consider diamagnetic fluids, and that the most important changes in the state variables are obtained between the rarefaction waves and between the shocks. In particular, the relativistic character of the flow in the $x$-direction increases when we consider fluids increasingly diamagnetic. Conversely, a transversal relativistic flow is generated  between the shocks for the paramagnetic case with $\chi_m=0.15$. We can also observe a transversal flow that moves in the opposite direction from the rest of the fluid in the solution with $\chi_{m}=-0.2$. As a final comment, we notice that it is more difficult for the code to deal with paramagnetic materials, because the solution for $\chi_m=0.15$ presents oscillations, specially between the shocks.

\subsubsection{Balsara 3}

\begin{figure*}
\begin{tabular}{cc}
\includegraphics[scale=0.7]{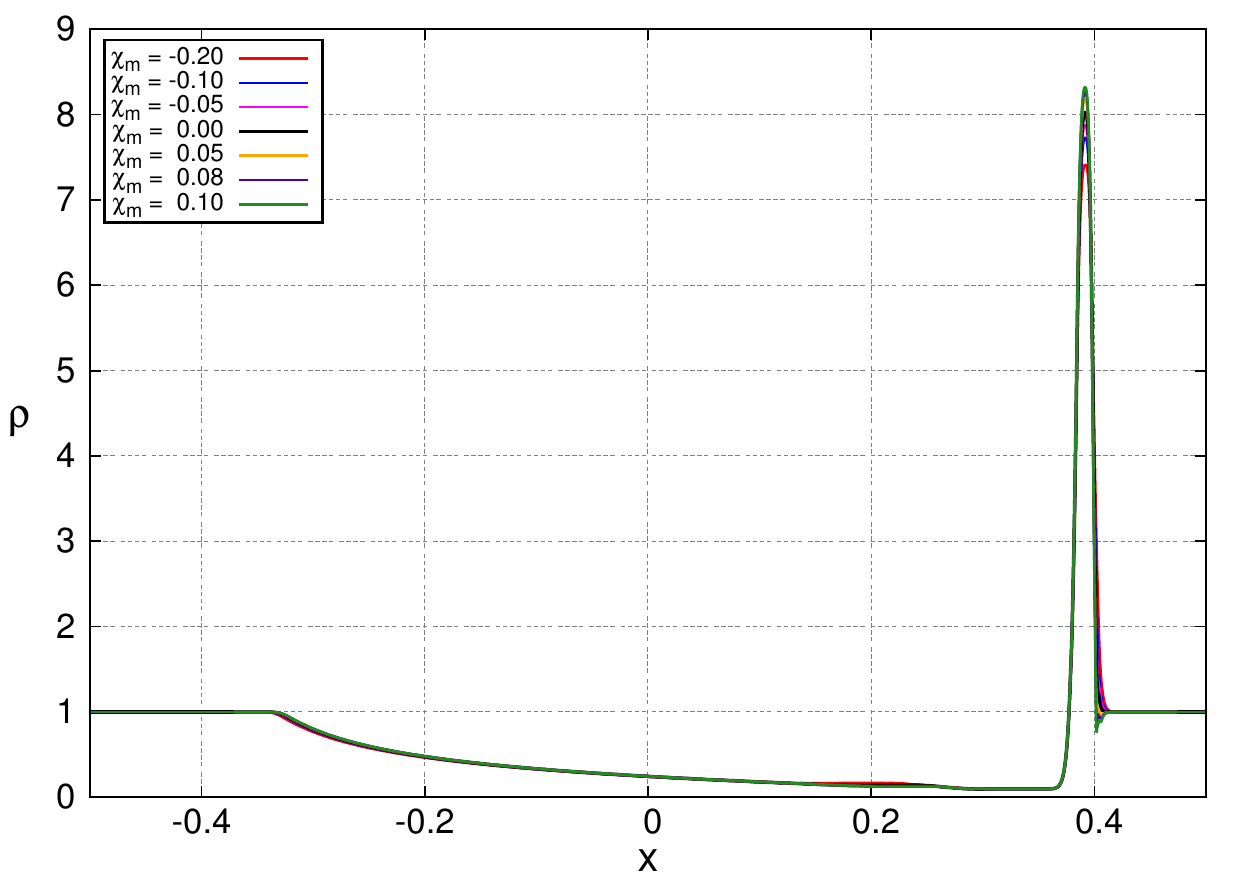} & \includegraphics[scale=0.7]{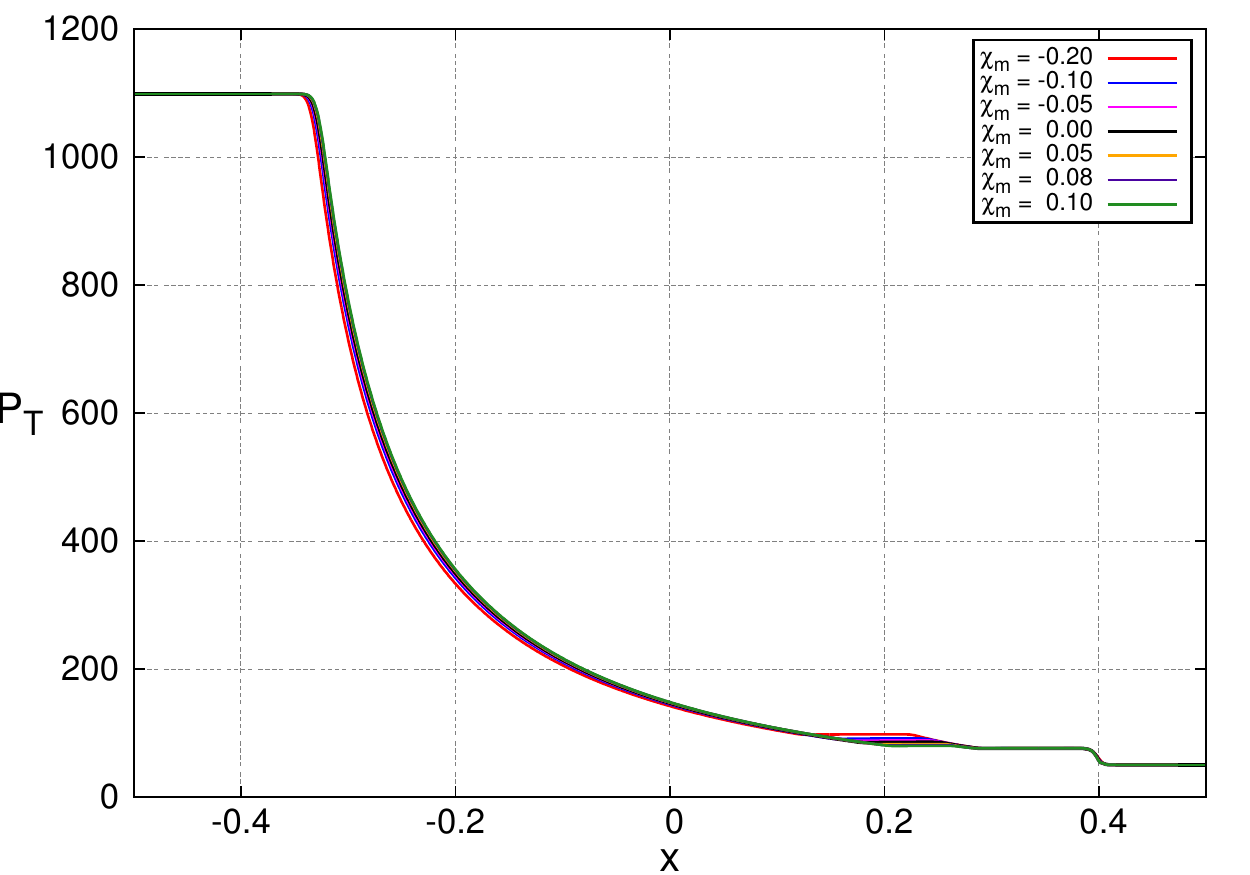}\\
\includegraphics[scale=0.7]{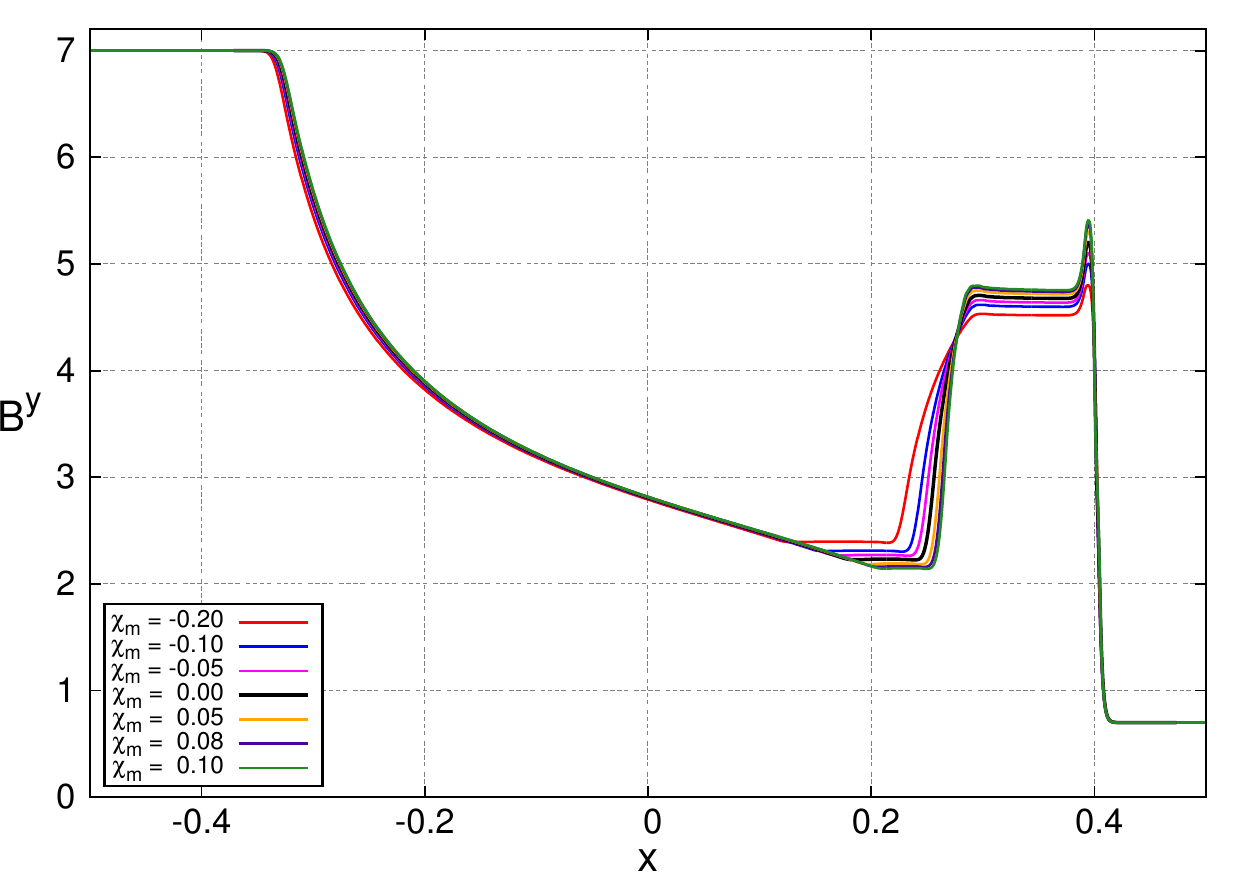} & \includegraphics[scale=0.7]{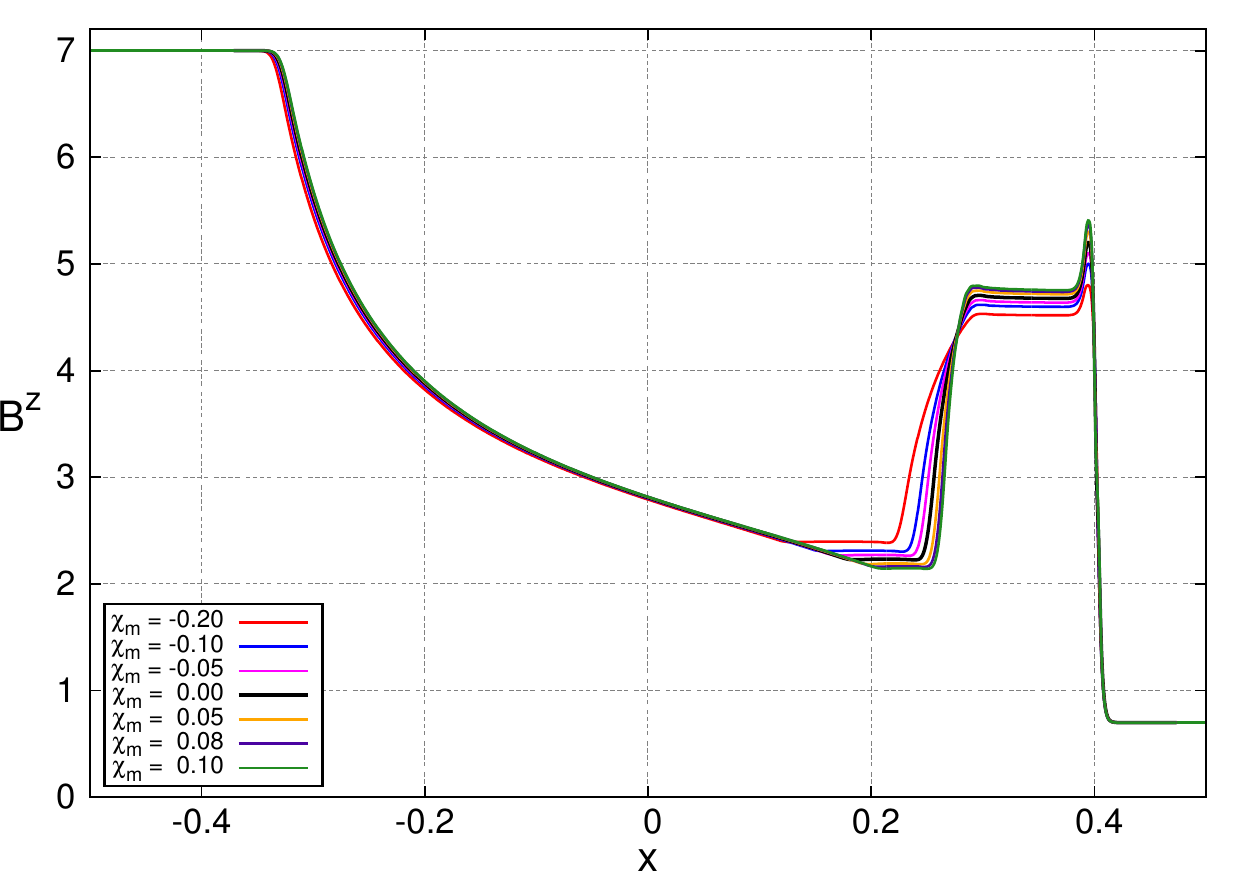}\\
\includegraphics[scale=0.7]{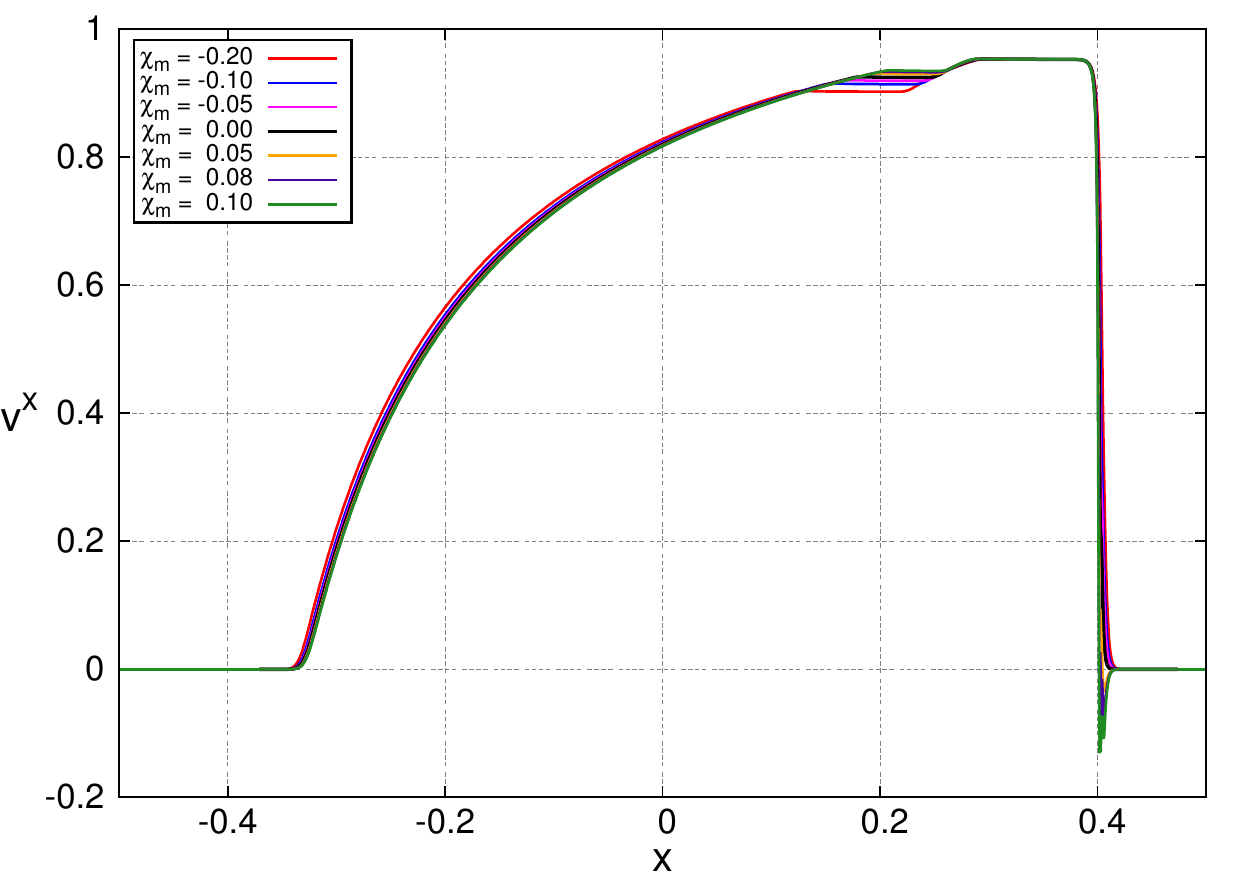} & \includegraphics[scale=0.7]{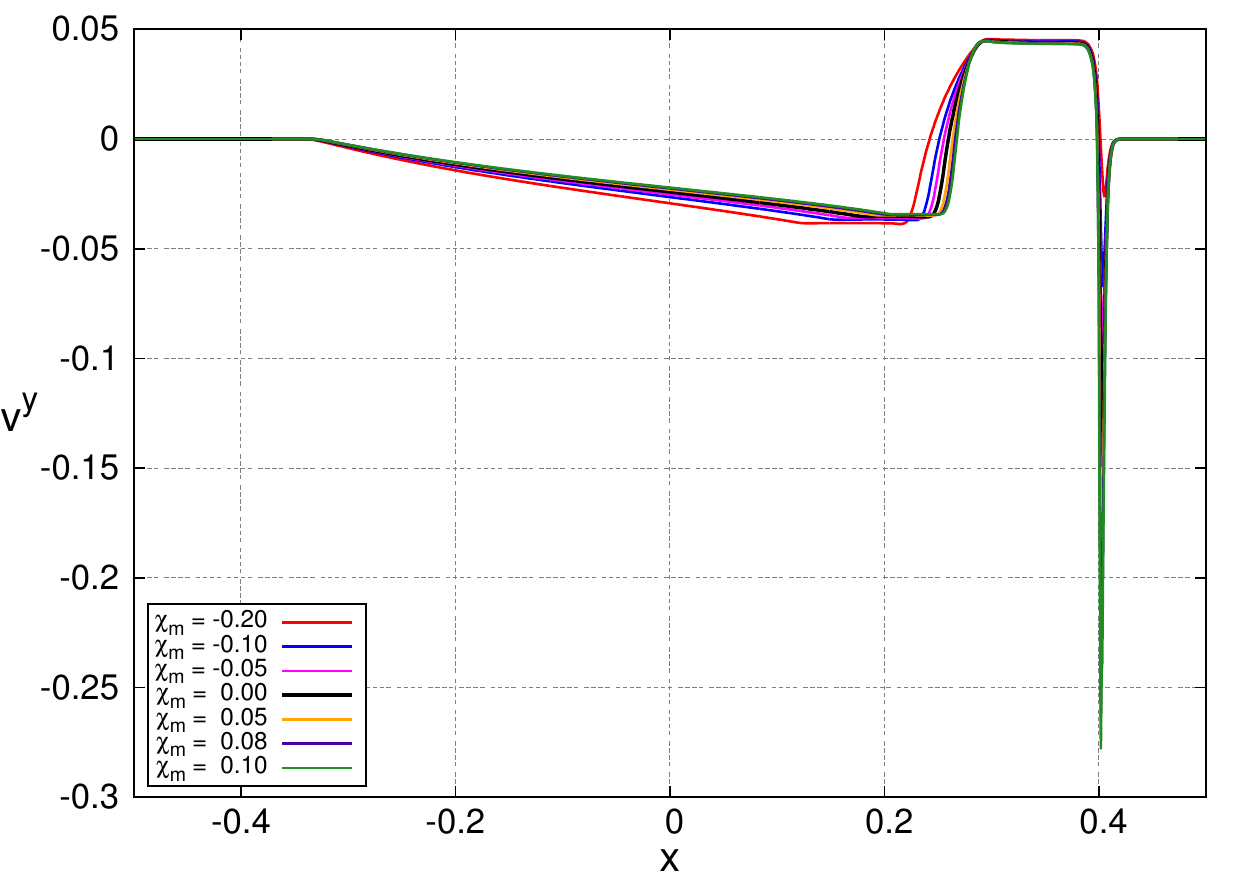}
\end{tabular}
\caption{Balsara 3 test at time $t=0.4$. We use a spatial resolution $\Delta x=1/1600$ and a Courant factor of 0.25.}
\label{test5}
\end{figure*}

The initial state of Balsara 3 is quite similar to that of Balsara 2, the main difference lies in the fact that now the left pressure is $10000$ times the pressure on the right. With this $\Delta p$, the range of values of $\chi_m$ that we can treat with the code is reduced. In particular, we use the magnetic susceptibilities $\chi_m=-0.20, -0.10, -0.05, 0.00, 0.05, 0.08, 0.10$. The wave structure of the solution, showed in figure \ref{test5} at time $t=0.4$, is the same as in the previous test. Nevertheless, since the pressure gradient is so big, it is difficult to capture the true constant state between the shocks (\cite{2007CQGra..24S.235G}) with the resolution that we utilized. Another consequence of the high pressure in the left state is that the magnetic susceptibility does not produce significant changes in the solution as compared with Balsara 2. The main change is that the slow rarefaction wave zone increases when we consider diamagnetic materials.

\subsubsection{Balsara 4}

\begin{figure*}
\begin{tabular}{cc}
\includegraphics[scale=0.7]{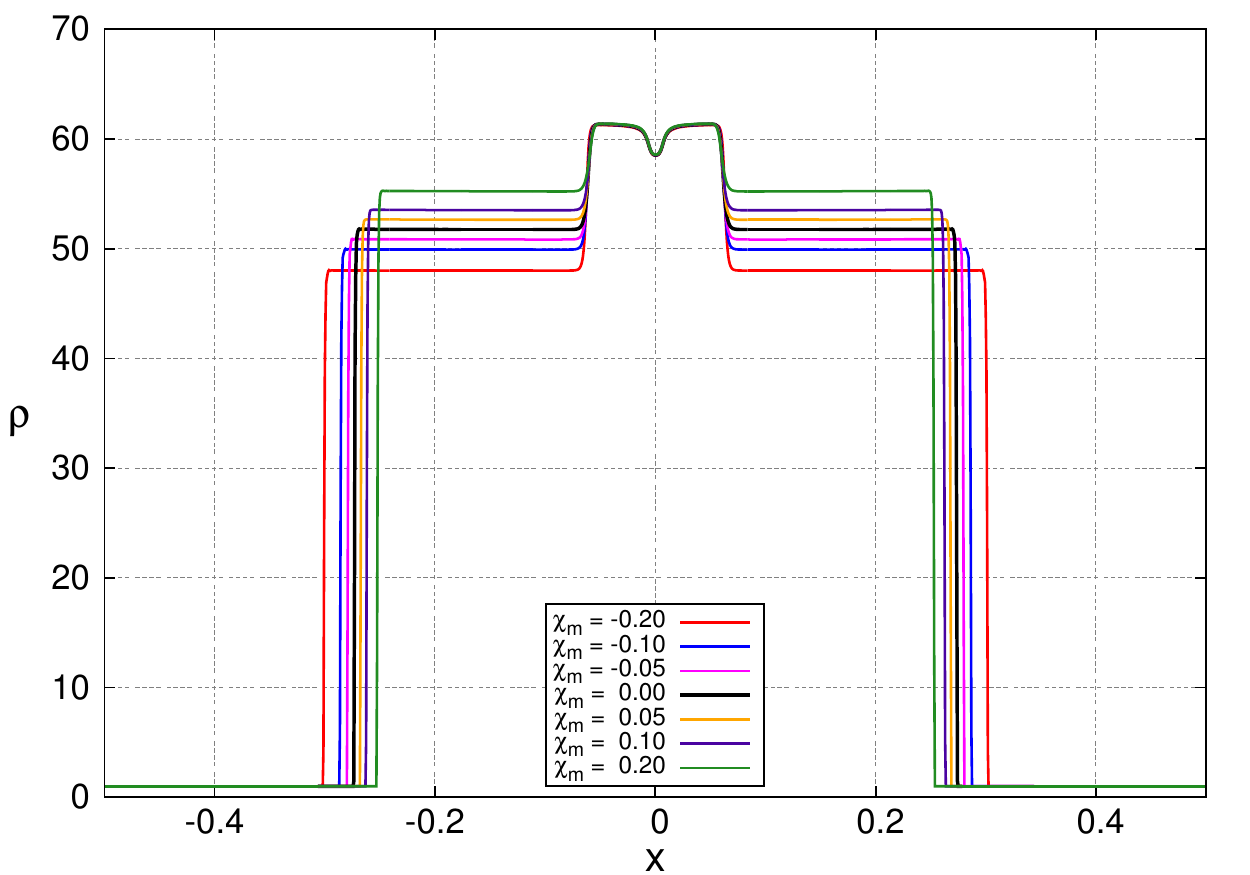} & \includegraphics[scale=0.7]{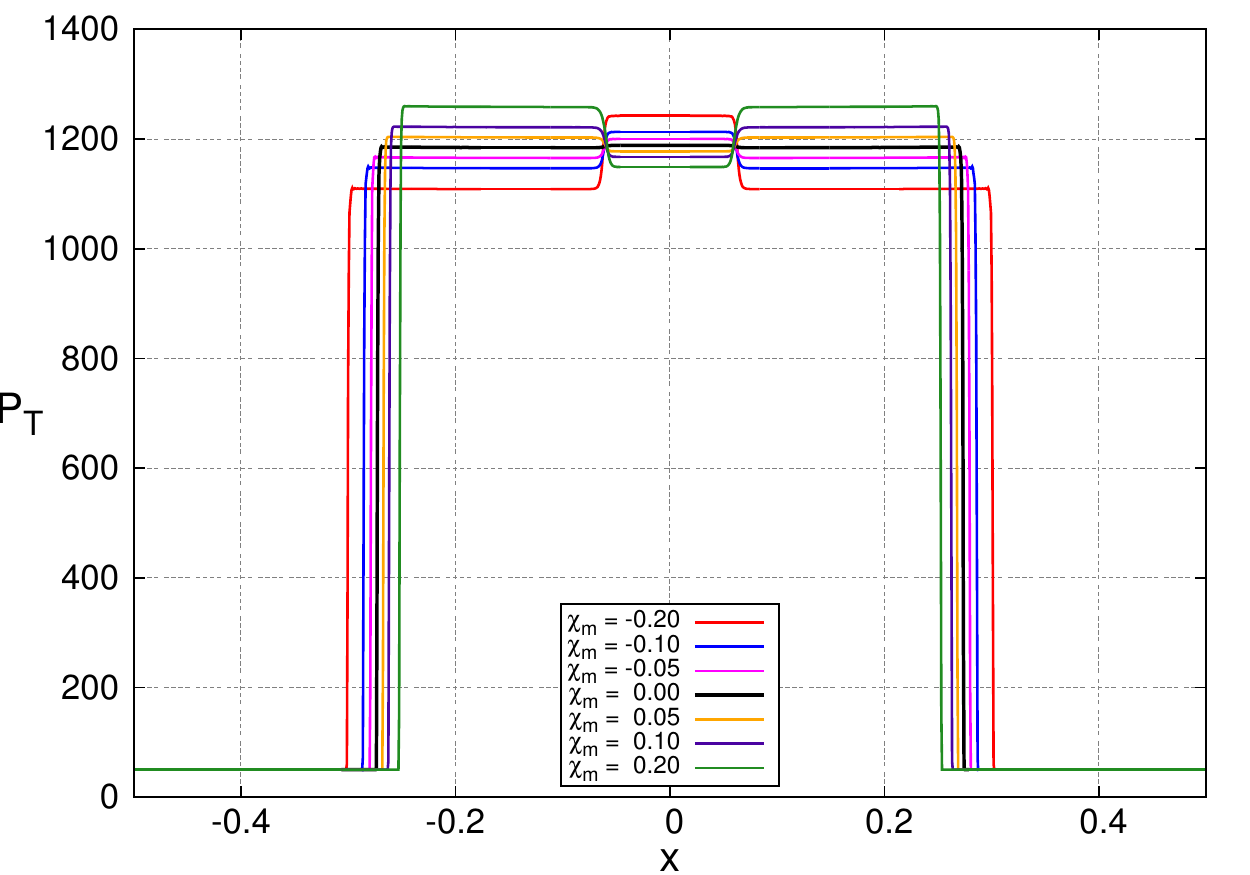}\\
\includegraphics[scale=0.7]{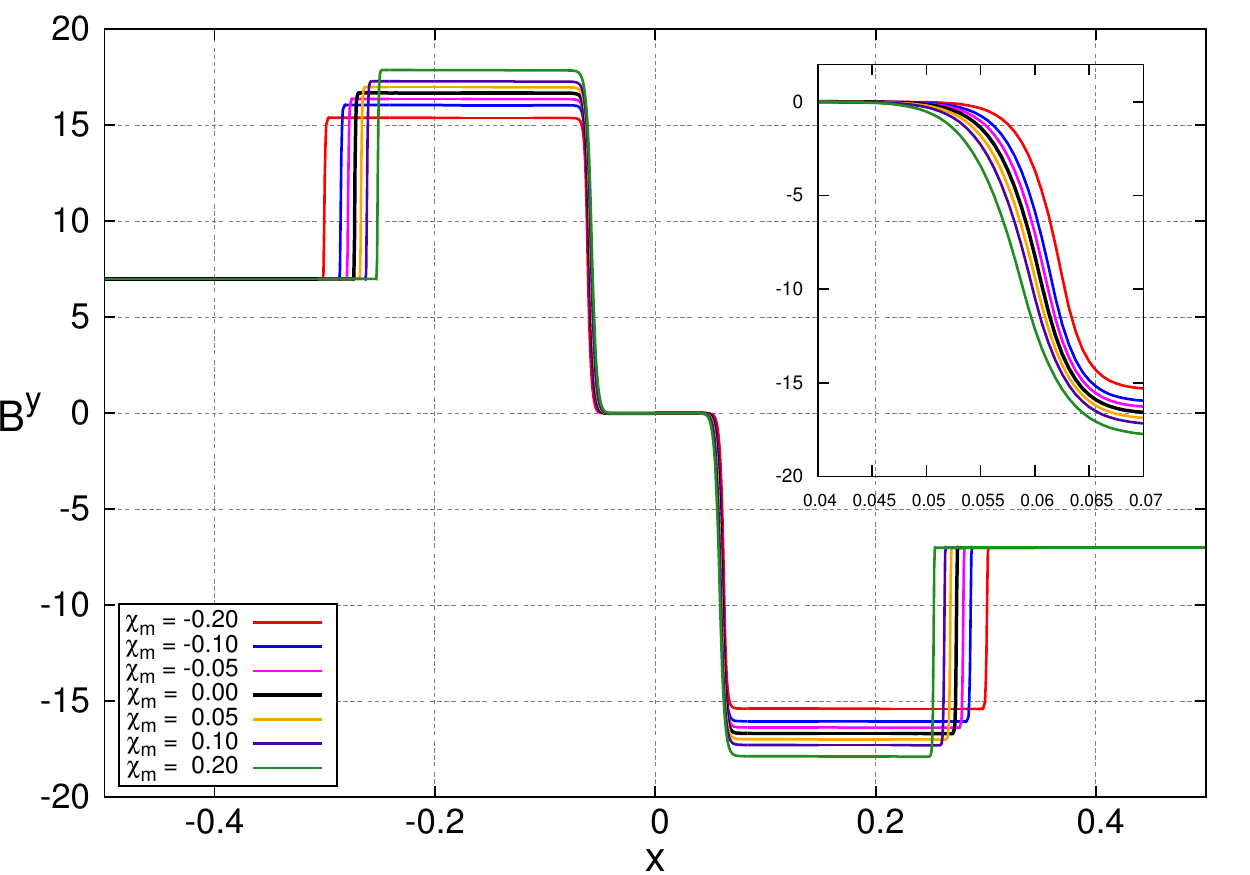} & \includegraphics[scale=0.7]{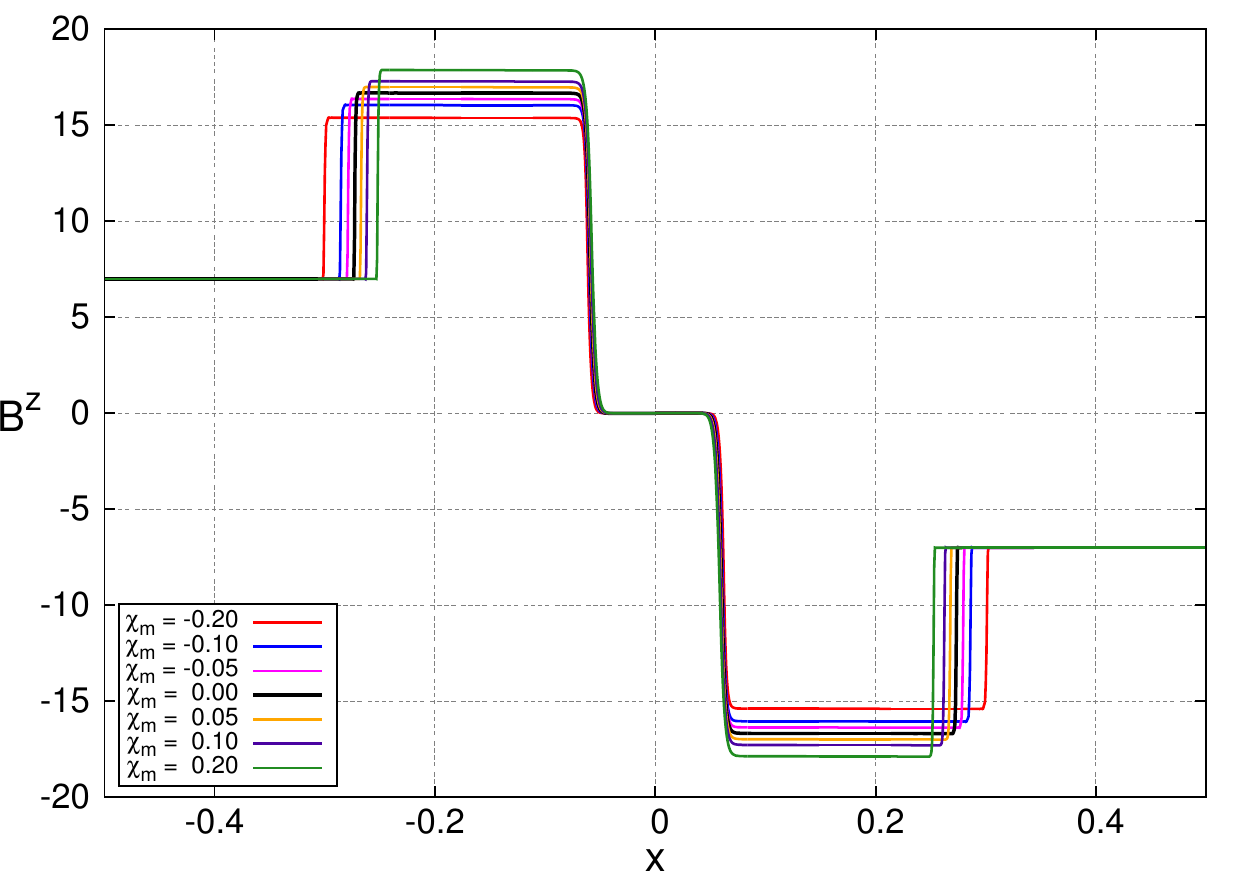}\\
\includegraphics[scale=0.7]{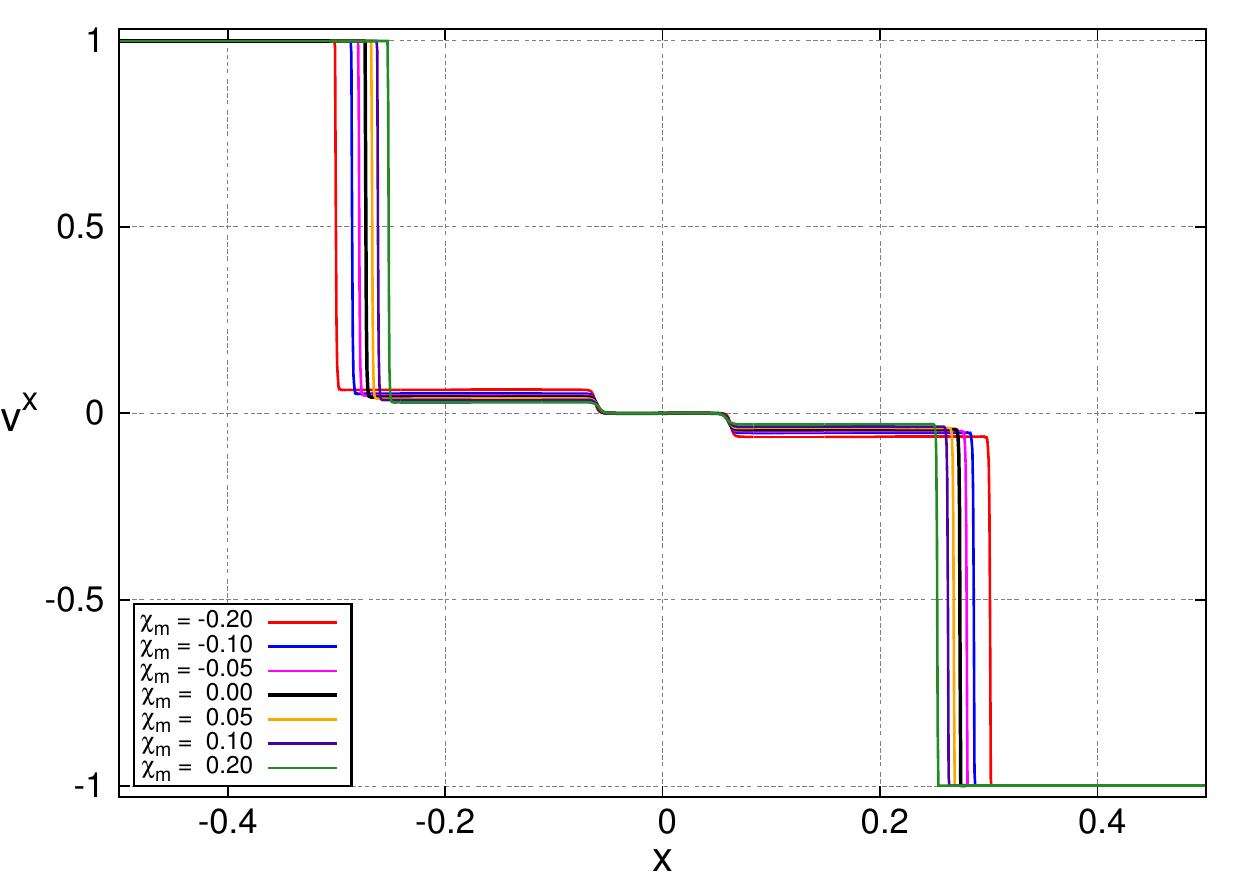} & \includegraphics[scale=0.7]{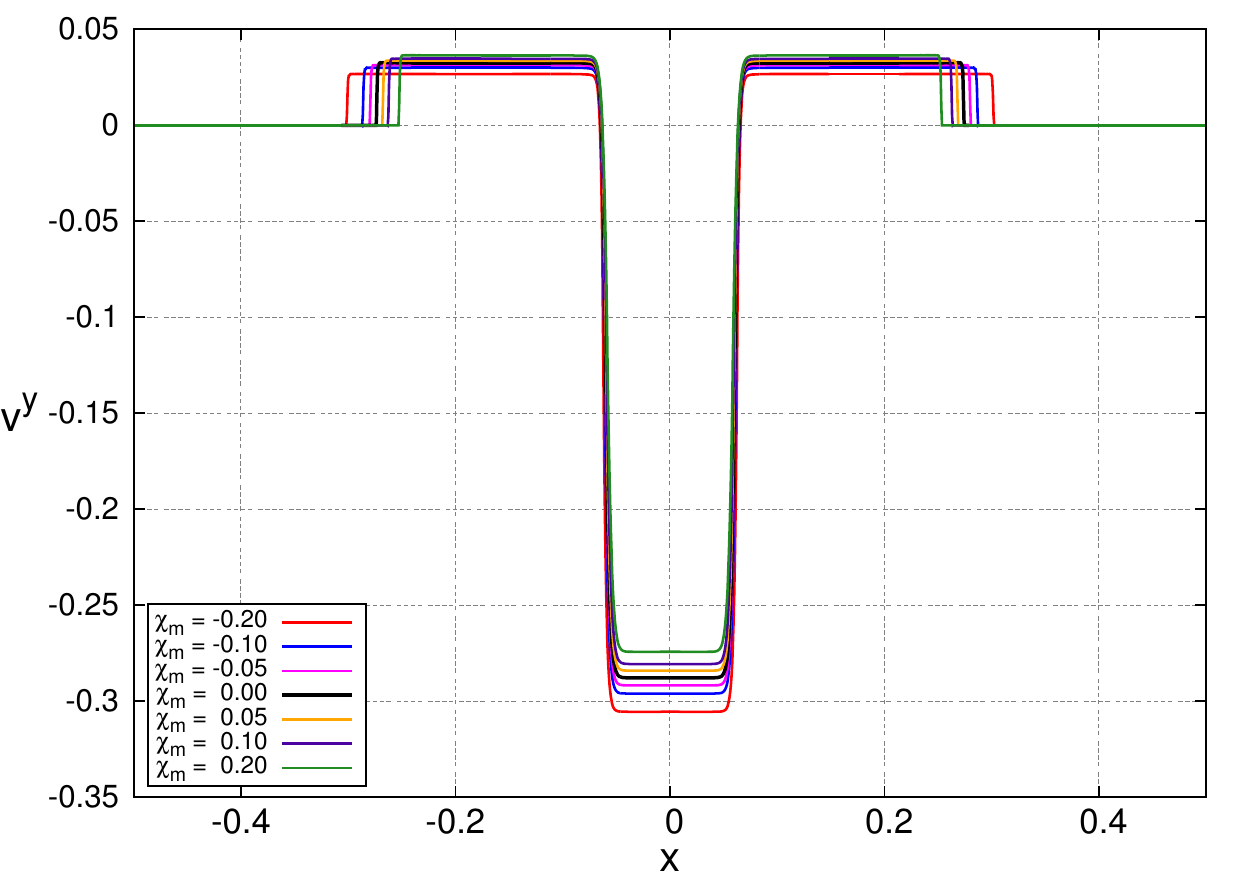}
\end{tabular}
\caption{Balsara 4 test at time $t=0.4$. We use a spatial resolution $\Delta x=1/1600$ and a Courant factor of 0.25.}
\label{test6}
\end{figure*}

The fourth test of Balsara is a head-on collision of two fluids with $\Gamma=5/3$, and a transversal magnetic field with opposite directions on both sides of the interface. This test is similar to the Komissarov collision but with a different adiabatic index and with the fluids moving against each other with a normal velocity of $0.999$ times the speed of light. In Figure \ref{test6} we show a snapshot of the numerical solution at time $t=0.4$ for the values $\chi_m=-0.20, -0.10, -0.05, 0.00, 0.05, 0.10, 0.20$. We observe in this numerical solution the same shock waves as in the Komissarov collision, and we notice again that the fast shocks travel faster in diamagnetic fluids. However, the slow shocks present a different behaviour because they move slower in materials with positive $\chi_m$. Finally, the maximum value of the rest mass density does not have an appreciate change when we vary the magnetic susceptibility.

\subsubsection{Balsara 5}

\begin{figure*}
\begin{tabular}{cc}
\includegraphics[scale=0.7]{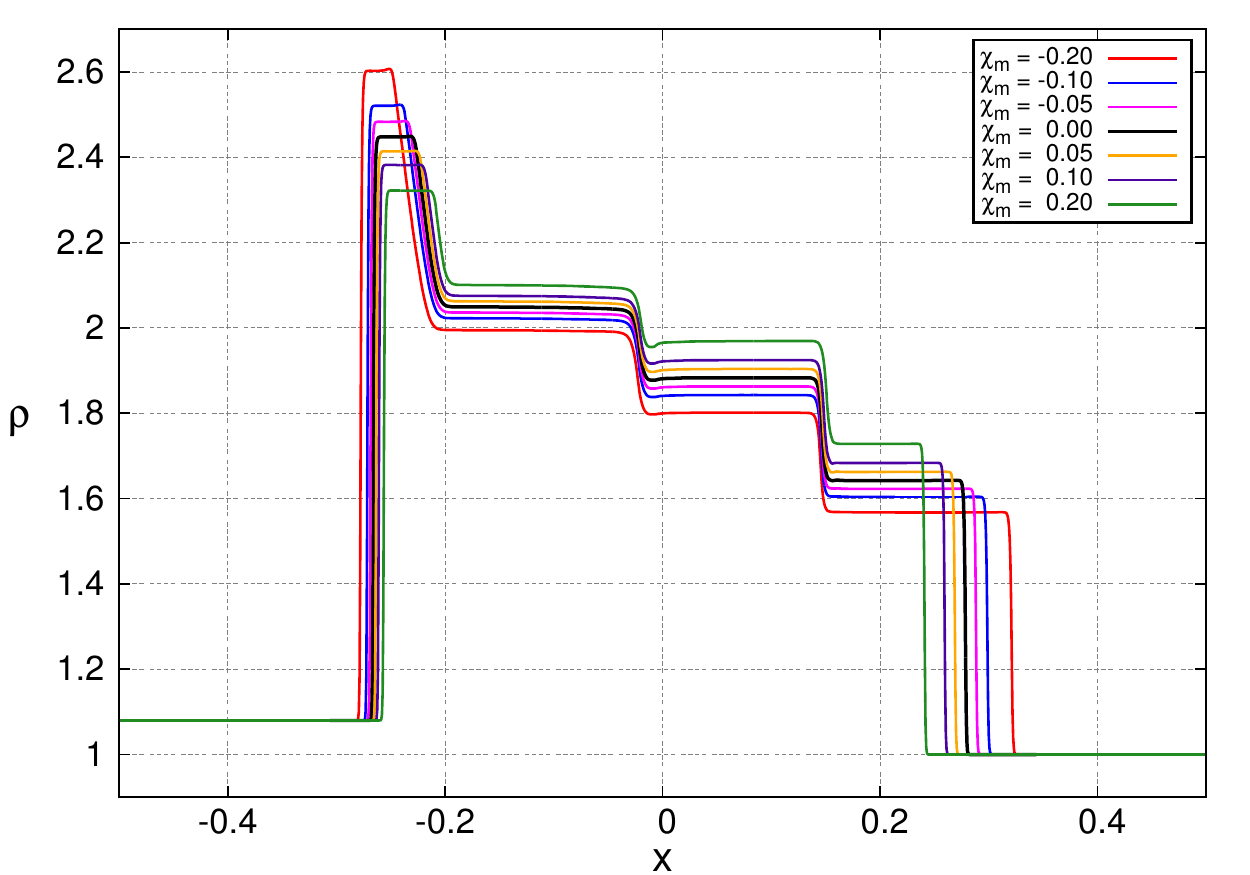} & \includegraphics[scale=0.7]{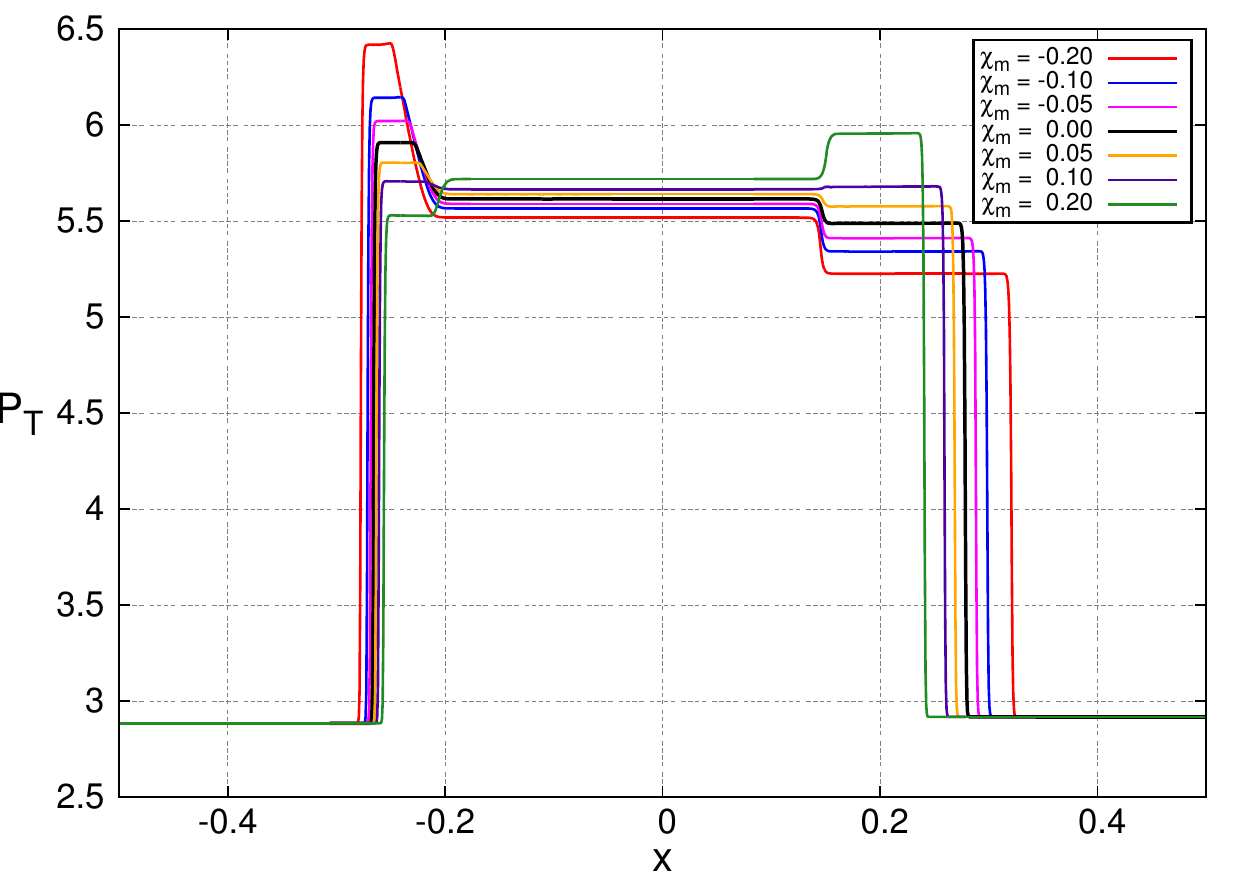}\\
\includegraphics[scale=0.7]{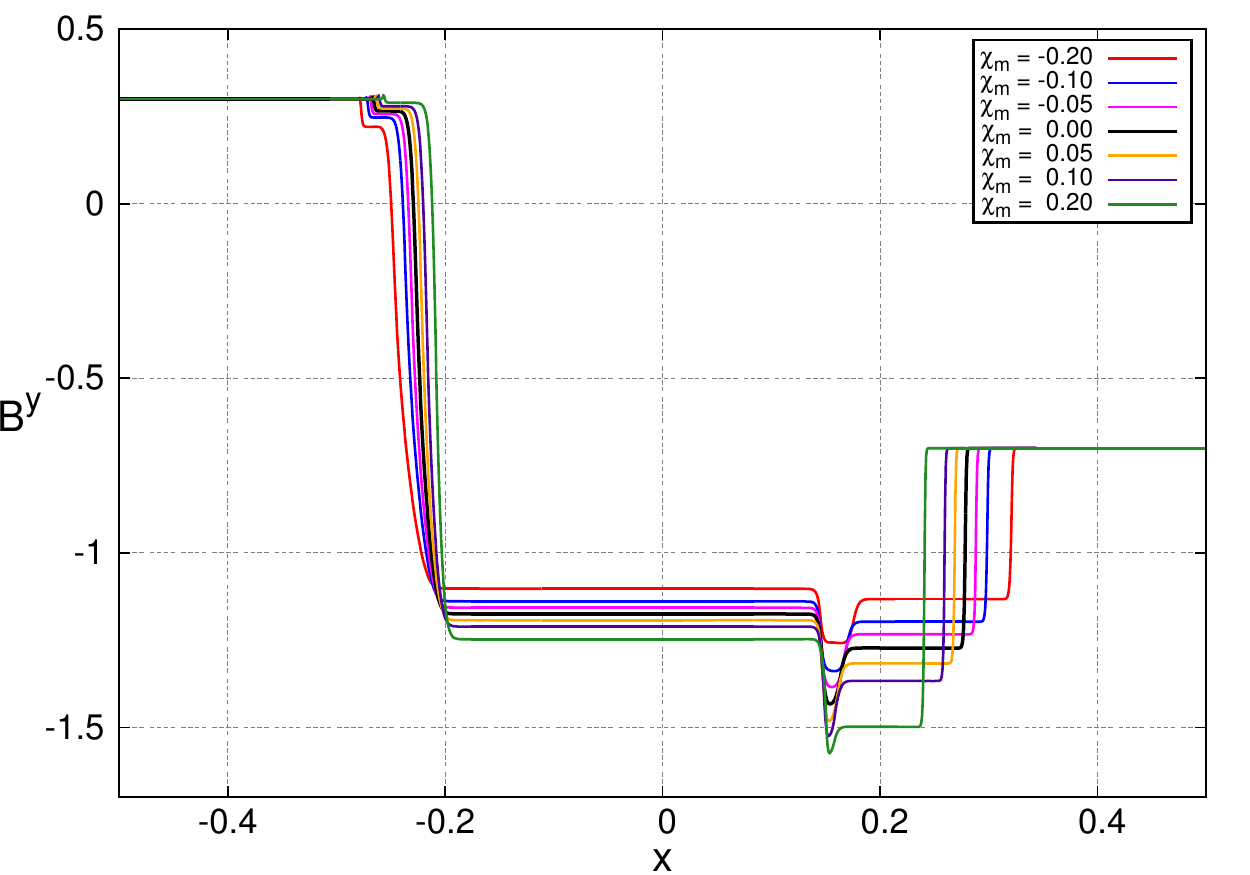} &
\includegraphics[scale=0.7]{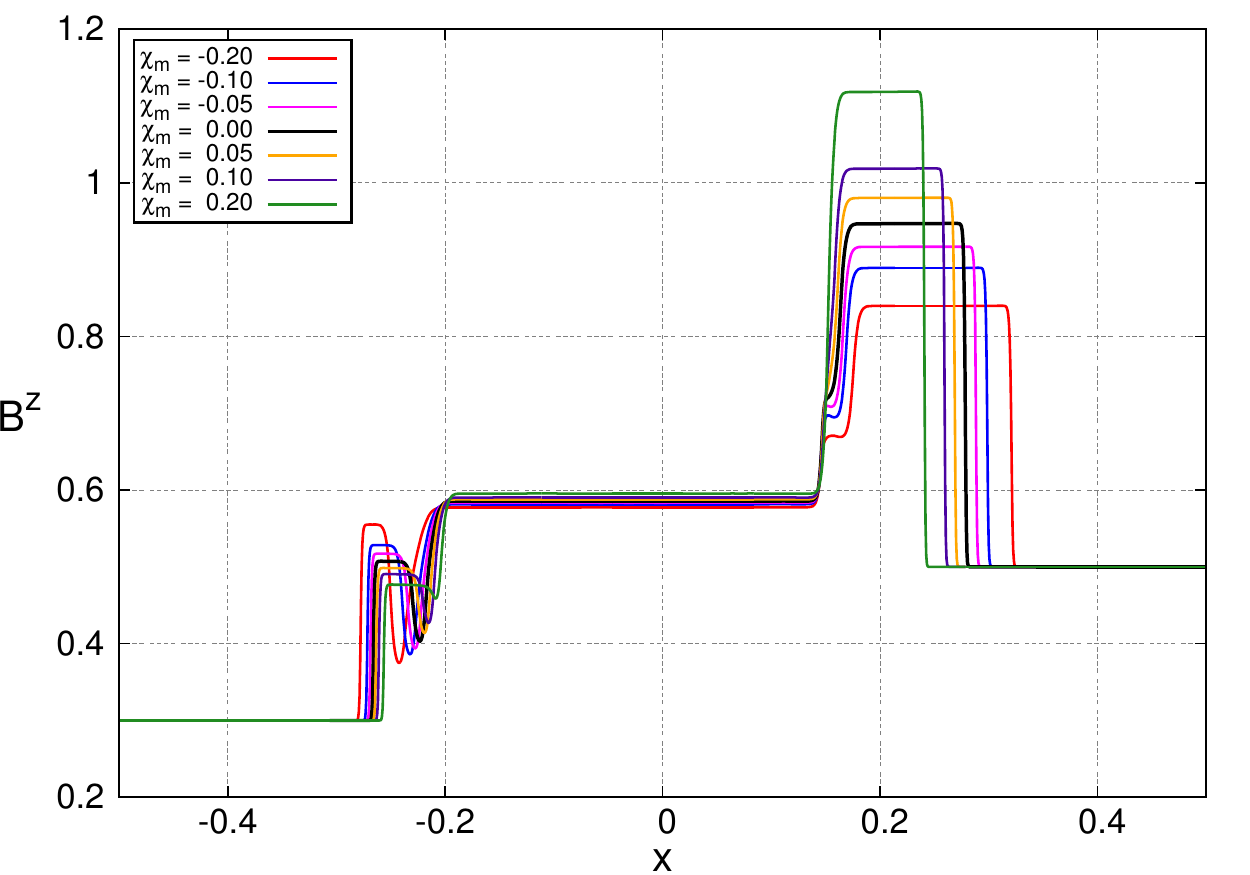}\\
\includegraphics[scale=0.7]{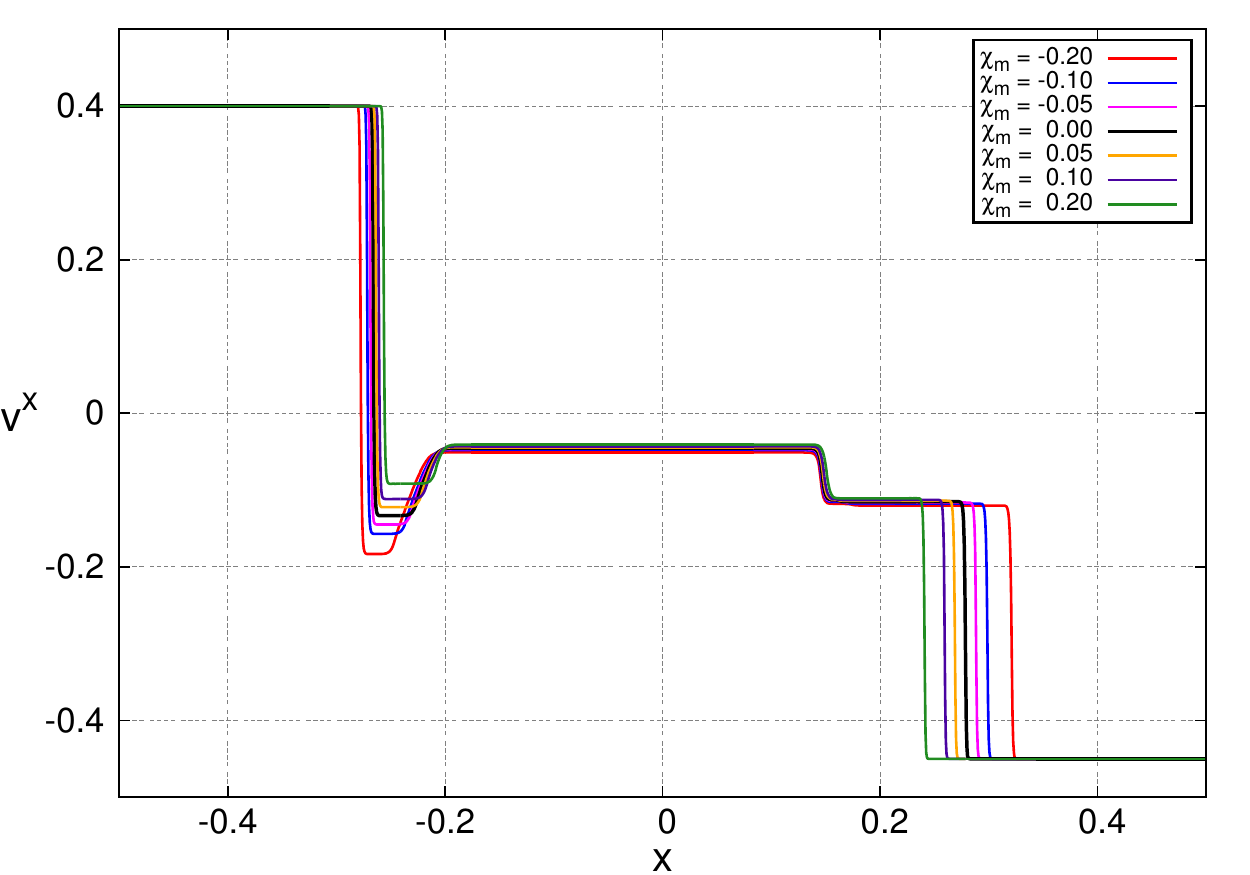} &
\includegraphics[scale=0.7]{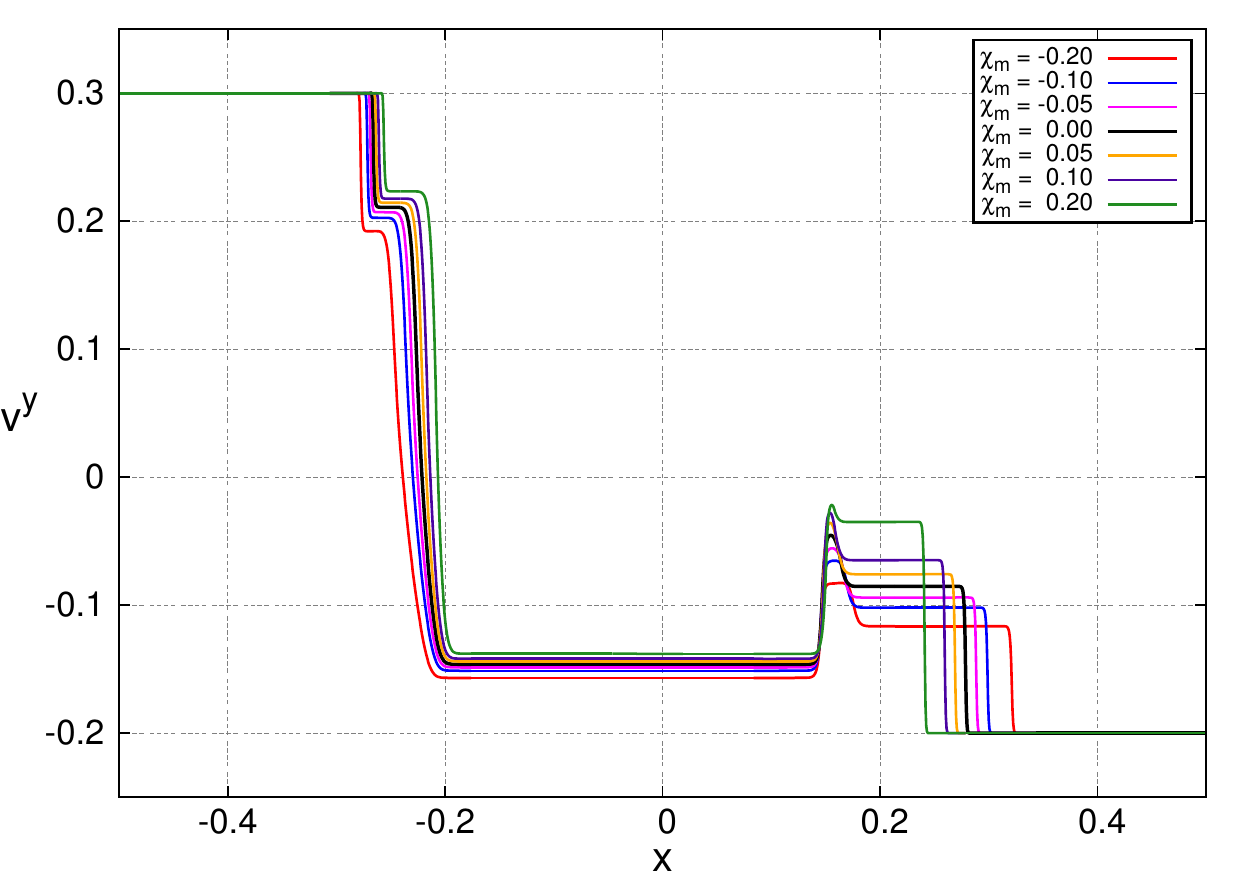}
\end{tabular}
\caption{Balsara 5 test at time $t=0.55$. We use a spatial resolution $\Delta x=1/1600$ and a Courant factor of 0.25.}
\label{test7}
\end{figure*}

The last test of Balsara is a collision of two fluids with an adiabatic index of $\Gamma=5/3$, but in this case the transversal components of the velocity and the magnetic field are discontinuous in the plane $x=0$ (see Table \ref{table1}). The numerical solution to this problem for the values $\chi_m=-0.20, -0.10, -0.05, 0.00, 0.05, 0.10, 0.20$, which is showed in Figure \ref{test7} at time $t=0.55$, is composed of four waves moving to the left: a fast shock, an Alfv$\grave{\text{e}}$n discontinuity, a slow rarefaction and a contact discontinuity; and three waves moving to the right: a slow shock, an Alfv$\grave{\text{e}}$n discontinuity, and a fast shock (\cite{2006JFM...562..223G}).

The numerical solutions with different magnetic susceptibilities show that the fast shocks, the fast rarefaction, and the Alfv$\grave{\text{e}}$n discontinuities move faster in diamagnetic fluids. Additionally, the maximum values of the rest mass density and the total pressure increases when we reduce the magnetic susceptibility. Finally, as in previous cases, we observe that from a positive value of $\chi_m$, the total pressure gradient across the rarefaction wave and the slow right-going shock changes sign.\\

\subsubsection{Generic Alfv$\grave{\text{e}}$n Wave}

\begin{figure*}
\begin{tabular}{cc}
\includegraphics[scale=0.7]{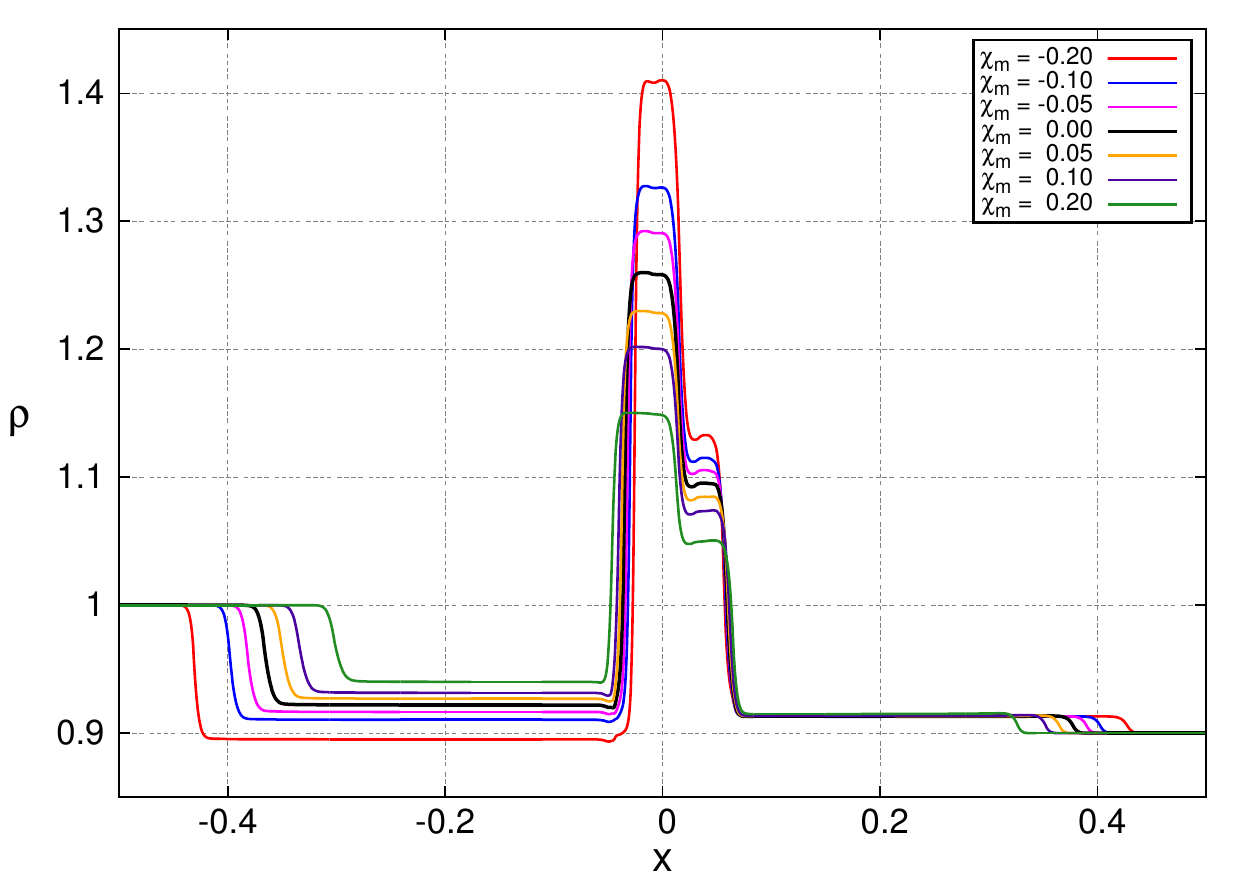} & \includegraphics[scale=0.7]{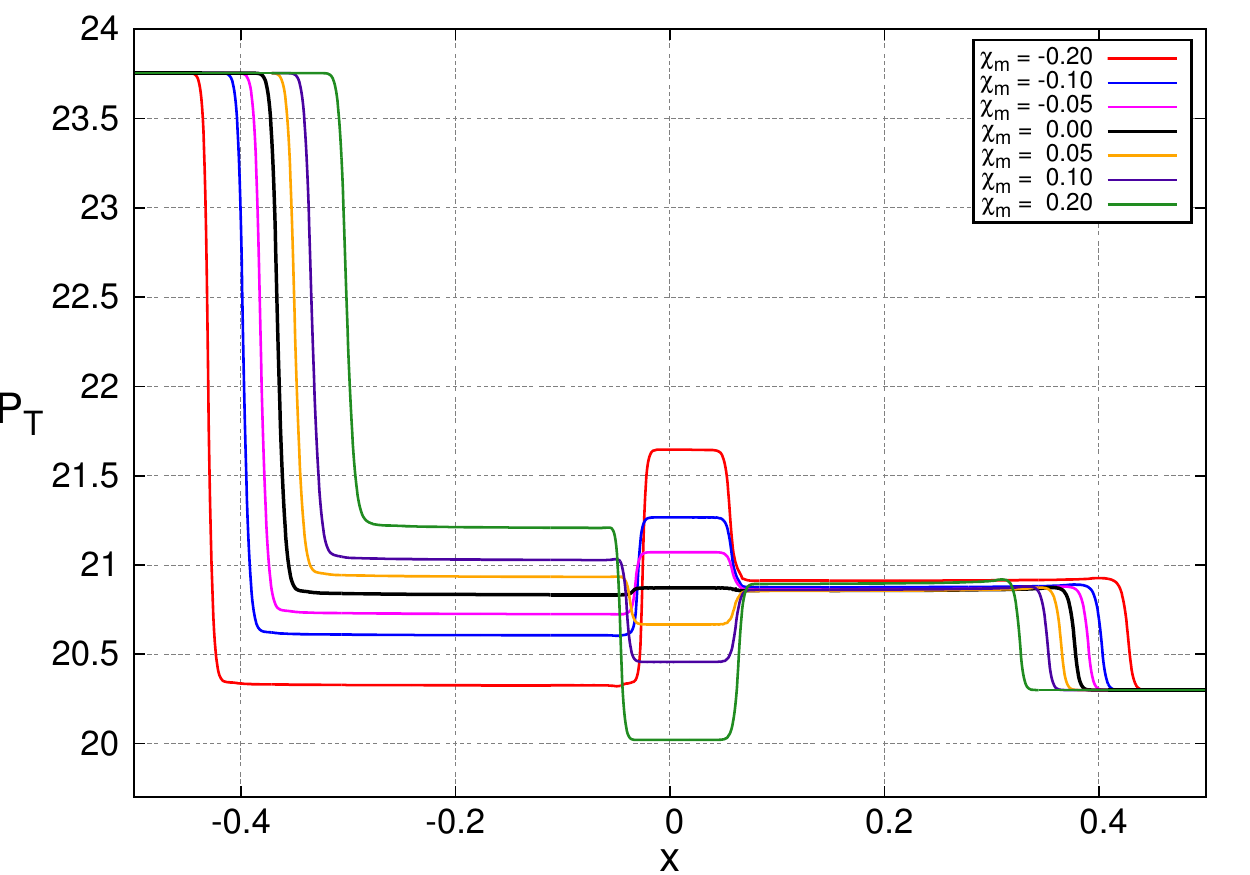}\\
\includegraphics[scale=0.7]{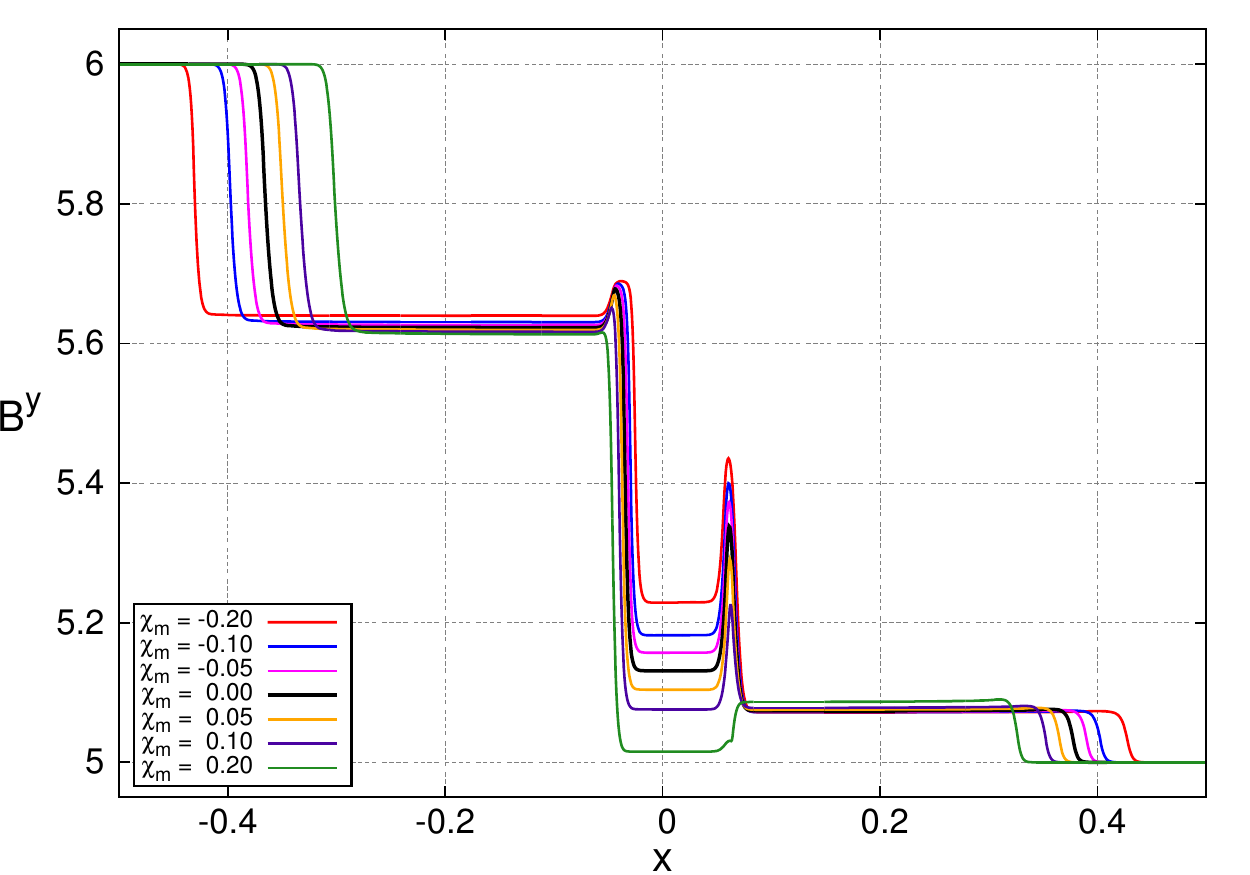} &
\includegraphics[scale=0.7]{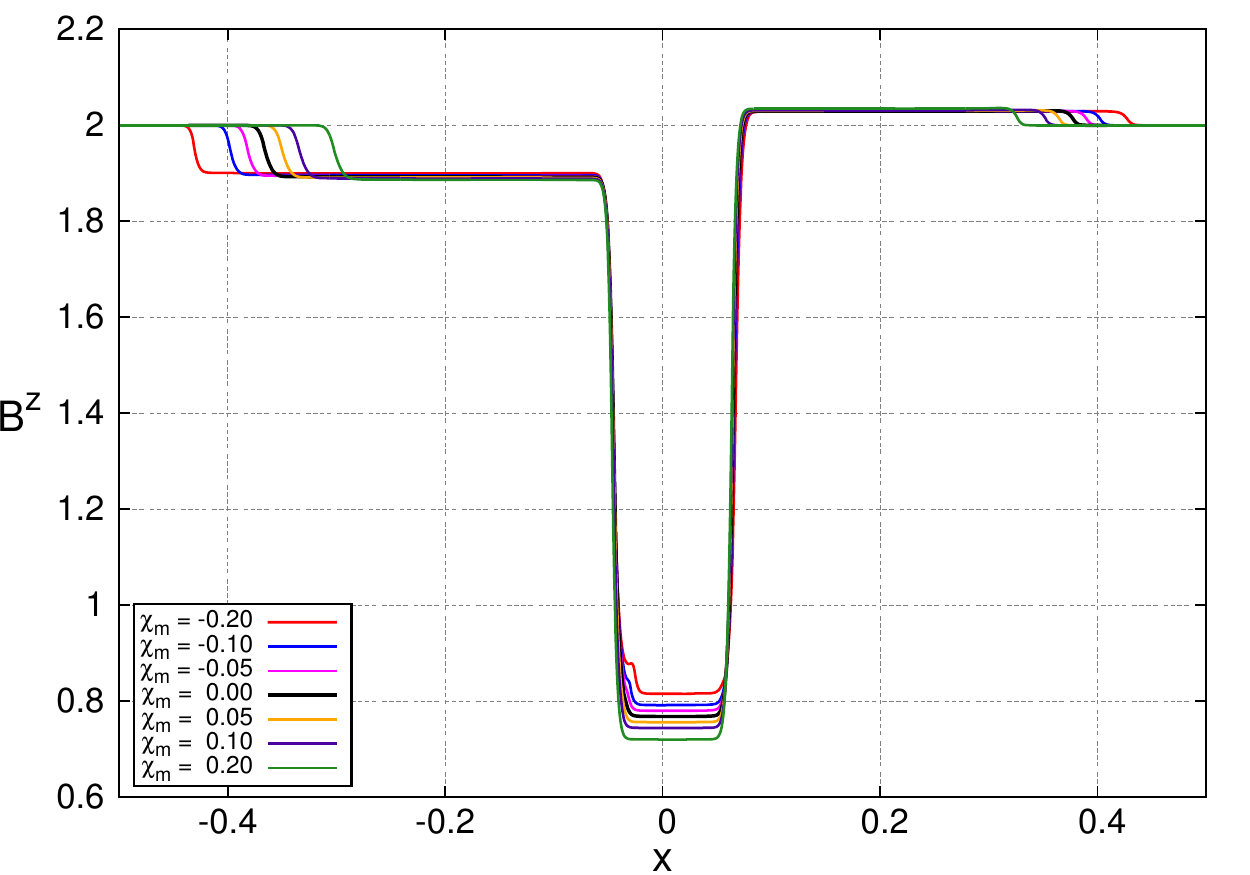}\\
\includegraphics[scale=0.7]{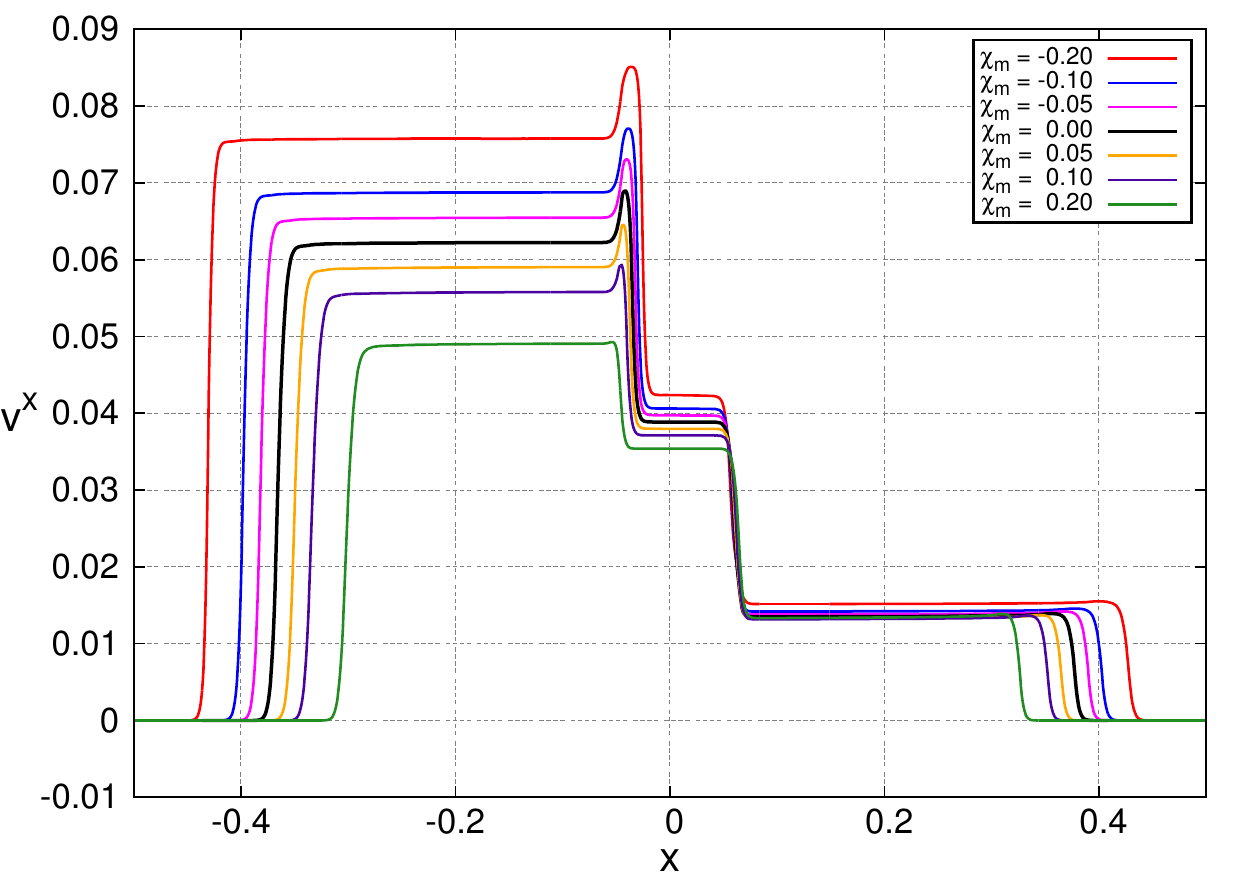} &
\includegraphics[scale=0.7]{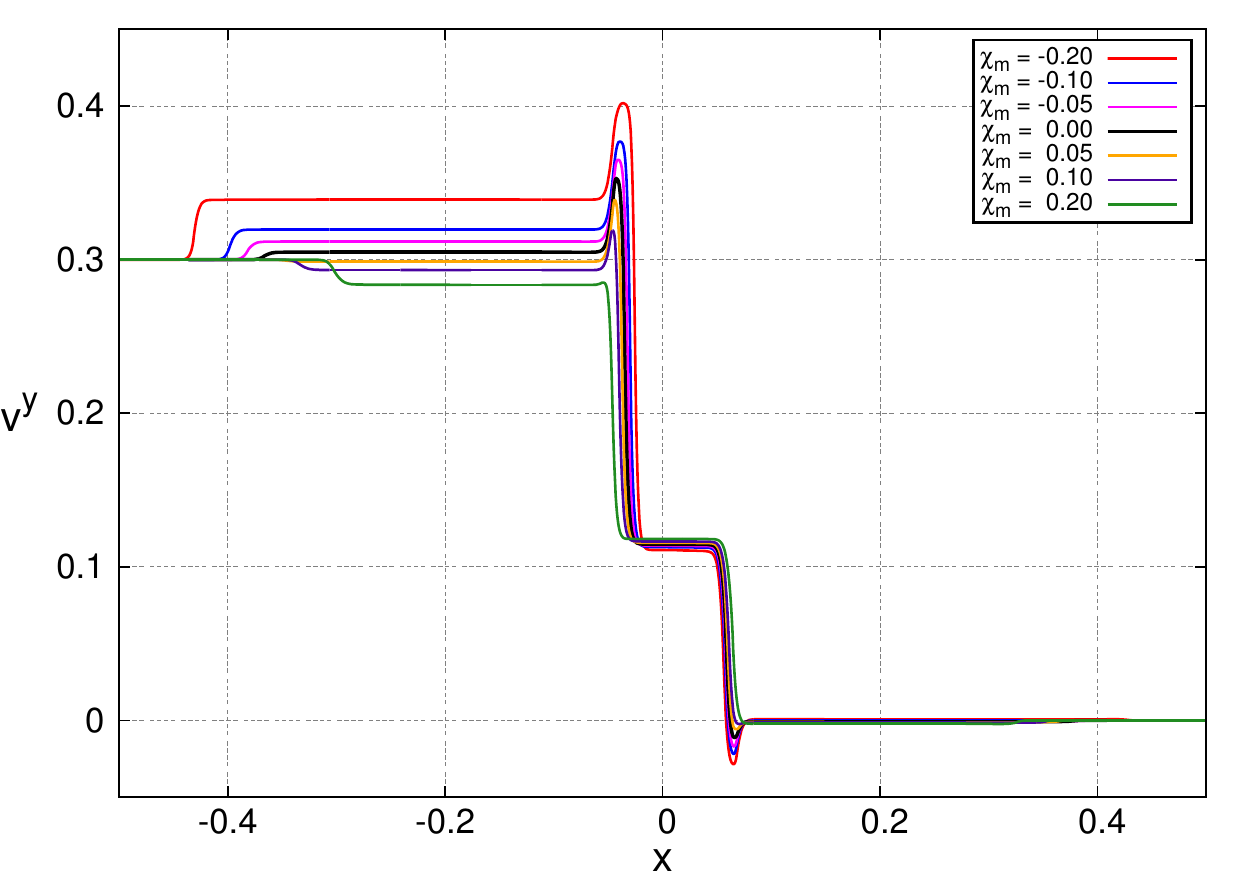}
\end{tabular}
\caption{Generic Alfv$\grave{\text{e}}$n test at time $t=0.4$. We use a spatial resolution $\Delta x=1/1600$ and a Courant factor of 0.25.}
\label{test8}
\end{figure*}

In this last test, we have a similar initial state as in Balasara 5: the transversal components of the velocity and the magnetic field are discontinuous at the interface. Nevertheless, in this case the fluid on the right is initially at rest, while in the left there is a purely transversal flow. The discontinuity in the magnetic field (magnitude and direction) is in turn higher in this test than in the previous one. The solution, which is showed in Figure \ref{test8} at time $t=0.4$ for $\chi_m=-0.20, -0.10, -0.05, 0.00, 0.05, 0.10, 0.20$, consists of three left-going waves: a fast rarefaction, an Alfv$\grave{\text{e}}$n discontinuity, and a slow shock; and of four right-going waves: a contact discontinuity, a slow shock, an Alfv$\grave{\text{e}}$n discontinuity, and a fast shock.

The solutions for different values of magnetic susceptibility exhibit a similar behaviour to that of the previous test. However, in this case the Alfv$\grave{\text{e}}$n waves seem to travel at the same speed regardless of $\chi_m$, and the constant state between the left-going 	Alfv$\grave{\text{e}}$n discontinuity and the left-going slow shock tends to occupy more $x$-domain in diamagnetic fluids. As a final comment about these last two tests, we notice that it is more difficult to capture numerically the Alfv$\grave{\text{e}}$n discontinuities in the paramagnetic materials than in those with diamagnetic properties.

\bgroup
\def\arraystretch{1.5}
\begin{table*}
\begin{center}
\caption{\label{table1} Initial conditions for the Riemann problems}
\begin{tabular}{|c|c|c|c|c|c|c|c|c|c|c|}\hline \hline
${\bf Test}$ & ${\bf State}$ & $\Gamma$ & $\rho_0$ & $p$ & $v^x$ & $v^y$ & $v^z$ & $B^x$ & $B^y$ & $B^z$ \\ \hline \hline
${\bf Shock ~ Tube ~ 1}$ & Left & 4/3 & 1.0 & 1000.0 & 0.0 & 0.0  & 0.0  & 1.0  & 0.0  & 0.0  \\ 
\cline{2-11}
& Right & 4/3 & 0.1 & 1.0 & 0.0  & 0.0  & 0.0  & 1.0 & 0.0 & 0.0  \\ \hline \hline
${\bf Collision}$ & Left & 4/3 & 1.0 & 1.0 & $5/\sqrt{26}$ & 0.0  & 0.0  & 10.0  & 10.0  & 0.0  \\ 
\cline{2-11}
& Right & 4/3 & 1.0 & 1.0 & $-5/\sqrt{26}$  & 0.0  & 0.0  & 10.0 & -10.0 & 0.0  \\ \hline \hline 
${\bf Balsara ~ 1}$ & Left & 2 & 1.0 & 1.0 & 0.0 & 0.0  & 0.0  & 0.5  & 1.0  & 0.0 \\ 
\cline{2-11}
& Right & 2 & 0.125 & 0.1 & 0.0  & 0.0  & 0.0  & 0.5 & -1.0 & 0.0  \\ \hline \hline
${\bf Balsara ~ 2}$ & Left & 5/3 & 1.0 & 30.0 & 0.0 & 0.0  & 0.0  & 5.0  & 6.0  & 6.0 \\ 
\cline{2-11}
& Right & 5/3 & 1.0 & 1.0 & 0.0  & 0.0  & 0.0  & 5.0 & 0.7 & 0.7  \\ \hline \hline
${\bf Balsara ~ 3}$ & Left & 5/3 & 1.0 & 1000.0 & 0.0 & 0.0  & 0.0  & 10.0  & 7.0  & 7.0  \\ 
\cline{2-11}
& Right & 5/3 & 1.0 & 0.1 & 0.0  & 0.0  & 0.0  & 10.0 & 0.7 & 0.7  \\ \hline \hline 
${\bf Balsara ~ 4}$ & Left  & 5/3 & 1.0 & 0.1 & 0.999 & 0.0  & 0.0  & 10.0  & 7.0  & 7.0  \\ 
\cline{2-11}
& Right & 5/3 & 1.0 & 0.1 & -0.999  & 0.0  & 0.0  & 10.0 & -7.0 & -7.0  \\ \hline \hline
${\bf Balsara ~ 5}$ & Left & 5/3 & 1.08 & 0.95 & 0.4 & 0.3  & 0.2  & 2.0  & 0.3  & 0.3  \\ 
\cline{2-11}
& Right & 5/3 & 1.0 & 1.0 & -0.45  & -0.2  & 0.2  & 2.0 & -0.7 & 0.5  \\ \hline \hline 
${\bf Generic ~ Alfven ~ Wave}$ & Left & 5/3 & 1.0 & 5.0 & 0.0 & 0.3  & 0.4  & 1.0  & 6.0  & 2.0  \\ 
\cline{2-11}
& Right & 5/3 & 0.9 & 5.3 & 0.0  & 0.0  & 0.0  & 1.0 & 5.0 & 2.0  \\ \hline \hline 
\end{tabular}
\end{center}
\end{table*} 
\egroup

\subsection{Magnetized Spherical Accretion of a Perfect Fluid with Magnetic Polarization}

In this section, we will analyze the ability of the code to maintain the stationary solution that describe the spherical accretion of a perfect fluid with magnetic polarization onto a Schwarzschild black hole, in presence of a radial magnetic field. Previously, in \cite{2003ApJ...589..458D}, the authors showed that the hydrodynamical solution obtained by \cite{1972Ap&SS..15..153M} remains the same when a radial magnetic field is added. This magnetized solution does not represents any real physical system (\cite{2006ApJ...637..296A}) but it is a useful nontrivial test in numerical GRMHD. Now, following \cite{2003ApJ...589..458D}, it is possible to show that when we consider the constitutive relation for a linear media (\ref{constituvive_relation}), and maintain the radial character of the magnetic field, the Michel solution is not affected. To prove this result, we compute the component
\begin{equation}
T^{r}_t=[\rho h+b^{2}(1-\chi)]u^{r}u_{t}-b^{r}b_{t}(1-\chi),
\label{component}
\end{equation}
which is the only one that is involved in the conservation of the energy flux $\nabla_{r}T^{r}_t=0$. Now, from the equations (\ref{transformations}), it is possible to show that $b^{r}b_{t}=b^{2}u^{r}u_{t}$, so the terms with magnetic field cancel each other. Therefore, we can use this test, but now in the context of the GRMHD with magnetically polarized matter to estimate the order of global convergence of the code with different magnetic susceptibilities.

The initial data for this test is the set of primitive variables $\vec{\mathcal{W}}=[\rho,v^r,p,B^r]^T$, where the rest mass density and the pressure are related by a polytropic EOS with an adiabatic index of $\Gamma=4/3$. The first three variables are given by the Michel solution, while the radial component of the magnetic field is chosen to satisfy the divergence-free condition. The hydrodynamical solution is completely determined by fixing the critical radius $r_c$, and the critical rest mass density $\rho_c = \rho(r_c)$. To carry out our simulations, we use the same parameters as in \cite{2007CQGra..24S.235G}, for which $r_{c}=8$ (we take $M_{BH}=1$) and $\rho_{c}=6.25\times 10^{-2}$. Now, we use Eddington$-$Finkelstein coordinates, so we can choose a spatial domain $r\in \left[1.9,20.9\right]$ with $N=50$ radial zones. On the other hand, the component $B^r$ at $t=0$ is obtained through the equation $B^{r}=C_{_{EH}}/(\sqrt{g_{rr}}r^{2})$, where $C_{_{EH}}=4\sqrt{\rho_{_{EH}}\beta_{_{EH}}}$. So we only need to define the magnetic field strength $\beta_{_{EH}}=b^{2}/\rho$, and the rest mass density $\rho_{_{EH}}$, at the event horizon $r_{_{EH}}=2$.

In the simulations we use the following positive values of magnetic susceptibility $\chi_{m}= 0.000, 0.001, 0.005, 0.008$, since we notice in the last section that it is more difficult to deal with paramagnetic materials. For each one of these values we compute the error
\begin{equation}
L_1=\frac{\sum_{i}|\rho(r_i)-\rho_{exact}(r_i)|}{\sum_{i} \rho_{exact}(r_i)},
\label{L-error}
\end{equation}
in the rest mass density with different values of $\beta_{_{EH}}$. In particular, we take $\beta_{_{EH}}=1,10,25$ which is equivalent to a ratio between the magnetic pressure and the gas pressure of $p_{m}/p=3.88, 38.80, 97.00,$ respectively. We found that, although the numerical solution is effectively maintained over time ($t > 4000$), the stationary state of the radial velocity deviates from the analytic solution when we increase the values of $\beta_{_{EH}}$ and $\chi_{m}$. This behaviour has already been reported by \cite{2005PhRvD..72d4014S}, but only for the case where the fluid is not magnetically polarized.

\begin{figure*}

\centering
\begin{center}
\subfigure[$\chi_m=0.000$]
{
\label{chi000}
\includegraphics[scale=0.86]{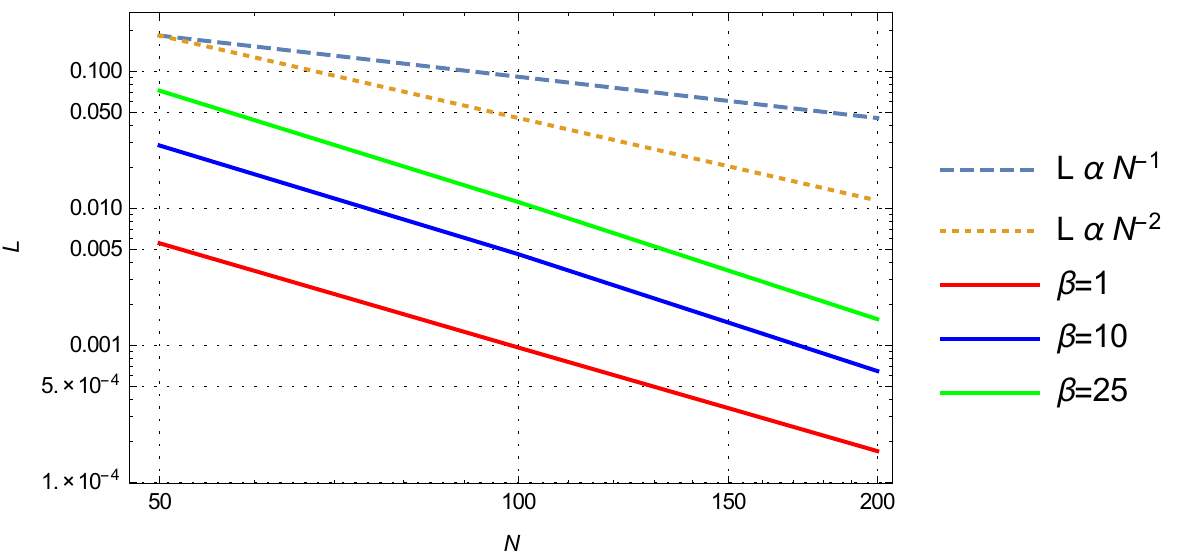}
}
\subfigure[$\chi_m=0.001$]
{
\label{chi001}
\includegraphics[scale=0.86]{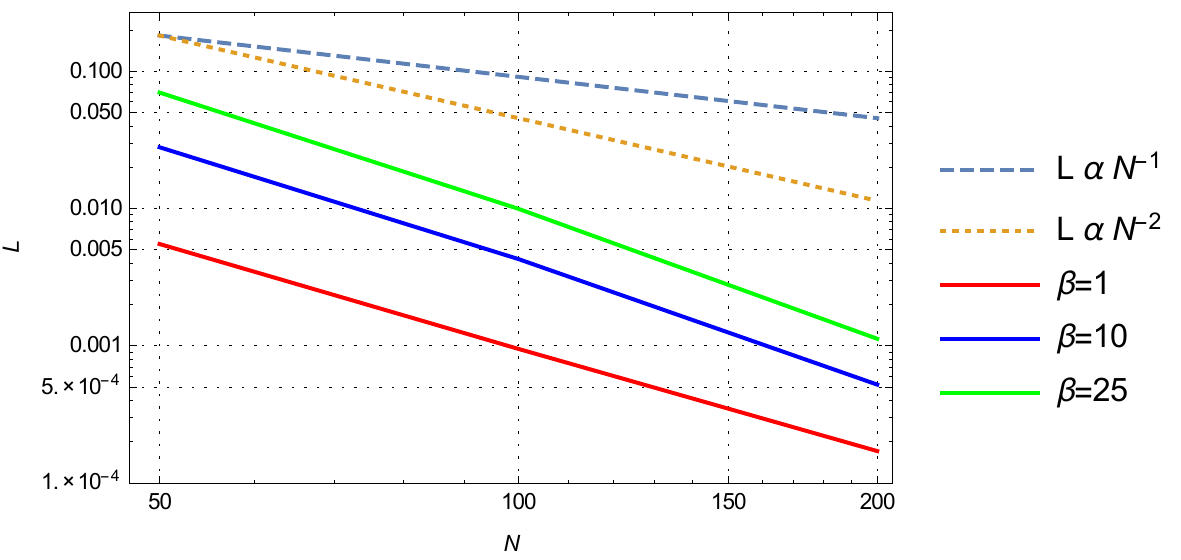}
}
\subfigure[$\chi_m=0.005$]
{
\label{chi005}
\includegraphics[scale=0.86]{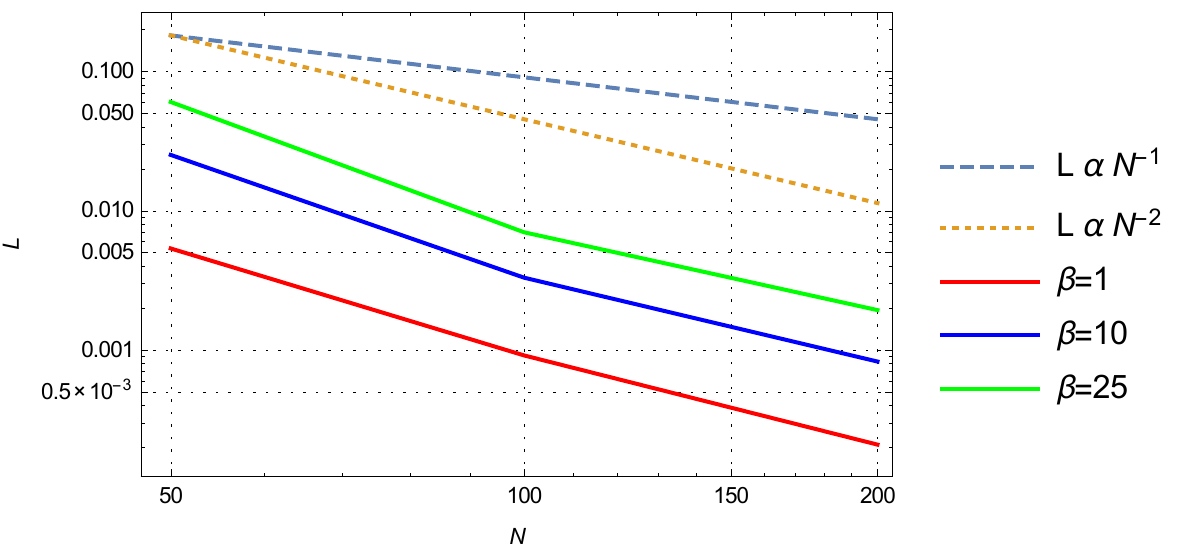}
}
\subfigure[$\chi_m=0.008$]
{
\label{chi008}
\includegraphics[scale=0.86]{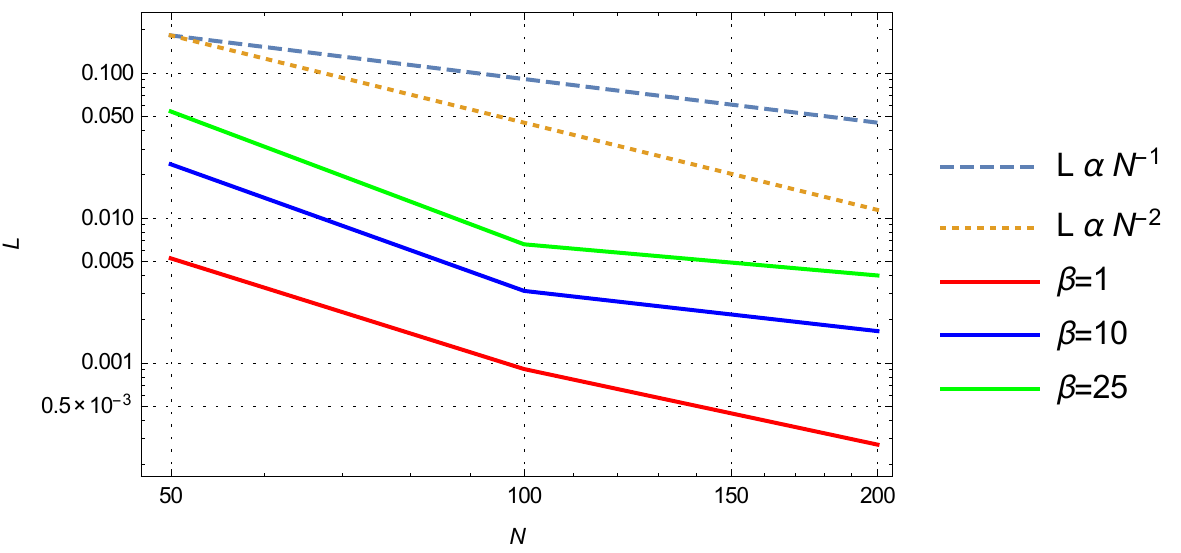}
}
\caption{$L_1$ norm of the relative error for the rest mass density as a function of the number of radial zones $N$, for different values of magnetic field strength $\beta$ at the event horizon. Each panel correspond to a different value of magnetic susceptibility $\chi_m$. The dotted lines indicate the first and second order of global convergence.\label{above}}
\end{center}
\end{figure*}

Now, with the aim of estimating the order of convergence, we present in figure \ref{above} the $L_1$ norm of the error for the rest mass density, at time $t=200$, as a function of the number of radial zones $N$. The Figure \ref{chi000} correspond to the GRMHD evolutions for the three values of $\beta_{_{EH}}$, without magnetic polarization, {\em i.e} with $\chi_m =0$. We notice from this plot that the convergence of the code is slightly greater that 2 for the magnetic field strengths that we considered. A similar result is obtained when we evolve the magnetized Michel solution with a magnetic susceptibility of $\chi_{m}=0.001$ (see Figure \ref{chi001}); Nevertheless, we notice in Figure \ref{chi005} that when $\chi_m$ is incremented to $0.005$, the convergence is reduced to second order for all the values of $\beta_{_{EH}}$ used in the simulations. The plots of Figure \ref{chi008} show that when the magnetic susceptibility of the fluid is $\chi_{m}=0.008$, the order of convergence is considerably reduced in the cases with $\beta_{_{EH}}=10$ and $\beta_{_{EH}}=25$.

We can say that the ability of the code for dealing with magnetically polarized fluids in strong magnetic and gravitational fields depend, not only on the ratio between the magnetic and gas pressure, but also on the magnetic susceptibility. The global convergence of the code is $\gtrsim$ 2 for $\chi_{m}\lesssim 0.005$ for all the magnetic field strength $\beta$ here considered. Nevertheless, when $\chi_m=0.008$ the global convergence of the code is between first and second order for $\beta=10$ and $\beta=25$. These results are interesting because a magnetic field strength of $\beta\approx 4$ corresponds to a large magnetic field of $\approx 10^{19}$ G (\cite{2007CQGra..24S.235G}), and the typical values for the magnetic susceptibility on the materials are around $10^{-5}$. Therefore, if we assume the strongest magnetic field in the universe, $\approx 10^{15}$ G, which could correspond to a magnetar, we can study with CAFE the magnetic polarization of realistic magnetic media in a strong gravitational field with high precision and second order of convergence. 

\begin{figure}
\begin{center}
\includegraphics[scale=0.68]{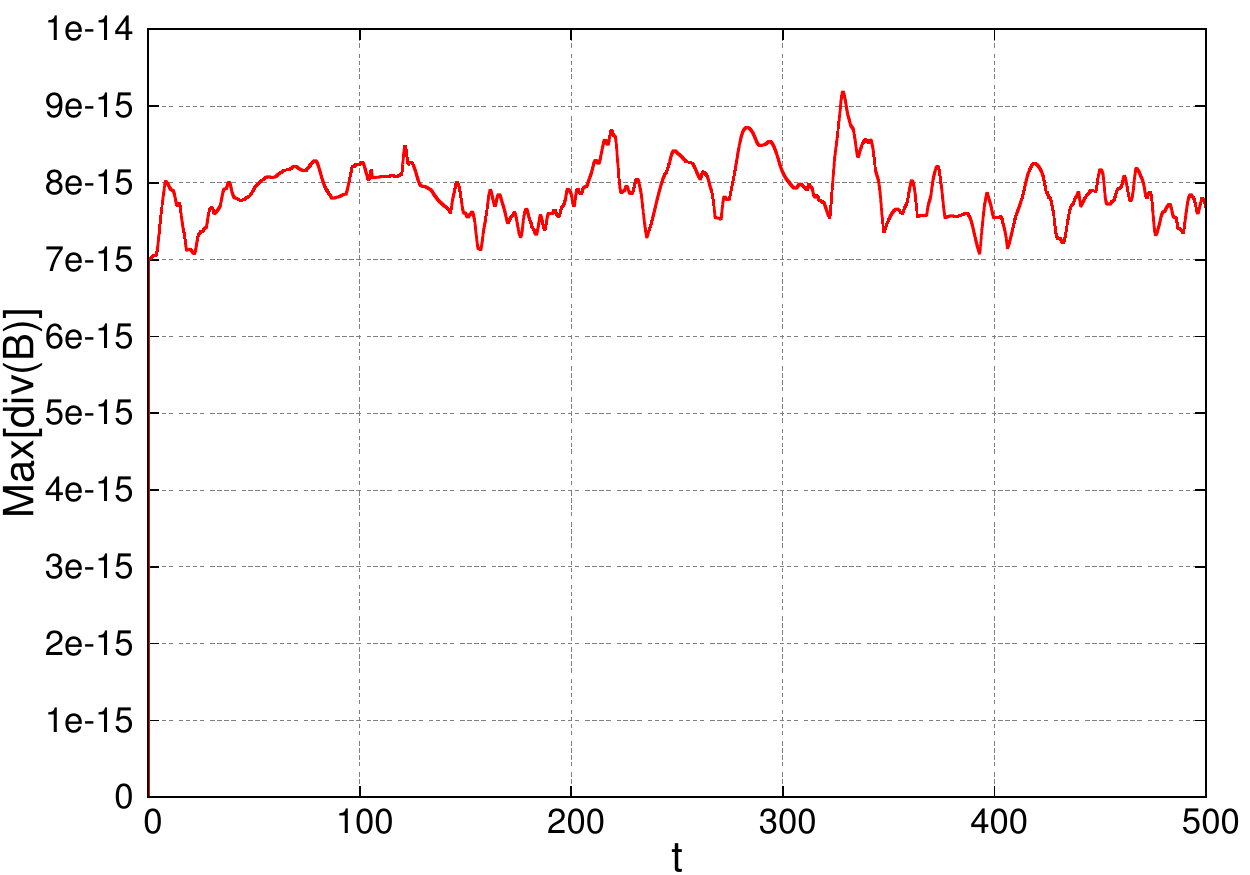}
\caption{Maximum of the magnetic field divergence, div($B$)$=\partial_{i}(\sqrt{\gamma}B^{i})$, during the evolution of the spherical accretion of a magnetically polarized fluid with $\chi_{m}=0.008$ in a magnetic field with $\beta_{EH}=25$. The free divergence condition for the magnetic field is satisfied numerically to machine precision during all the evolution.}
\label{div}
\end{center}
\end{figure}

As a final comment, as we mention in Sec. \ref{sec4},  CAFE preserves the free divergence constraint of the magnetic field to machine precision in all the simulations that we present in this work. We show in Figure \ref{div} the evolution in time of the maximum value of $\partial_{i}(\sqrt{\gamma}B^{i})$ for the hardest spherical accretion test, in which the fluid has a magnetic susceptibility of $\chi_{m}=0.008$ and the magnetic field is such that $\beta_{EH}=25$. In this figure we note, that after the first time step, the magnetic field divergence oscillates between $7\times 10^{-15}$ and $1\times 10^{-14}$.

\section{Concluding Remarks}
\label{conclutions}

In this paper we have presented for the first time the conservative form of the ideal GRMHD equations for a magnetically polarized fluid endowed with a magnetic field, and around a strong gravitational field. With the aim of solving numerically the last system of equations, we have also computed its eigenvalue structure by following the Anile procedure. We have found that in the particular case, where the magnetic polarization vector is in the same direction as the magnetic field, the speed of the material waves is independent of the magnetic susceptibility $\chi_m$, and is the same as in the usual GRMHD case. Nevertheless, the Alfv$\grave{\text{e}}$n eigenvalues and the bounds to the magnetosonic speeds depend on the magnetic susceptibility. The constitutive relation (\ref{constituvive_relation}), that we use to compute the eigenvalues, allows us to study the role of the diamagnetic and paramagnetic fluids in astrophysical scenarios. Now, in order to compute the primitive variables from the conservative ones, we have generalized the primitive variable recovery proposed by \cite{2006MNRAS.368.1040M} to include the magnetic polarization of the material. 

With the new numerical and theoretical results implemented in the CAFE code, we have carried out the first 1D shock tubes simulations with magnetically polarized fluids in the Minkowski spacetime. When the magnetic susceptibility is zero, the numerical solutions are the same as those obtained in the usual GRMHD tests (\cite{2006JFM...562..223G, 2015ApJS..218...24L}), but when $\chi_m\neq 0$, we obtained significant differences. For instance, we found that the propagation speed of the fastest waves in the solutions is greater in diamagnetic materials that in paramagnetic ones. This behaviour is interesting because it is independent of the initial configuration of the problem. Another remarkable results is that the magnetic susceptibility considerably increases the relativistic character of the flows in some problems, such as the Komissarov collision or the Balsara 2. Moreover, in the Balsara 2 and the Balsara 3 tests we noticed that the magnetic polarization can reverse the direction of the flows as compared with the solutions where $\chi_m=0$. Additionally, we also found that the total pressure gradient across the waves can be reversed due to the paramagnetic character of the fluids. this behaviour clearly appear for instance in the Balsara 4 and the Generic Alfv$\grave{\text{e}}$n tests. All these differences between the case with $\chi_m=0$ and the cases with $\chi_m\neq 0$ are more evident when the magnetic pressure dominates over the gas pressure. Additionally, it is important to mention that the ability of the code for dealing with magnetically polarized fluids is reduced with increasingly $|\chi_m|$, in particular, it is more difficult to numerically evolve paramagnetic fluids.

On the other hand, with the aim of testing the code in the strong gravitational field regime, we have presented for the first time the magnetized Michel accretion of a magnetically polarized fluid. We have showed that with the constitutive relation (\ref{constituvive_relation}), the Michel solution is not affected, so we can use it to test the code. Now, we have only considered paramagnetic fluids because it is more difficult to maintain in time the stationary solution when $\chi_m>0$. From the simulations, we have found that the solution is effectively maintained over time ($t > 4000$), and that the global convergence of the code is 2 for $\chi_{m}\lesssim 0.005$ and for all the magnetic field strength $\beta$ that we considered. Nevertheless, when $\chi_m=0.008$ and $\beta\geq 10$, the global convergence of the code is reduced to a value between first and second order. This results are interesting because even the strong magnetic fields in the magnetars are within $\beta\leq 4$ (\cite{2001ApJ...552L..35Z}), and the typical values for the magnetic susceptibility on the materials are around $10^{-5}$.\\

\section*{Acknowledgments}

O. M. P. wants to thanks the financial support from COLCIENCIAS under the program Becas Doctorados Nacionales 647 and Universidad Industrial de Santander. F.D.L-C and G. A. G. were supported in part by VIE-UIS, under Grant No. 2314 and by COLCIENCIAS, Colombia, under Grant No. 8863.

\appendix
\section{Primitive Variable Recovery}
\label{AppendixA}

In order to write down the primitive variables $\rho$, $v^{i}$, $p$, and $B^{k}$, in terms of the conservative ones $D$, $S_{j}$, $\tau$, and $B^{k}$, we need to solve the $5\times 5$ algebraic system (\ref{restmass_euler}-\ref{energy_euler}). Unfortunately, the GRMHD with magnetically polarized matter shares the same feature as the GRMHD: it is not possible to find $\vec{\mathcal{W}}(\vec{U})$ in a closed form (\cite{2006ApJ...641..626N}). In this paper we follow the method proposed by \cite{2006MNRAS.368.1040M} to reduce the five non-linear equations to only one by setting $Z=\rho hW^{2}$. Considering the constitutive relation (\ref{constituvive_relation}), we compute the scalar $S^{2}=\gamma^{ij}S_{i}S_{j}$ from the equation (\ref{momentum_euler}). The resulting expression takes the form,
\begin{equation}
S^{2}=(Z+\tilde{\chi}B^{2})^{2}(1-W^{-2})-\tilde{\chi}(2Z+\tilde{\chi}B^{2})\left(\frac{\vec{B}\cdot\vec{S}}{Z}\right)^{2},
\label{s2}
\end{equation} 
where we have defined $\tilde{\chi}=1-\chi$. The conservative variable $\tau$ can also be written in terms of $Z$ as
\begin{equation}
\tau=Z+\tilde{\chi}B^{2}-p-D+\frac{1-2\tilde{\chi}}{2W^{2}}B^{2}+\frac{1-2\tilde{\chi}}{2}\left(\frac{\vec{B}\cdot\vec{S}}{Z}\right)^{2}.
\label{tau}
\end{equation}
In the last two expressions we have used the fact that $\alpha^{2}(b^{0})^{2}=W^{2}(\vec{B}\cdot\vec{S})^{2}/Z^{2}$. Additionally, with the ideal gas equation of state, $p=\rho\epsilon(\Gamma-1)$, we can replace the thermodynamic pressure in (\ref{tau}) by
\begin{equation}
p=\frac{Z-WD}{W^{2}}\frac{\Gamma-1}{\Gamma}.
\label{pres}
\end{equation}    
Therefore, we have reduced the $5\times 5$ system of equations to a pair of equations for the unknowns $Z$ and $W$. Nevertheless, from (\ref{s2}), we can find the following expression for the Lorentz factor in terms of $Z$ and the conservative variables
\begin{equation}
W(Z)=\left[1-\frac{(\vec{B}\cdot\vec{S})^{2}\tilde{\chi}(2Z+\tilde{\chi}B^{2})+S^{2}Z^{2}}{(Z+\tilde{\chi}B^{2})^{2}Z^{2}}\right]^{-1/2}.
\label{lorentz}
\end{equation}
In this way, the equation (\ref{tau}), with $p$ and $W$ given in (\ref{pres}) and (\ref{lorentz}), respectively, becomes a non-linear equation for the unknown $Z$. To solve this equation, CAFE uses a Newton-Raphson algorithm together with the bisection method for finding the roots of the function
\begin{equation}
f(Z)=Z+\tilde{\chi}B^{2}-(D+\tau)-p+\frac{1-2\tilde{\chi}}{2W^{2}}B^{2}+\frac{1-2\tilde{\chi}}{2}\left(\frac{\vec{B}\cdot\vec{S}}{Z}\right)^{2}=0.
\label{f}
\end{equation}
Those roots correspond to the values of $Z$. 

Once $Z$ has been obtained, it is possible to compute the Lorentz factor with the equation ($\ref{lorentz}$), and therefore the rest mass density is obtained as $\rho=D/W$. Finally, the thermodynamic pressure is computed from the equation of state (\ref{pres}), and the components of the spatial velocity, $v^{i}$ through the equation
\begin{equation}
v^{i}=\frac{S^{i}+\tilde{\chi}(\vec{B}\cdot\vec{S})B^{i}/Z}{Z+\tilde{\chi}B^{2}},
\label{vi}
\end{equation}
which is obtained from (\ref{momentum_euler}). Note that when the magnetic susceptibility is zero {\em i.e} when $\tilde{\chi}=1$, all the expressions in this Appendix reduce to those of the GRMHD (\cite{2007CQGra..24S.235G}). 



\end{document}